\documentclass[twocolumn,aps,superscriptaddress,prb,longbibliography]{revtex4-1}
\setcitestyle{numbers,square}
\usepackage{amsmath,amssymb}
\usepackage{tabularx}
\usepackage{graphicx}
\usepackage{bmpsize}
\usepackage{bm}
\usepackage{color}
\usepackage{amssymb}
\usepackage[version=3]{mhchem}
\usepackage{newtxmath}
\usepackage{physics}
\usepackage{hyperref}
\usepackage{mathtools}
\usepackage{xcolor}

\DeclareMathAlphabet{\mathpzc}{OT1}{pzc}{m}{it}
\newcommand{\nn}{\nonumber}

\newcommand{\red}[1]{\textcolor{red}{#1}}

\newcommand{\kawakami}[1]{\textcolor{purple}{ #1}}

\begin{document}
\title{
Electromagnetic response in dipole superfluids: vortex lattices and singular domain walls
}
\author{Kazuki Yamamoto}
\affiliation{Department of Physics, Osaka University, Toyonaka, Osaka 560-0043, Japan}
\author{Takuto Kawakami}
\affiliation{Department of Physics, Osaka University, Toyonaka, Osaka 560-0043, Japan}
\author{Mikito Koshino}
\affiliation{Department of Physics, Osaka University, Toyonaka, Osaka 560-0043, Japan}
\date{\today}

\begin{abstract}
Among the most significant macroscopic quantum phenomena in condensed matter physics is the Meissner effect observed in superconductivity, which arises from the unique interaction between superfluids of charged particles and electromagnetic fields. However, superfluids can also emerge from particles possessing distinct electromagnetic properties. In particular, there has been growing interest in superfluids composed of charge-neutral particles with magnetic or electric dipole moments, such as Bose-Einstein condensates of magnons or excitons. Despite this interest, the electromagnetic response of dipole superfluids, including potential analogs or contrasts to the Meissner effect, remains poorly understood.
In this work, we develop a Ginzburg-Landau phenomenological theory to describe magnetic and electric dipole superfluids subjected to pseudo-magnetic fields induced by geometric phases. For magnetic dipole superfluids interacting with the Aharonov-Casher (AC) phase, we find that they form vortex lattices with sharply localized pseudo-magnetic fields along hexagonal domain walls, leading to singular and discontinuous change of physical variables at these boundaries. For electric dipole superfluids influenced by the He-McKellar-Wilkens (HMW) phase, in contrast, we identify vortex lattices where the pseudo-magnetic field and supercurrent are concentrated at vortex cores, resembling superconductors. These results reveal strikingly different electromagnetic responses in dipole superfluids, opening new directions for exploring superfluid systems with unconventional electromagnetic responses.

\end{abstract}
\maketitle
\section{Introduction}
\begin{figure}
    \centering
    \includegraphics[width=85mm]
    {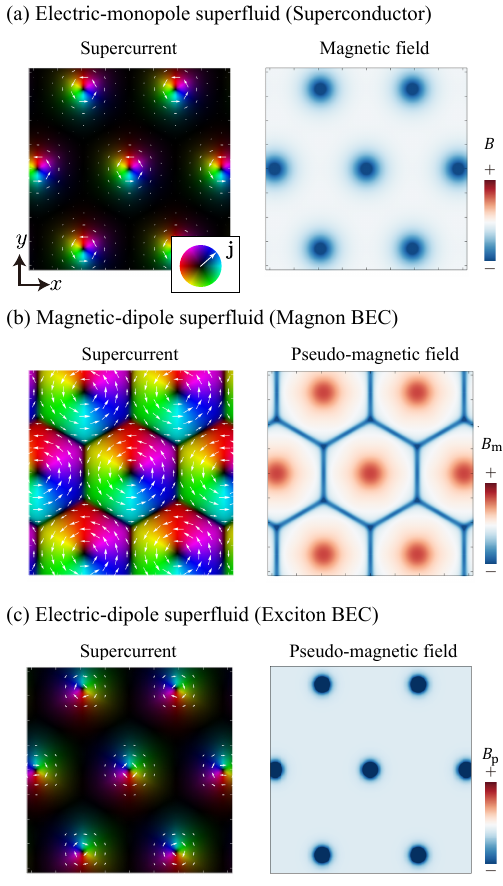}
    \caption{
    Vortex lattices in (a) a type-II superconductor (electric-monopole superfluid) under a magnetic field, (b) a magnetic-dipole superfluid under a pseudo-magnetic field, and (c) an electric-dipole superfluid under a pseudo-magnetic field.
    In all cases, we apply an external (pseudo) magnetic field
    in the negative $z$ direction. 
    The left panel plots the monopole supercurrent (not including negative charge) in (a), the magnetic-dipole supercurrent in (b), and the electric-dipole supercurrent in (c)
    with arrows and colors indicating the magnitude and direction. 
    The right panel depicts the distribution of the real magnetic field in (a), and the pseudo-magnetic field in (b)and (c).
    }
    \label{figure:comparison}
\end{figure}

When a magnetic field is applied to a superconductor, an electric supercurrent flows to expel the external magnetic field, a phenomenon known as the Meissner effect. 
This arises from a unique interaction between the condensate of charged particles and the electromagnetic field.
However, condensates can also form from particles with different electromagnetic properties. Of particular interest is a superfluid composed of charge-neutral particles that possess a magnetic dipole moment. 
Recent attention has focused on such systems, including spin-triplet excitonic insulators ~\cite{Ikeda2023,Jiang2020,Sun2013,Yang2024,Nishida2019,Wang2019,Liu2012,Sethi2021,Sethi2023,Wei2012} and magnon Bose-Einstein condensates (BECs)~\cite{Demokritov2006,Bozhko2016,Bozhko2019,Olsson2020,Divinskiy2021,Nikuni2000,Ruegg2003,Giamarchi2008,Aczel2009,Sonin2010,Zapf2014,Yuan2018,Esaki2024,Kimura2016,Kimura2017,Kimura2020,Sakurai2020}.

In general, magnetic dipoles experience a geometrical phase known as the Aharonov-Casher (AC) phase~\cite{Aharanov1984}, analogous to the Aharonov-Bohm (AB) phase observed for charged particles in magnetic fields. 
In the AC effect, an electric field acts as an effective vector potential, and hence spatial variations in the electric field generate a pseudo-magnetic field~\cite{Meier2003,Nakata2017,Nakata2017_Landu_level,Wang2024,SuWang2017}, which plays a role analogous to the real magnetic field in the AB effect. 
The AC effect has been observed in a wide range of systems, 
including neutrons~\cite{Wagh1996}, atomic systems~\cite{Sangster1995}, mesoscopic systems~\cite{Avishai2019,Avishai2023_review,Konig2006,Shekhter2022,Yamane2022,Balatsky1993}, Josephson vortices~\cite{Grosfeld2011,Elion1993,Seidov2021}, and spin waves~\cite{Zhang2014,Serha2023}.
In particular, the AC effect serves as a fundamental basis for spin-current-driven electric polarization, enabling multiferroicity in magnetic materials.~\cite{Katsura2005,Liu2011,Go2024,Hirsch1999,Sun2004,Tokura2014,Solovyev2021}

This raises a natural question: When magnetic dipoles condense into a superfluid, how do they respond to a pseudo-magnetic field (a spatially varying electric field) induced by the AC effect?
The interaction between magnetic-dipole superfluids and the AC phase has been explored from various theoretical perspectives~\cite{Nakata2014,Nakata2021,Wu2022,Wang2013,Bao2013,Sun2011}.
In this work, we propose a Ginzburg-Landau phenomenological theory to describe dipole superfluids in external electromagnetic fields, revealing behavior that starkly contrasts with that of superconductors. We find that, under a pseudo-magnetic field, a dipole superfluid spontaneously forms a vortex lattice similar to a type-II superconductor, though the field texture is nearly the inverse of what is observed in conventional type-II superconductors.
Figure \ref{figure:comparison} compares the vortex lattices in (a) a type-II superconductor (electric-monopole superfluid) under a magnetic field and (b) a magnetic-dipole superfluid under a pseudo-magnetic field, which constitutes the main result of this work. 
In each figure, the left panel illustrates the distribution of the supercurrent, and the right panel depicts the distribution of the real magnetic field in (a) and the pseudo-magnetic field in (b).
In the superconductor [Fig.~\ref{figure:comparison}(a)], 
the supercurrent circulates only in the vicinity of the vortices, where the magnetic field is concentrated. 
In contrast, in the dipole superfluid, the circulation of the supercurrent extends to the hexagonal boundaries separating the vortices, where the supercurrent abruptly changes direction and pseudo-magnetic field is sharply localized. These boundaries are singular planes for physical variables: supercurrent, electric polarization, order parameter derivatives are discontinuous, and pseudo-magnetic field and electric charge density diverge.
The formation of these singular walls occurs abruptly when the coupling strength between the dipole superfluid and the electromagnetic field exceeds a specific critical value.

Finally, we investigate superfluids of {\it electric} dipoles, where electric and magnetic roles are reversed. Representative physical systems are exciton BECs~\cite{Jerome1967,Kohn1967,Butov1994,Khveshchenko2001,Eisenstein2004,Cercellier2007,Min2008,Anshul2017,Li2017,WangZefang2019,Luis2019,Ma2021,Huang2024,Balatsky2004,Dubinkin2021,Jiang2015}, where excitons are charge-neutral and possess an electric-dipole moment. When a magnetic field is applied to the electric-dipole, it experiences an geometric phase, called He-McKellar-Wilkens (HMW) phase~\cite{HeMcKellar1993,Wilkens1994}, that is the electric-dipole counterpart of AC phase in the magnetic-dipoles. We formulate the Ginzburg-Landau theory and obtain vortex lattices in an analogous way to the case of magnetic-dipole superfluids. Interestingly, we find that the behaviors are significantly different from those in magnetic-dipole superluids, where the supercurrent and the pseudo-magnetic field is concentrated only at vortex cores as illustrated in Fig.~\ref{figure:comparison}(c),
resembling the behavior observed in superconductors.


The paper is organized as follows.
In Sec.\ref{sec: general theory of spin SF}, we develop the phenomenological theory of magnetic-dipole superfluid and derive the self-consistent equations.
In Sec.\ref{sec:single}, we explore the single vortex state by numerically solving these equations, showing that the vortex state becomes thermodynamically stable above a critical pseudo-magnetic field.
In Sec.\ref{sec:vortex lattice}, we investigate the vortex lattice state where single vortices are periodically arranged.
In Sec.~\ref{sec:electric_dipole}, we extend the formulation to electric-dipole superfluids interacting with a magnetic field, following a parallel approach to that of magnetic-dipole superfluids. 
In Sec.~\ref{sec:discussion}, we compare the corresponding theories of magnetic and electric dipole superfluids, as well as monopole superfluids (superconductors), and elucidate the origins of their different behaviors.
Finally, a brief conclusion is presented in Sec.~\ref{sec:conclusion}.


\section{Phenominological theory of magnetic-dipole superfluid}\label{sec: general theory of spin SF}

In this section, 
we construct a phenomenological theory 
for the magnetic-dipole  superfluid including the AC phase.
We assume that bosons with a constant magnetic moment $\vb*{\mu}=\mu\vb{e}_z$ form a Bose-Einstein condensate,
where $\vb{e}_z$ is a unit vector along the direction of the magnetic dipole moment.
The considered situation applies to a spin-triplet excitonic insulator and a magnon Bose-Einstein condensate, with $\mu = g\mu_B$, where $g \simeq 2$ is the Lande g-factor and $\mu_B$ is the Bohr magneton.

For the sake of comparison, we also present the phenomenological Ginzburg-Landau theory for conventional superconductors (electric-monopole superfluid) in Appendix \ref{sec:glsc}
to be directly compared with the following formulation for magnetic-dipole superfluids.

\subsection{Aharonov-Casher phase}
\label{sec:AC}
\begin{figure}
    \centering
    \includegraphics
    {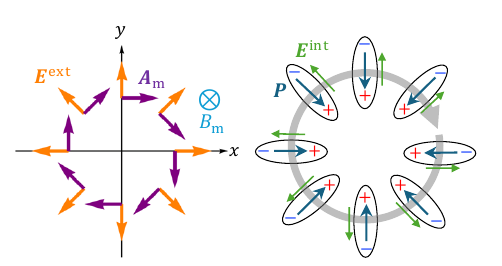}
    \caption{
Schematic figures of externally applied electric field (left) and the electromagnetic response 
   of magnetic dipoles  (right).
    An external electric field $\vb{E}^\mathrm{int}$ induce effective vector potential $\vb{A}_\mathrm{m}$ of the AC phase, resulting negative pseudo-magnetic field $B_\mathrm{m}<0$.
    The magnetic dipole experiences Lorentz force from the pseudo-magnetic field and perform cyclotron motion (circular arrow in the right panel). The moving magnetic dipole then induces electric polarization $\vb{P}$, and generate additional electric field $\vb{E}^\mathrm{int}$, showing positive feedback effect $\vb{E}^\mathrm{ext}\vdot\vb{E}^\mathrm{int}>0$.
    }
    \label{figure:classical_AC}
\end{figure}

When a magnetic dipole moment $\vb*{\mu}=\mu\vb{e}_z$  moves in an electric field $\vb{E}$, it acquires the Aharonov-Casher (AC) phase \cite{Aharanov1984},
\begin{equation}
    \theta_{\mathrm{AC}}=-\frac{1}{\hbar}\int \vb{A}_\mathrm{m}\vdot d\vb{r},\label{equation:AC phase}
\end{equation}
where an effective vector potential is defined by
\begin{equation}
\vb{A}_{\rm{m}}=g^{}_{\rm AC}\vb{E}\cross\vb{e}_z\label{equation:Am}.
\end{equation}
The constant $g^{}_{\rm AC}(>0)$ represents the strength of the spin-orbit coupling, and depends on the detail of the material. In vacuum, the $g^{}_{\rm AC}$ is $\mu/c^2$, where $c$ is the speed of light.
A simple derivation of the AC phase is presented in Appendix \ref{sec:app_AC}.
For magnons in a typical magnetic insulator, $g^{}_{\rm AC}$ greatly enhances and is typically $10^6$ times as large as in vacuum~\cite{Katsura2005,Liu2011}.

Being perpendicular to $\vb{e}_z$ [Eq.~\eqref{equation:Am}], $\vb{A}_\mathrm{m}$  becomes a two dimensional vector with $A_\mathrm{m}^x$ and $A_\mathrm{m}^y$ components. 
The corresponding pseudo-magnetic field along $z$-direction is then written as
\begin{equation}
    B_\mathrm{m}=
    (\grad \times \vb{A}_\mathrm{m})_z = -g^{}_{\rm AC} \qty(\pdv{E^x}{x}+\pdv{E^y}{y}).
    \label{equation:Bm}
\end{equation}
Unlike the usual vector potential in AB phase, the effective vector potential $\vb{A}_\mathrm{m}$ does not have a gauge degrees of freedom, since it is directly expressed by the electric field.

For instance, an electric field distribution 
\begin{equation}
   \vb{E}=\frac{\mathcal{E}}{2}(x,y,-2z) 
\label{eq_E_exp}
\end{equation}
gives $\vb{A}_\mathrm{m}= 
(-g^{}_{\rm AC}\mathcal{E}/2)r\,\vb{e}_\phi$
in the cylindrical coordinate $(r,\phi,z)$,
resulting in a uniform pseudo-magnetic field $B_\mathrm{m}=-g^{}_{\rm AC}\mathcal{E}$.
Such an electric field distribution can be realized by an appropriate arrangement of external electric charges. For instance, if we put a pair of point charges $+Q$ at $\bm{r}_\pm=(0,0,\pm l)$, the desired electric field with $\mathcal{E}\approx Q/(2\pi\varepsilon_0 l^3)$ is obtained near the origin.

\subsection{Positive feedback effect}\label{sec:positive feedback}
When a magnetic-dipole system is placed in an external electric field that induces a pseudo-magnetic field, the system responds with positive feedback, amplifying the applied electric field.
Here we explain this effect through a qualitative argument based on the classical mechanics.

We consider a situation where a magnetic-dipole polarized to $z$ axis is placed in an external electric field $\bm{E}^{\rm ext}$
given by Eq.~\eqref{eq_E_exp}.
$\bm{E}^{\rm ext}$ is radially distributed in the $xy$-plane 
as shown in Fig.~\ref{figure:classical_AC}.
It induces an effective vector potential in the azimuthal direction,
resulting in a negative uniform pseudo-magnetic field $B_\mathrm{m}<0$ in $z$ direction.
According to the discussion in the appendix \ref{sec:app_AC}, the Hamiltonian of a magnetic dipole is expressed by
$H=(\vb{p}+\vb{A}_\mathrm{m})^2/(2m^*)$,
where $m^*$ and $\vb{p}$ represent the mass and the momentum of a magnetic dipole.
Consequently, the dipole experiences a Lorentz force $\vb{F}=-\vb{v}\times\vb{B}_\mathrm{m}$ due to the pseudo-magnetic field 
and undergoes cyclotron motion~\cite{Meier2003,Nakata2017,Nakata2017_Landu_level}, as depicted by the gray arrow in Fig.~\ref{figure:classical_AC}.

Generally, the motion of the magnetic dipole induces an electric dipole $\bm{p} = g_\mathrm{AC}^{}\vb{v}\times\vb{e}_\mathrm{z}$ perpendicular to its velocity,
as a reciprocal phenomenon to the AC  effect ~\cite{Katsura2005,Liu2011,Go2024,Hirsch1999,Sun2004,Tokura2014}
(see Appendix \ref{sec:app_AC} for a brief derivation).
When multiple magnetic dipoles exhibit the same circular motion,
the macroscopic electric polarization $\bm{P}$ (the density of $\bm{p}$)
arises in the same direction,
and it generates an additional electric field $\vb{E}^\mathrm{int} = -\bm{P}/\varepsilon_0$.
As shown in Fig.~\ref{figure:classical_AC},
$\bm{P}$ and $\bm{E}^{\rm ext}$ 
are oppositely oriented unlike in usual dielectric materials, indicating a negative dielectric constant.
As a result, the induced electric field $\vb{E}^\mathrm{int}$ aligns with $\vb{E}^\mathrm{ext}$, i.e., the system exhibits a positive feedback effect ($\vb{E}^\mathrm{ext}\vdot\vb{E}^\mathrm{int}>0$).
In the following sections, we demonstrate that
the positive feedback effect plays a crucial role in the magnetic-dipole superfluid.

\subsection{Ginzburg-Landau theory}
\label{sec:GP}

We consider charge neutral Bose particles with a magnetic moment $\vb*{\mu}=\mu\vb{e}_z$ forming a Bose-Einstein condensate.
Utilizing an analogy to the Ginzburg-Landau theory of superconductivity or rotating atomic BEC, the free energy $F_\mathrm{GL}$ of the magnetic-dipole superfluid is given by
\begin{align}
    F_\mathrm{GL}[\Psi,\vb{A}_\mathrm{m}] &=\int d^3\vb{r}
    \Big[
    \frac{1}{2m^*}\big|(-i\hbar\grad+\vb{A}_{\rm{m}}(\vb{r}))\Psi(\vb{r})\big|^2
    \nonumber\\
    &\hspace{2.3cm}
    -\alpha\abs{\Psi(\vb{r})}^2+\frac{\beta}{2}\abs{\Psi(\vb{r})}^4\Big],
    \label{equation:GP_Hamiltonian}
\end{align}
where $m^*$ is the effective mass,
and $\Psi(\vb{r})=|\Psi(\vb{r})|\exp \qty(i\theta(\vb{r}))$ is the order parameter, or the macroscopic wavefunction of the magnetic-dipole superfluid. We assume $\alpha>0$ and $\beta>0$ for the stability of the BEC.
The superfluid satisfies the Ginzburg-Landau (GL) equation
\begin{equation}
    \fdv{F_\mathrm{GL}}{\Psi^*}=\frac{1}{2m^*}\qty(-i\hbar\grad+\vb{A}_{\rm{m}})^2\Psi
    -\alpha\Psi+\beta\abs{\Psi}^2\Psi=0.
    \label{equation:GL_equation}
\end{equation}



The first term in the integral of Eq.~\eqref{equation:GP_Hamiltonian}
is the kinetic energy density, and it should correspond to
$n^*(m^*v^{*2}/2)$ where $n^* = \abs{\Psi}^2$ is the the superfluid density
and $v^*$ is the velocity.
Then, the associated supercurrent density, $\vb{j}_{\rm{m}} = n^* v^*$, is written as
\begin{equation}
    \vb{j}_{\rm{m}}
    =\frac{\hbar}{m^*}\abs{\Psi}^2\qty(\grad\theta+\frac{1}{\hbar}\vb{A}_{\rm{m}}) 
    =\fdv{F_\mathrm{GL}}{\vb{A}_\mathrm{m}}.\label{equation:jm}
\end{equation}
Eq.~\eqref{equation:jm} has an equivalent form to a conventional superconductor under mangnetic field, except that the vector potential is replaced with the effective vector potential $\vb{A}_\mathrm{m}$ for the AC effect.
The supercurrent flows in the direction of the effective vector potential $\vb{A}_\mathrm{m}$, which is perpendicular to both the electric field $\vb{E}$ and the magnetic dipole moment $\parallel\vb{e}_z$.

As argued in the previous section, 
a current of magnetic-dipole moment induces an electric polarization 
~\cite{Katsura2005,Liu2011,Go2024,Hirsch1999,Sun2004,Tokura2014,Solovyev2021}. In the present case,
the polarization $\vb{P}$ (density of electric dipole moment) 
induced by the magnetic-dipole supercurrent $\vb{j}_{\rm{m}}$ is given by
\begin{equation}
 \vb{P}=g^{}_{\rm AC}\,\vb{j}_{\rm{m}}\cross \vb{e}_z.\label{equation:P}
\end{equation}
A brief derivation of this relation is presented in Appendix \ref{sec:app_AC}.
By definition, $\vb{P}$ does not have a $z$-component.
The electric field can then be expressed as
\begin{equation}\label{equation:etot}
\vb{E}=\vb{E}^{\mathrm{ext}} - \frac{\vb{P}}{\varepsilon_0},
\end{equation}
where $\vb{E}^{\mathrm{ext}}$ represents the external electric field contributed from electric charges outside the system, and $\varepsilon_0$ is the permittivity of free space. 
For simplicity, we assume the polarization of the superfluid is entirely generated by the supercurrent, ignoring the polarization of normal state.
In standard terms of electromagnetism,
$\vb{E}^{\mathrm{ext}}$ corresonds to the 
the electric displacement field $\vb{D}=\varepsilon_0 \vb{E}^{\mathrm{ext}}$,
and Eq.~\eqref{equation:etot} is written as
$\vb{E} = (\vb{D}-\vb{P})/\varepsilon_0$.

From Eqs.~\eqref{equation:Am},\eqref{equation:P}, and \eqref{equation:etot}, the effective vector potential is written as
\begin{equation}\label{equation:atot}
\vb{A}_{\mathrm{m}}=\vb{A}^{\mathrm{ext}}_{\mathrm{m}}+\frac{g_{\rm AC}^2}{\varepsilon_{0}}\vb{j}^{\parallel}_{\mathrm{m}},
\end{equation}
where $\vb{A}^{\mathrm{ext}}_{\mathrm{m}}=g^{}_{\rm AC}\vb{E}_{\mathrm{ext}}\times \vb{e}_z$
and $\vb{j}^{\parallel}_{\mathrm{m}}$ is the $xy$-component of $\vb{j}_{\mathrm{m}}$.
The corresponding pseudo-magnetic field $B_\mathrm{m}=(\grad\times\vb{A}_{\mathrm{m}})_z$ is given by
\begin{align}
   {B}_\mathrm{m} 
 &= B^{\mathrm{ext}}_{\mathrm{m}} + \frac{g_{\rm AC}^2}{\varepsilon_0}
 (\grad\times\vb{j}^{\parallel}_{\mathrm{m}})_z \nonumber\\
 & = B^{\mathrm{ext}}_{\mathrm{m}} + \frac{g^{}_{\rm AC}}{\varepsilon_0}\rho, 
    \label{equation:Bm_int}
\end{align}
where 
$\rho=-\div\vb{P}$ is the polarization charge density.
When we consider the response of the system to an external electric field $\vb{E}^{\rm ext}$, 
we solve the GL equation [Eq.~\eqref{equation:GL_equation}] self-consistently with Eq~\eqref{equation:jm} and \eqref{equation:atot}.

The same set of equations can also be derived by variation of the total Helmholtz free energy including the electrostatic field. 
For the magnetic-dipole superfluid, it is given by
\begin{equation}~\label{equation:free energy}
    F[\vb{D}(\vb{r}), T] = F_\mathrm{GL} + \int d\vb{r}\,\,\frac{\vb{D}(\vb{r})^2-\vb{P}(\vb{r})^2}{2\varepsilon_0},
\end{equation}
as a functional of the external field $\vb{D}(\vb{r})=\varepsilon_0 \vb{E}^{\mathrm{ext}}(\vb{r})$.
Equation \eqref{equation:free energy} is one of the principal results of our paper.
It can be derived by considering 
a quasi-static process where the external electric field is slowly introduced to the dipole superfluid. 
The detailed calculation is provided in Appendix \ref{sec:free_energy_magnetic_dipole}.
It is straightforward to check that 
Eqs.~\eqref{equation:GL_equation} and \eqref{equation:atot} are obtained by the equilibrium conditions $\delta F/\delta\Psi=0$ and $\delta F/\delta\vb{A}_\mathrm{m}=0$, respectively (See, Appendix \ref{sec:free_energy_magnetic_dipole}).


\subsection{Dimensionless equations}
\label{sec:dimless}

To highlight the essential scales of the system and relevant parameters, we rewrite the equations in a dimensionless form. 
Similar to the standard Ginzburg-Landau framework for superconductors, we normalize the position $\vb{r}$ as $\Tilde{\vb{r}}=\vb{r}/\xi$ and the spatial derivative as $\Tilde{\grad}=\xi\grad$, where
\begin{equation}
    \xi=\sqrt{\frac{\hbar^2}{2m^*\alpha}}
\end{equation}
is the coherence length of the superfluid.
The order parameter is normalized as 
$\Tilde{\Psi}=\Psi/\Psi_\infty$ 
in units of a uniform BEC value $\Psi_\infty=\sqrt{\alpha/\beta}$, the effective vector potential as $\Tilde{\vb{A}}_\mathrm{m}=\vb{A}_\mathrm{m}/(\hbar/\xi)$, 
and energy functional as $\Tilde{F}_{\mathrm{GL}}=F_\mathrm{GL}/(\alpha^2\xi^3/\beta)$.
As a result, the superfluid free energy [Eq.~\eqref{equation:GP_Hamiltonian}] is expressed in a dimensionless form as
\begin{equation}
    \Tilde{F}_\mathrm{GL}=\int d^3\Tilde{r}\,|(\Tilde{\grad}+i\Tilde{\vb{A}}_{\rm{m}})\Tilde{\Psi}|^2-|\Tilde{\Psi}|^2+\frac{1}{2}|\Tilde{\Psi}|^4.
    \label{equation:dimensionless H}
\end{equation}
Eq.~\eqref{equation:GL_equation} becomes
\begin{equation}
    -(\Tilde{\grad}+i\Tilde{\vb{A}}_{\rm{m}})^2\Tilde{\Psi}-\Tilde{\Psi}+|\Tilde{\Psi}|^2\Tilde{\Psi}=0.\label{equation:dimensionless GL eq}
\end{equation}

The supercurrent [Eq.~\eqref{equation:jm}], polarization [Eq.~\eqref{equation:P}], and polarization charge density are normalized as  
\begin{align}
\Tilde{\vb{j}}_\mathrm{m} &=
|\Tilde{\Psi}|^2 (\Tilde{\grad}\theta+\Tilde{\vb{A}}_{\rm{m}}), 
\label{equation:dimensionless jm} \\
\Tilde{\vb{P}} &=\Tilde{\vb{j}}_{\mathrm{m}}\times \vb{e}_z, \label{equation:dimensionless P}\\
\Tilde{\rho} &=-\Tilde{\vb{\grad}}\cdot \Tilde{\vb{P}},
\end{align}
where $\Tilde{\vb{j}}_{\mathrm{m}}=\vb{j}_{\mathrm{m}}/j_0$,~$\Tilde{\vb{P}}=\vb{P}/(g^{}_{\rm AC} j_0)$ and $\Tilde{\rho}=\rho/(g^{}_{\rm AC} j_0/\xi)$ with $j_0=\hbar\Psi_{\infty}^2/(m^\ast \xi)$.
The effective vector potential, Eq.~\eqref{equation:atot}, becomes
\begin{equation}
\Tilde{\vb{A}}_{\mathrm{m}}=\Tilde{\vb{A}}_{\mathrm{m}}^{\mathrm{ext}}
+\eta \, \Tilde{\vb{j}}_{\mathrm{m}} \label{equation:dimensionless Am}
\end{equation}
with 
\begin{equation}\label{equation:eta}
    \eta=\frac{g_{\rm AC}^2\Psi_\infty^2}{m^*\varepsilon_0}.
\end{equation}
The parameter $\eta$ is a dimensionless quantity that characterizes the strength of the electrostatic coupling in the superfluid.
The pseudo-magnetic field [Eq.~\eqref{equation:Bm_int}] is written as 
\begin{align}
\Tilde{B}_{\mathrm{m}} & = \Tilde{B}^{\mathrm{ext}}_{\mathrm{m}} + \eta (\Tilde{\vb{\grad}}\times\Tilde{\vb{j}}_{\mathrm{m}})_z \nonumber \\
& =\Tilde{B}^{\mathrm{ext}}_{\mathrm{m}}-\eta \Tilde{\rho}, \label{equation:dimensionless Bm} 
\end{align}
where $\Tilde{B}_{\mathrm{m}}=B_{\mathrm{m}}/(\hbar{\xi}^2)$.
In this unit, the pseudo-magnetic flux penetrating a given area becomes equal to the AC phase integrated along the boundary,
and hence the dimensionless flux quanta is  given by $\Tilde{\phi}^0_m=2\pi$.

Finally, the total free energy of Eq.\eqref{equation:free energy} is normalized as
\begin{equation}
\Tilde{F}= \Tilde{F}_\mathrm{GL}[\tilde{\Psi},\Tilde{\vb{A}}_\mathrm{m}] + \eta\int d^3\Tilde{r}~\big(\Tilde{\vb{D}}^2-\Tilde{\vb{P}}^2\big),
\end{equation}
where $\Tilde{F}=F/(\xi^3\alpha^2/\beta)$, $\Tilde{\vb{D}}=\vb{D}/(g_\mathrm{AC}^{}j_0)$ and $\Tilde{\vb{E}}=\vb{E}/(g_\mathrm{AC}^{}j_0/\varepsilon_0)$.

Equations ~\eqref{equation:dimensionless GL eq}, ~\eqref{equation:dimensionless jm} and \eqref{equation:dimensionless Am} are a set of the self-consistent equations. 
The latter two equations can be readily solved for $\Tilde{\vb{j}}_\mathrm{m}$ and $\Tilde{\vb{A}}_{\rm{m}}$ as
\begin{align}
&    \Tilde{\vb{j}}_\mathrm{m} =
\frac{|\Tilde{\Psi}|^2}{1-\eta |\Tilde{\Psi}|^2}
(\Tilde{\grad}\theta+\Tilde{\vb{A}}^\mathrm{ext}_{\rm{m}})
\label{equation:jm_self_consistent},
\\
& \Tilde{\vb{A}}_\mathrm{m} =
\frac{1}{1-\eta |\Tilde{\Psi}|^2}
(\eta|\Tilde{\Psi}|^2
\Tilde{\grad}\theta+\Tilde{\vb{A}}^\mathrm{ext}_{\rm{m}}).
\label{equation:Am_self_consistent}
\end{align}
An important observation is that the presence of $\eta \,(>0)$ enhances the supercurrent and the vector potential by decreasing the denominator $1-\eta |\Tilde{\Psi}|^2$.
This is the positive feedback effect 
argued in Sec.~\ref{sec:positive feedback}.
The equations also imply a singular behavior when the denominator vanishes, i.e., $\eta |\Tilde{\Psi}|^2 = 1$.
This will be discussed in the Sec.~\ref{sec:single}.

\begin{figure*}[t]
    \centering
    \includegraphics[width=0.95\linewidth]{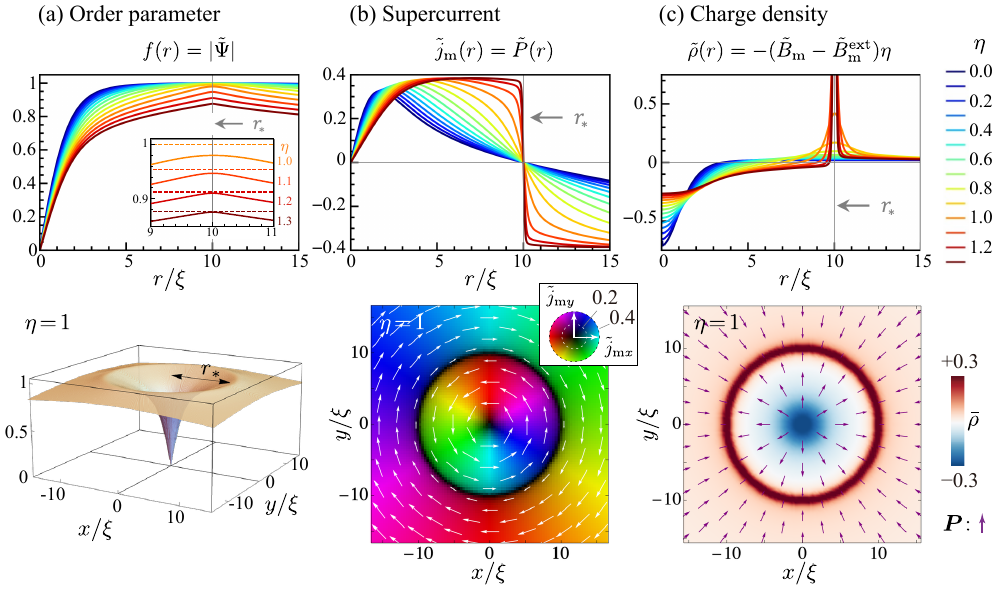}
    \caption{
Self-consistent solution of a magnetic-dipole superfluid with a single vortex of $n=1$ under $\Tilde{B}_\mathrm{m}^\mathrm{ext}=-0.02 (r_*=10)$.
In top panels, we plot
(a) absolute order parameter $f(r)=|\tilde{\Psi}(r)|$,
(b) azimuthal component of the supercurrent $\Tilde{j}_\mathrm{m}(r)$ (equal to the radial electric polarization, $\Tilde{P}(r)$),
and (c) polarization charge density $\Tilde{\rho}(r)$ (equal to the internal pseudo-magnetic field),
where different colors correspond to different $\eta$'s. 
Inset of (a) is a magnified plot near $r_*=10$,
where dashed lines indicate the value of $\eta^{-1/2}$.
The lower panels display the two-dimensional plots 
of  (a) $f(r)$, (b) $\Tilde{\vb{j}}_\mathrm{m}(r)$ and (c) $\Tilde{\rho}(r)$ on the $xy$ plane, at the parameter of $\eta=1$. The arrows and colors in (b) indicate the direction and magnitude of the supercurrent $\Tilde{\vb{j}}_\mathrm{m}(r)$.
In (c), the arrows represent the polarization $\Tilde{\vb{P}}(r)$.
    }
    \label{figure:single vortex}
\end{figure*}

\begin{figure}[t]
    \centering
    \includegraphics[width=1.\linewidth]{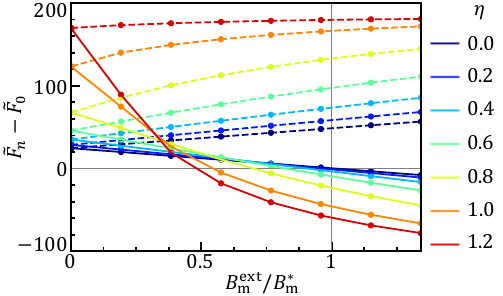}
    \caption{
 Relative total free energy of vortex states $n=1$ (solid curves) and $n=-1$ (dashed curves) compared to the uniform $n=0$ state, as a function of the external pseudo-magnetic field $B_\mathrm{m}^\mathrm{ext}$.
    Colors represent the value of $\eta$. 
    }
    \label{figure:Bm dependence}
\end{figure}

\section{Single-vortex state}\label{sec:single}

\begin{figure*}[t]
    \centering
    \includegraphics[width=0.9\linewidth]
    {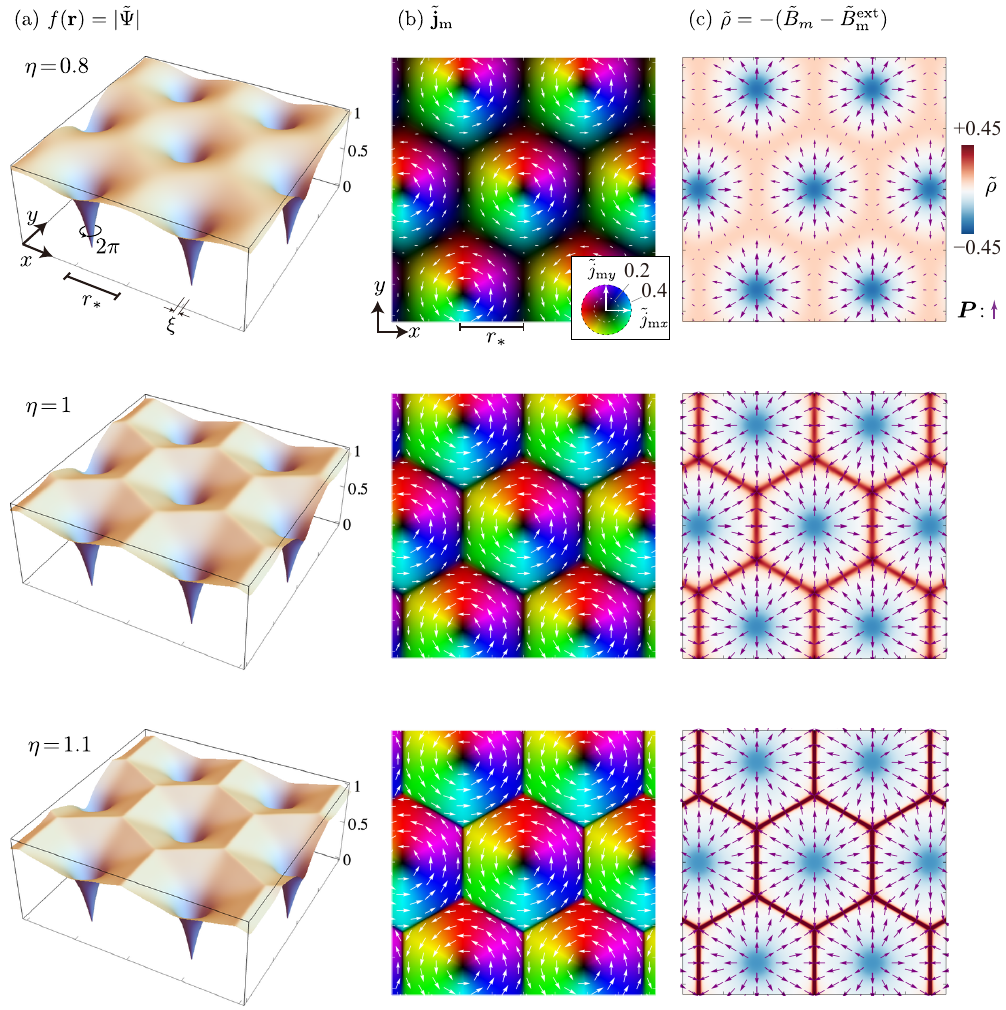}
    \caption{Self-consistent vortex lattice of a magnetic-dipole superfluid for (a) $\eta=0.8$, (b) 1 and (c)  $1.1$, in $\Tilde{B}_\mathrm{m}^\mathrm{ext}=-0.02$.
    In each figure, the left, middle and right columns
present spatial maps of the absolute order parameter $f(\vb{r})=|\tilde{\Psi}(\vb{r})|$, the supercurrent $\Tilde{\vb{j}}_\mathrm{m}(\vb{r})$, and the charge density $\Tilde{\rho}(\vb{r})\,(=-\eta \Tilde{B}^\mathrm{int}_\mathrm{m}(\vb{r})$), respectively.
%
    }
    \label{figure:vortex lattice}
\end{figure*}


In the following, we consider an infinitely-large magnetic-dipole superfluid under an external field of $\Tilde{\vb{A}}_{\rm{m}}^\mathrm{ext}=(-|\Tilde{B}_\mathrm{m}^\mathrm{ext}|r/2)\vb{e}_\phi$, 
which gives a uniform pseudo-magnetic field $B_\mathrm{m}^\mathrm{ext}$ (see, Sec. \ref{sec:AC}).
Here we set $\Tilde{B}_\mathrm{m}^\mathrm{ext}=-0.02$ (along $-z$ direction).
We assume a superfluid solution of the form 
$\Tilde{\Psi}(r,\phi,z)=f(r)e^{in\phi}$
with a single vortex of $n=+1$ along the $z$-axis.
We solve the set of self-consistent equations 
[Eqs.~\eqref{equation:GL_equation}, \eqref{equation:jm} and \eqref{equation:atot}] 
by a numerical iteration (See Appendix \ref{sec:technical_detail} for detail).
The results for different $\eta$ parameters are summarized in Fig.~\ref{figure:single vortex}.
Here we plot
(a) the absolute order parameter $f(r)=|\tilde{\Psi}(r)|$,
(b) the azimuthal component of the supercurrent $\Tilde{j}_\mathrm{m}(r)$, which is equal to the radial component of electric polarization, $\Tilde{P}(r)$,
and (c) polarization charge density $\Tilde{\rho}(r)$, which is related to the internal pseudo-magnetic field $\Tilde{B}^{\rm int}_{\rm m}=\Tilde{B}_{\rm m}-\Tilde{B}^{\rm ext}_{\rm m}$ by 
$\Tilde{\rho}= -\Tilde{B}^{\rm int}_{\rm m}/\eta$.
The lower panels display the two-dimensional plots of the distribution of
$|\Tilde{\Psi}|$,
$\Tilde{\vb{j}}_\mathrm{m}$, 
$\Tilde{\rho}$ on the $xy$ plane at $\eta=1$.
In the plot of $\Tilde{\vb{j}}_\mathrm{m}$, arrows and colors indicate the direction and magnitude of the supercurrent.

First, we focus on $\eta=0$ (blue curve in Fig.~\ref{figure:single vortex}), where the electrostatic feedback is absent.
In this case, the supercurrent [Eq.~\eqref{equation:jm_self_consistent}] is given by
$\Tilde{\vb{j}}_\mathrm{m} =
|\Tilde{\Psi}|^2(\Tilde{\grad}\theta+\Tilde{\vb{A}}^\mathrm{ext}_{\rm{m}})$ where
$\Tilde{\grad}\theta =(1/r)\vb{e}_\phi$. Its azimuthal component then becomes
$\Tilde{j}_\mathrm{m}(r) = f(r)^2(1/r - |\Tilde{B}_\mathrm{m}^\mathrm{ext}|r/2)$, which vanishes at the radius $r_*\equiv \sqrt{2/|\Tilde{B}_\mathrm{m}^\mathrm{ext}|}$
($=10$ in the present case).
The $r_*$ is the radius such that the pseudo-magnetic flux penetrating the region $r<r_*$ 
is equal to a quantum flux $\phi^0_{\rm m}$.

As the parameter $\eta$ is increased from zero, we observe that
the vanishing point of  $\Tilde{\vb{j}}_\mathrm{m}$ still remains at $r=r_*$, while its slope at $r=r_*$ becomes gradually steeper, as seen in Fig.~\ref{figure:single vortex}(b).
This is expected from  Eq.~\eqref{equation:jm_self_consistent} where the supercurrent amplitude is enhanced by
the denominator $1-\eta |\Tilde{\Psi}|^2$.
In $\eta\gtrsim 1$, interestingly, $\Tilde{\vb{j}}_\mathrm{m}(r)$ becomes a step-like function, i.e., the direction of the azimuthal supercurrent abruptly changes at $r=r_*$. 
Consequently, the pseudo-magnetic field exhibits a sharp peak according to the relation $B_{\rm m}-B_{\rm m}^{\rm ext} \propto \grad \times \vb{j}_\mathrm{m}$ [Eq.~\eqref{equation:dimensionless Bm}], which is clearly observed in Fig.~\ref{figure:single vortex}(c).
In terms of electric quantities, this corresponds to a high magnitude of $\grad \cdot \vb{P}$,
i.e., a sharply-localized polarization charge density at $r=r_*$.

The step-like behavior of $\Tilde{j}_\mathrm{m}(r)$ in $\eta \gtrsim 1$ can be explained by vanishing denominator
$1-\eta |\Tilde{\Psi}|^2$  in Eq.~\eqref{equation:jm_self_consistent}.
In the present case, Eq.~\eqref{equation:jm_self_consistent}
is written as
\begin{equation}
    \Tilde{j}_\mathrm{m}(r)=
    \frac{1}{\frac{1}{f(r)^2}-\eta}
    \qty(\frac{1}{r}-\frac{\abs{B_\mathrm{m}^\mathrm{ext}}}{2}r).\label{equation:jm phase transition}
\end{equation}
In the numerical solutions presented in  Fig.~\ref{figure:single vortex}, the $f(r)$ of $\eta \gtrsim 1$ exhibits a kink structure in the vicinity of $r_*$, which is well described by 
\begin{equation}
  f(r)=
  \eta^{-1/2} -a|r-r_*|
  \quad (a>0).\label{equation:approximate f}
\end{equation}
Then the denominator of Eq.~\eqref{equation:jm phase transition} is written as
$\frac{1}{f(r)^2}-\eta \approx 2\eta^{3/2}a|r-r_*|$ in the lowest order in $r-r_*$.
Noting the bracket in Eq.~\eqref{equation:jm phase transition} can be approximated by 
 $(-2/r_*^2)(r-r_*)$, we end up with a step function
\begin{equation}
    \Tilde{j}_\mathrm{m}(r)=
    \frac{1}{\eta^{3/2}a r_*^2}\mathrm{sgn}(r_*-r),
\end{equation}
where $\mathrm{sgn}(x) \equiv x/|x|$.
Here we note that the maximum of $f(r)$ is $\eta^{-1/2}$
[Eq.~\eqref{equation:approximate f}], i.e.,
the $f(r)$ is tuned in such a way that the denominator of Eq.~\eqref{equation:jm phase transition} vanishes right at $r=r_*$. This cancels with the original zero of the supercurrent, resulting in a step function.
In the actual numerical result, the kink of $f(r)$ at $r=r_*$
is rounded so that the maximum is slightly lower than $\eta^{-1/2}$ 
as observed in the inset of Fig.~\ref{figure:single vortex}(a),
which prevents the denominator from completely vanishing. Consequently, the step structure in $\Tilde{j}_\mathrm{m}$ is smoothed into a continuous, yet rapidly-changing function.


It should also be noted that the total flux inside the radius $r=r^*$ remains quantized even when the internal field $B_{\rm m}^{\rm int}=B_{\rm m}-B_{\rm m}^{\rm ext}$ is included. This can be understood by the following consideration:
By applying the Stokes theorem to Eq.~\eqref{equation:dimensionless Bm}, the flux of the internal field in $r<r_*$ becomes
$\int_{r<r_*}B_{\rm m}^{\rm int} d^2 r \propto \oint_{r=r_*} \vb{j}_{\rm m} \cdot d \vb{r}$,
which vanishes since $\vb{j}_{\rm m} =0$ at $r=r_*$.
Therefore, the total flux inside $r_*$ is contributed solely by the external field $B_{\rm m}^{\rm ext}$, which is quantized by definition.
Zero flux of $B_{\rm m}^{\rm int}$ inside the radius $r_*$
is equivalent to a zero net charge within the region.



To demonstrate the energetic advantage of the vortex state in pseudo-magnetic fields, we investigate the total free energy $F$ with and without a vortex.
Here we consider a finite system of the disc with radius $R=40$, and obtain the self-consistent solutions with no vortex ($n=0$), a single vortex ($n=1$), a single anti-vortex ($n=-1$) at the center of the disc.
For the obtained solutions, we calculate the total free energy of the system $F$ given by Eq.~\eqref{equation:free energy}.

Figure \ref{figure:Bm dependence} plots the relative total free energy of $n=1$ (solid curves) and $n=-1$ (dashed curves) compared to the $n=0$ state, as a function of the external pseudo-magnetic field $B_\mathrm{m}^\mathrm{ext}$.
The different colors correspond to different $\eta$.
Let us first consider the case of $\eta=0$ (blue curve), which corresponds to a standard superfluid without the electrostatic feedback effect. 
At $\Tilde{B}_\mathrm{m}^\mathrm{ext}=0$,
the uniform solution ($n=0$) is the ground state, while
the vortex states ($n=\pm1$) are degenerate in higher energy.
When $\Tilde{B}_\mathrm{m}^\mathrm{ext}$ is increased, the relative energy of the $n=1$ state decreases, and it finally becomes the lowest at $B=B_\mathrm{m}^*(R) \equiv (1+2\log R)/(1+R^2)$.
The critical field $B_\mathrm{m}^*(R)$ can be regarded as a pseudo-magnetic field at which the system of size $R$ allows a single vortex to enter.
The anti-vortex state ($n=-1$) is always higher in energy as naturally expected.


When we increase $\eta$, the critical field at which the $n=1$ state becomes stable monotonically decreases, i.e., the $n=1$ vortex state is stabilized in a smaller field.
It is stark contrast to the typical behaviors of superconductivity,
where
the critical field $H_{c1}$ increases as the coupling to the electromagnetic field increases~\cite{tinkhambook}.
This corresponds to a transition from type II regime with the formation of the vortices to the type I regime where the Meissner effect occurs.



\section{Vortex lattice}\label{sec:vortex lattice}

When the pseudo-magnetic field is much greater than $B_\mathrm{m}^*(R)$, we expect that multiple vortices enter the superfluid, forming of a vortex lattice.
Here we demonstrate the emergence of a vortex lattice in a dipole superfluid by numerically solving Eqs.~\eqref{equation:dimensionless GL eq}, \eqref{equation:dimensionless jm} and~\eqref{equation:dimensionless Am} self-consistently. 
We assume a triangular lattice with primitive lattice vectors $\vb{a}_1=a(1,0)$ and $\vb{a}_2=a(1/2,\sqrt{3}/2)$
where $a$ is the lattice constant.
The lattice constant $a$ is determined by the requirement that the unit cell contains exactly a quantum flux $2\pi$, giving $a^2=2\pi r_*^2/\sqrt{3}$. 

We assume a periodic current profile $\vb{j}_{\mathrm{m}} (\vb{r}+\vb{a}_i) = \vb{j}_{\mathrm{m}}(\vb{r})$, which ensures $\vb{A}^{\mathrm{int}}_{\mathrm{m}}(\vb{r}) = \vb{A}^{\mathrm{int}}_{\mathrm{m}}(\vb{r}+\vb{a}_i)$ by Eq.~\eqref{equation:dimensionless Am}. 
Since the periodic vector potential can only give zero flux on average, the total pseudo-magnetic flux is contributed solely by the external field $B_{\rm m}^\mathrm{ext}$.
Therefore, we can assume that the total vector potential, including the external field, obeys the periodicity,
\begin{equation}
\vb{A}_{\mathrm{m}}(\vb{r}+\vb{L}) = \vb{A}_{\mathrm{m}}(\vb{r}) + \frac{B^{\mathrm{ext}}_{\mathrm{m}}}{2}\vb{e}_z\times\vb{L}
\label{equation:periodicAm}
\end{equation}
where $\vb{L}(l_1,l_2)=l_1 \vb{a}_1+l_2 \vb{a}_2$ is a lattice translation vector with integer $l_i$. 
The order parameter then satisfies the magnetic Bloch condition,  
\begin{equation}
\Psi(\vb{r}+\vb{L}) = \Psi(\vb{r}) \exp\left(-i\frac{B^{\mathrm{ext}}_{\mathrm{m}}}{2}(\vb{e}_z\times\vb{L})\cdot \vb{r} + i\pi (l_1+l_2+l_1l_2) \right). \label{equation:periodicpsi}
\end{equation}
The single-valuedness of the order parameter is guaranteed when unit cell spanned by $\vb{a}_1$ and $\vb{a}_2$ accommodates an integer mutiple of flux quanta.
We solve the self-consistent equation by a numerical iteration with the periodic boundary condition of Eqs.\eqref{equation:periodicAm} and \eqref{equation:periodicpsi}, starting from an initial state with a single vortex per unit cell.

Figure \ref{figure:vortex lattice} summarizes the obtained superfluid vortex lattice for (a) $\eta=0.8$, (b) 1 and (c)  $1.1$, in $\Tilde{B}_\mathrm{m}^\mathrm{ext}=-0.02$.
In each figure, the left, middle and right columns
present spatial maps of the absolute order parameter $f(\vb{r})$, the supercurrent $\Tilde{\vb{j}}_\mathrm{m}(\vb{r})$, and the charge density $\Tilde{\rho}(\vb{r})\,(=-\eta \Tilde{B}^\mathrm{int}_\mathrm{m}(\vb{r})$), respectively.
We observe that the single vortex structure argued in the previous section are periodically arranged in a consistent manner. The supercurrent $\vb{j}_{\rm m}$ circulates in the counterclockwise direction in each cell [Fig.~\ref{figure:vortex lattice}(b)],
or equivalently, the polarization $\vb{P}$ is arranged radially with respect to the vortex.
When $\eta$ exceeds 1, significantly, the supercurrent distribution extends upto the edge of the honeycomb unit cell, and its direction abruptly changes across the domain wall  as in the single vortex case. As a result, the polarization charge (i.e., the internal pseudo-magnetic field) is sharply concentrated at the domain wall, forming a bold honeycomb domain-wall network as seen in Fig.~\ref{figure:vortex lattice}(c).
We also have negative polarization charge localized around the vortex cores, which exactly cancels with those on the domain wall, resulting in a zero net charge.
Equivalently, the integral of $B^\mathrm{int}$ over a unit cell is vanishing, so that the total pseudo-magnetic flux remains per unit cell equal to that of the external field $B_\mathrm{m}^\mathrm{ext}$.

We have also systematically performed the same self-consistent calculations for different angles ranging from 60$^\circ$ (triangular lattice) to 90$^\circ$ (square lattice).
We find that the triangular vortex lattice always has the lowest free energy for all values of $\eta$.


\section{Electric-dipole superfluids}
\label{sec:electric_dipole}

The above argument for magnetic-dipole superfluids can be directly extended to electric dipoles, where the roles of electric and magnetic fields reversed.
Such a situation can be realized in excitonic condensates, where excitons (bound states of an electron and a hole) are charge-neutral and possess an electric dipole moment \cite{Jerome1967,Kohn1967,Butov1994,Khveshchenko2001,Eisenstein2004,Cercellier2007,Min2008,Anshul2017,Li2017,WangZefang2019,Luis2019,Ma2021,Huang2024,Balatsky2004,Dubinkin2021,Jiang2015}.
In the following, we present a phenomenological theory for electric-dipole superfluids interacting with a magnetic field, paralleling the discussion of magnetic-dipole superfluids in the previous sections.
However, the electric-dipole superfluids show distinct behavior, owing to its negative feedback effect, while the magnetic-dipole superfluids show positive feedback effect. 
\subsection{He-McKellar-Wilkens phase}
While the magnetic dipole interacts with the electric field through the AC phase, the electric dipole $\vb{p} = p \vb{e}_z$ interacts with the magnetic flux density $\vb{B}$ by means of the He-McKellar-Wilkens (HMW) phase \cite{HeMcKellar1993,Wilkens1994,Wei1995,Dowling1999,Lepoutre2012,Chen2013,Wood2016,Jiang2015,Dubinkin2021},
\begin{equation}
    \theta_{\mathrm{HMW}}=\frac{1}{\hbar}\int \vb{A}_\mathrm{p}\vdot d\vb{r},\label{equation:HMW phase}
\end{equation}
where $\vb{A}_\mathrm{p}$ is the effective vector potential given by
\begin{equation}
    \vb{A}_\mathrm{p} = g_{\mathrm{HMW}}^{} \,
    \vb{B}\times\vb{e}_z,\label{equation:Ap}
\end{equation}
where $g_{\mathrm{HMW}}^{} = p$ for vaccum.
A simple derivation of HMW phase in presented in Appendix~\ref{sec:app_HMW}.
The corresponding pseudo-magnetic field along $z$-direction is then written as
\begin{equation}
    B_\mathrm{p}=
    (\grad \times \vb{A}_\mathrm{p})_z 
    = -g_{\mathrm{HMW}}^{} \, \qty(\pdv{B^x}{x}+\pdv{B^y}{y}).
    \label{equation:Bp}
\end{equation}
Here the pseudo-magnetic field is given by a spatially-dependent magnetic field.
In the following, $\vb{B}$ and $\vb{B}^{\rm ext}$ represent real magnetic fields, not pseudo-magnetic fields.
\subsection{Negative feedback effect}\label{sec:negative feedback}

\begin{figure}
    \centering
    \includegraphics
    {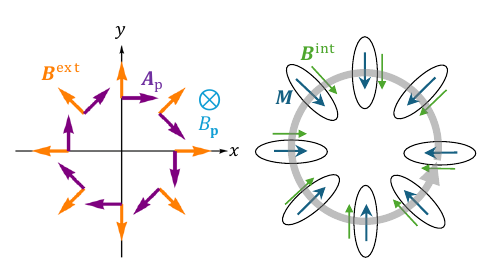}
    \caption{
    Schematic figures of externally applied magnetic field (left) and the electromagnetic response 
   of electric dipoles  (right),
   which correspond to Fig.~\ref{figure:classical_AC} for magnetic dipoles.
    An external magnetic field $\vb{B}^\mathrm{ext}$ (orange) induce effective vector potential $\vb{A}_\mathrm{p}$ (purple) of HMW phase, resulting negative pseudo-magnetic field $B_\mathrm{p}<0$.
    The electric dipole experiences Lorentz force from the pseudo-magnetic field and perform cyclotron motion (gray). The moving electric dipole induces a magnetization $\vb{M}$, and they generate additional magnetic field $\vb{B}^\mathrm{int}$, showing negative feedback effect $\vb{B}^\mathrm{ext}\vdot\vb{B}^\mathrm{int}<0$.
    }
    \label{figure:classical_HMW}
\end{figure}

In contrast to the magnetic dipole case, 
the electric dipole exhibits the negative feedback against the external field.
Figure \ref{figure:classical_HMW} shows the schematic figure 
illustrating the response of electric dipoles,
corresponding to Fig.~\ref{figure:classical_AC} for magnetic dipoles.
Here we apply radially-distributed magnetic field $\bm{B}^{\rm ext}$, which gives the pseudo-magnetic field  $\bm{B}_p$ in the $-z$ direction.
According to the Hamiltonian 
$H=(\vb{p}-\vb{A}_\mathrm{p})^2/(2m^*)$ (Appendix \ref{sec:app_HMW}),
the electric-dipole experiences a Lorentz force $\vb{F}=+\vb{v}\times\vb{B}_\mathrm{p}$ due to the pseudo-magnetic field, leading to the cyclotron motion.
The motion of the electric dipoles generates a magnetic-dipole $\bm{\mu} = -g_\mathrm{HMW}^{}\vb{v}\times\vb{e}_\mathrm{z}$
(see Appendix \ref{sec:app_HMW} for brief explanation),
giving magnetization $\bm{M}$ (the density of $\bm{\mu}$).
$\bm{M}$ and $\bm{B}_{\rm ext}$ are oppositely oriented,
indicating a diamagnetic response. This aligns with the negative dielectric response in the magnetic dipole case (Sec.\ref{sec:positive feedback}).
The magnetization contributes to the additional magnetic field
$\vb{B}^\mathrm{int} = \mu_0 \bm{M}$.
Since $\vb{B}^\mathrm{ext}\vdot\vb{B}^\mathrm{int}<0$, the feedback is negative.

This negative feedback effect contrasts with the positive feedback effect in the magnetic-dipole system [Sec.~\ref{sec:positive feedback}].
The opposite features are due to the fact that 
the electric field in the magnetic-dipole case corresponds to $H$-field (not $B$-field) in the electric-dipole case,
and that generally $H$-field and $B$-field induced by magnetic dipoles are oppositely directed (e.g., in magnetic materials).
Indeed, if we replace  $+$ and $-$ charges of induced electric dipole in Fig.~\ref{figure:classical_AC}
with $N$ and $S$ magnetic monopoles, we reproduce the situation of Fig.~\ref{figure:classical_HMW}, where the induced $H$-field is oriented outward, similar to $\vb{E}^\mathrm{int}$ in Fig.~\ref{figure:classical_AC}. Therefore the induced $B$-field points inward.

\subsection{Ginzburg-Landau theory}
We assume that the Ginzburg-Landau free energy functional for electric-dipole superfluids is given by 
\begin{align}
F_{\mathrm{GL}}[\Psi,\vb{A}_\mathrm{p}] = \int d^3 r \Big[\frac{1}{2m^*}\big|(-i\hbar\vb{\grad}-\vb{A}_\mathrm{p})\Psi(\vb{r})\big|^2 \nonumber\\ 
- \alpha |\Psi(\vb{r})|^2 + \frac{\beta}{2} |\Psi(\vb{r})|^4 \Big],
  \label{equation:GP_Hamiltonian_electric_dipole}
\end{align}
where $m^*$ is the effective mass of a quasiparticle having an electric dipole
and $\Psi(\vb{r})$ is the order parameter.
The Ginzburg-Landau equation is written as
\begin{equation}\label{equation:gl_HMW}
    \frac{\delta F_{\mathrm{GL}}}{\delta\Psi^\ast}
    = \frac{1}{2m^*}(-i\hbar\grad-\vb{A}_\mathrm{p})^2\Psi - \alpha\Psi+\beta|\Psi|^2\Psi=0.
\end{equation}
The supercurrent is expressed as 
\begin{equation}\label{equation:jp}
    \vb{j}_{\mathrm{p}}
    =\frac{\hbar}{m^*}|\Psi|^2(\grad\theta-\frac{1}{\hbar}\vb{A}_\mathrm{p})
    =-\frac{\delta F_{\mathrm{GL}}}{\delta \vb{A}_\mathrm{p}}.
\end{equation}

Just as the current of a magnetic dipole induces electric polarization, 
the current of electric dipole $\vb{j}_\mathrm{p}$ induces the magnetization by
\begin{equation} \vb{M}=g_{\mathrm{HMW}}^{} \, \vb{e}_z\times\vb{j}_{\mathrm{p}}.\label{equation:M}
\end{equation}
A brief derivation of this relation is presented in Appendix~\ref{sec:app_HMW}.
The magnetic flux density can be expressed as
\begin{equation}
    \vb{B}=\vb{B}^\mathrm{ext}+\mu_0\vb{M}.\label{equation:btot}
\end{equation}
where $\vb{B}^{\mathrm{ext}}$ is the external magnetic flux density, and $\mu_0$ is the permeability. 
In the standard notation of electromagnetism, 
$\vb{B}^{\mathrm{ext}}$ corresponds to
the magnetic field $\vb{H}= \vb{B}^{\mathrm{ext}}/\mu_0$, and Eq.~\eqref{equation:btot} is written as $\vb{B} = \mu_0(\vb{H}+\vb{M})$.
From Eqs.~\eqref{equation:Ap}, \eqref{equation:M}, and \eqref{equation:btot}, the total magnetic field induces the effective vector potential 
\begin{equation}\label{equation:Ap_tot}
    \vb{A}_{\mathrm{p}}=\vb{A}^{\mathrm{ext}}_{\mathrm{p}}
    +\mu_0g_{\mathrm{HMW}}^2\,\vb{j}_{\mathrm{p}},
\end{equation}
with $\vb{A}^{\mathrm{ext}}_{\mathrm{p}}= g_{\mathrm{HMW}}^{} \,\vb{B}^{\mathrm{ext}}\times \vb{e}_z$.



Equations \eqref{equation:gl_HMW} and \eqref{equation:Ap_tot} can alternatively be derived by variation of the total free energy including the magnetic field.
In the present case, it is appropriate to use the Gibbs free energy $G[\vb{H}(\vb{r}),T]=F[\vb{B}(\vb{r}),T]-\int d\vb{r}~\vb{B}(\vb{r})\cdot\vb{H}(\vb{r})$ rather than the Helmholtz free energy $F[\vb{B}(\vb{r}),T]$, since the external magnetic field $\vb{H}$ is fixed.
According to the argument of Appendix \ref{sec:free_energy_electric_dipole}, the Gibbs free energy is written as
\begin{equation}
    G[\vb{H}(\vb{r}),T]=F_\mathrm{GL}+ \int d\vb{r}~ \qty(\frac{\vb{B}(\vb{r})^2}{2\mu_0}-\vb{B}(\vb{r})\cdot\vb{H}(\vb{r})).\label{equation:Gibbs}
\end{equation}
It is straightforward to check that 
Eqs.~\eqref{equation:gl_HMW} and \eqref{equation:Ap_tot} are obtained by the equilibrium conditions $\delta G/\delta\Psi=0$ and $\delta G/\delta\vb{A}_\mathrm{p}=0$, respectively
(See,  Appendix \ref{sec:free_energy_electric_dipole}).

Similar to the argument for magnetic-dipole systems, the equations \eqref{equation:jp} and \eqref{equation:Ap_tot} can be expressed in a dimensionless form as,
\begin{align}
\Tilde{\vb{j}}_\mathrm{p} =
|\Tilde{\Psi}|^2 (\Tilde{\nabla}\theta-\Tilde{\vb{A}}_{\rm{p}}),
\quad
\Tilde{\vb{A}}_{\mathrm{p}} =\Tilde{\vb{A}}_{\mathrm{p}}^{\mathrm{ext}}
+\eta_\mathrm{p} \, \Tilde{\vb{j}}_{\mathrm{p}}, \label{equation:dimensionless jp and Ap}
\end{align}
where the variables with tilde represent dimensionless quantities
and we defined a dimensionless parameter 
\begin{equation}
    \eta_\mathrm{p}=\frac{\mu_0g_\mathrm{HMW}^2\Psi_\infty^2}{m^*},
\end{equation}
which corresponds to $\eta$ for magnetic-dipole supserfluids.
The two equations Eqs.~\eqref{equation:dimensionless jp and Ap} can be solved for $\Tilde{\vb{j}}_\mathrm{p}$ and $\Tilde{\vb{A}}_{\rm{p}}$ as
\begin{align}
&    \Tilde{\vb{j}}_\mathrm{m} =
\frac{|\Tilde{\Psi}|^2}{1+\eta_\mathrm{p} |\Tilde{\Psi}|^2}
(\Tilde{\grad}\theta-\Tilde{\vb{A}}^\mathrm{ext}_{\rm{p}})
\label{equation:jp_self_consistent},
\\
& \Tilde{\vb{A}}_\mathrm{p} =
\frac{1}{1+\eta_\mathrm{p}|\Tilde{\Psi}|^2}
(\eta_\mathrm{p}|\Tilde{\Psi}|^2
\Tilde{\grad}\theta+\Tilde{\vb{A}}^\mathrm{ext}_{\rm{p}}).
\label{equation:Ap_self_consistent}
\end{align}
In contrast to the corresponding formula for the 
magnetic-dipole case, Eqs.~\eqref{equation:jm_self_consistent} and
\eqref{equation:Am_self_consistent},
$\eta_\mathrm{p}$ enters with positive sign in the denominator
$1+\eta_\mathrm{p} |\Tilde{\Psi}|^2$.
Therefore the $\eta_\mathrm{p}>0$ 
gives s a negative feedback reducing the applied field.
This is consistent with the qualitative argument in Sec.~\ref{sec:negative feedback}.

\subsection{Self-consistent vortex solutions}

\begin{figure*}[t]
    \centering
    \includegraphics[width=1\linewidth]{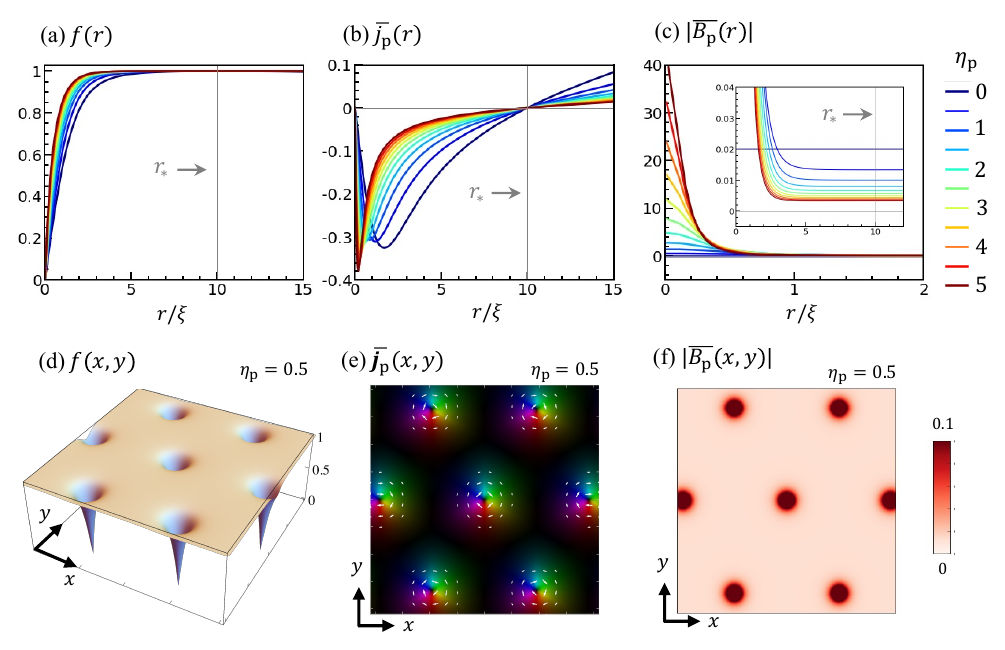}
    \caption{
In top panels, self-consistent solutions of an electric-dipole superfluid with a single vortex of $n=-1$ under $\Tilde{B}_\mathrm{p}^\mathrm{ext}=-0.02~(r_*=10)$ are summarized.
We plot
(a) absolute order parameter $f(r)=|\tilde{\Psi}(r)|$,
(b) azimuthal component of the supercurrent $\Tilde{j}_\mathrm{p}(r)$ (equal to the radial magnetization, $\Tilde{M}(r)$),
and (c) absolute value of total pseudo-magnetic field $|\Tilde{B_\mathrm{p}}(r)|$,
where different colors correspond to different $\eta_\mathrm{p}$'s. 
Inset of (c) is a magnified plot near $|\Tilde{B_\mathrm{p}}(r)|=0$.
In lower panels, self-consistent vortex lattice of an electric-dipole superfluid  under $\Tilde{B}_\mathrm{m}^\mathrm{ext}=-0.02~(r_*=10)$ are summarized.
We display the two-dimensional plots of  
of  (d) $f(x,y)=|\Psi(x,y)|$, (e) $\Tilde{\vb{j}}_\mathrm{p}(x,y)$ and (f) $|\Tilde{B_\mathrm{p}}(x,y)|$ on the $xy$ plane, at the parameter of $\eta=0.5$. The arrows and colors in (e) indicate the direction and magnitude of the supercurrent $\Tilde{\vb{j}}_\mathrm{p}(x,y)$.
    }
    \label{figure:electric-dipole-vortex}
\end{figure*}

When an electric-dipole superfluid is subjected to a pseudo-magnetic field $B_p$ (a spatially-dependent magnetic field), the system also forms a vortex lattice, while the electro-magnetic field texture is completely distinct from the case of magnetic-dipole superfluids due to the negative feedback effect.

Here, we perform analyses of a single vortex state and a vortex lattice in the electric-dipole superfluid, aligning with the corresponding calculations for the magnetic-dipole cases.~[Secs. \ref{sec:single} and \ref{sec:vortex lattice}].
First, we consider a single vortex state 
under an external field of $\Tilde{\vb{A}}_{\rm{p}}^\mathrm{ext}=(-|\Tilde{B}_\mathrm{p}^\mathrm{ext}|r/2)\vb{e}_\phi$ with $\Tilde{B}_\mathrm{p}^\mathrm{ext}=-0.02$.
We assume an anti-vortex state $\Tilde{\Psi}(r,\phi,z)=f(r)e^{-i\phi}$, where the negative sign in the phase factor corresponds to the negative sign of the effective vector potential in Eq.~\eqref{equation:jp}.
In Fig.~\eqref{figure:electric-dipole-vortex}, we plot
(a) the absolute value of the order parameter $f(r)=|\tilde{\Psi}(r)|$, (b) the azimuthal component of the supercurrent $\Tilde{j}_\mathrm{p}(r)$, which is equal to the radial component of magnetization, $\Tilde{M}(r)$,
and (c) the abolute value of pseudo-magnetic field $|\Tilde{B_\mathrm{p}}(r)|$,
for different $\eta_\mathrm{p}$ parameters.


\red{
}

As seen in Fig.~\ref{figure:electric-dipole-vortex}(b), the supercurrent is suppressed owing to the negative feedback effect 
(i.e. the denominator $1+\eta_\mathrm{p} |\Tilde{\Psi}|^2$ in Eq.~\eqref{equation:jp_self_consistent}).
This behavior is stark contrast to magnetic-dipole superfluids, where the supercurrent is amplified by the positive feedback effect.
Furthermore, as is shown in Fig.~\ref{figure:electric-dipole-vortex}(c), the pseudo-magnetic field is sharply concentrated at the vortex core, whereas away from the core ($r\gg \xi$), it converges to a nonzero constant, 
\begin{equation}
    \Tilde{B}_\mathrm{p}(r) \approx \frac{1}{1+\eta_\mathrm{p}}\Tilde{B}_\mathrm{p}^\mathrm{ext}\quad(r\gg \xi).
    \label{equation:weak Meissner}
\end{equation}
This equation is obtained by taking the curl of Eq.~\eqref{equation:Ap_self_consistent} with $|\tilde{\Psi}(r)|^2=1$.
It is worth noting that the total flux inside the radius $r^*\equiv\sqrt{2/|\Tilde{B}_\mathrm{p}^\mathrm{ext}|}(=10)$ is always quantized,  as demonstrated by the same reasoning presented in Sec.~\ref{sec:single}.

We also calculate vortex lattice states of electric-dipole superfluids, in the same manner as described in Sec.~\ref{sec:vortex lattice}.
In Fig.~\ref{figure:electric-dipole-vortex}, we show the plots of the distributions of
(d) $f(x,y)=|\Tilde{\Psi}(x,y)|$,
(e) $\Tilde{\vb{j}}_\mathrm{p}(x,y)$, 
(f) $|\Tilde{B_\mathrm{p}}(x,y)|$ on the $xy$ plane at $\eta=0.5$.
We observe that the single-vortex structures are periodically arranged in a consistent manner; the pseudo-magnetic field is primarily concentrated around the vortices and remains nearly constant away from them at the value given by Eq.~\eqref{equation:weak Meissner}.


\section{Discussion}\label{sec:discussion}
We have shown that magnetic and electric dipole superfluids under a pseudo-magnetic field form vortex lattices with distinctly different electromagnetic field textures. The magnetic-dipole system exhibits a positive feedback effect for the applied field, resulting in discontinuous domain walls that separate neighboring vortices. In contrast, the electric-dipole system shows a negative feedback effect, leading to a strong suppression of the pseudo-magnetic field away from the vortex core.
The latter behavior resembles the vortex lattice in type-II superconductors (electric-monopole superfluids), but with a significant difference: the magnetic field in superconductors decays exponentially away from the vortex, whereas the pseudo-magnetic field in electric-dipole superfluids converges to a constant value proportional to $\propto 1/(1+\eta_p)$.
In this section, we discuss the origins of these different behaviors among the three distinct classes of systems --- superconductors, magnetic-dipole superfluids, and electric-dipole superfluids --- through their key equations.



In the three cases,
the superfluid free energy $F_{\rm GL}$ [Eqs. \eqref{equation:GP_Hamiltonian},  \eqref{equation:GP_Hamiltonian_electric_dipole}, \eqref{equation:GP_Hamiltonian_superconductor}] 
and the supercurrent
[Eqs. \eqref{equation:jm}, \eqref{equation:jp}, \eqref{equation:sc_j}] 
share the same functional form with respect to the corresponding vector potential, i.e., $\vb{A}$ for superconductors, and $\vb{A}_\mathrm{m}$ ($\vb{A}_p$) for magnetic (electric)-dipole superfluid. 
Accordingly, the supercurrent and vector potential obey the same relationship.
When the phase $\theta$ is constant, in particular, 
we have the London-type equation
$\vb{j} \propto -\vb{A}$,
$\vb{j}_\mathrm{m} \propto \vb{A}_\mathrm{m}$, and
$\vb{j}_\mathrm{p} \propto -\vb{A}_\mathrm{p}$.

However, a difference lies in 
the relationship between the vector potential and the real magnetic/electric field:
\begin{align}
  &  \grad \times \vb{A} = \vb{B}
    \quad \mbox{(electric monopole),}
    \nonumber\\
   & \vb{A}_{\rm{m}}=g^{}_{\rm AC}\vb{E}\cross\vb{e}_z
    \quad \mbox{(magnetic dipole).}
    \nonumber\\
    & 
   \vb{A}_\mathrm{p}=g_{\mathrm{HMW}}^{} \, \vb{B}\times\vb{e}_z
   \quad \mbox{(electric dipole),}
    \label{equation:A_vs_B}
\end{align}
As an inverse effect, the relationship between the supercurrent and the induced magnetization / polarization is also different, as
\begin{align}
  &  \curl\vb{M} = \vb{j}
    \quad \mbox{(electric monopole),}
    \nonumber\\
   &  \vb{P}=g^{}_{\rm AC}\,\vb{j}_{\rm{m}}\cross \vb{e}_z
   \quad \mbox{(magnetic dipole),}
    \nonumber\\
    & 
   \vb{M}= - g_{\mathrm{HMW}}^{} \, \vb{j}_{\mathrm{p}}\times\vb{e}_z
   \quad \mbox{(electric dipole),}
   \label{equation:M_vs_j}
\end{align}
Note that Equations ~\eqref{equation:A_vs_B} and \eqref{equation:M_vs_j} represent different aspects of the same phenomenon. In fact,
Eq.~\eqref{equation:M_vs_j} can be derived from the variation of the free energy with the same form, under condition of Eq.~\eqref{equation:A_vs_B} (See, Appendix \ref{sec:toal_free_energy}).


In Eqs.~\eqref{equation:A_vs_B} and \eqref{equation:M_vs_j},
the crucial difference between monopoles and dipoles is in the existence/absence of the spatial derivative $\grad$.
For the monopole case, in particular, 
these relations
with London's equation $\vb{j} \propto \vb{A}$ lead to
$\grad\times\grad\times\vb{A} \propto \vb{A}$ or  $\grad^2\vb{B}\propto \vb{B}$, resulting in an
exponential decay of the magnetic field (Meissner effect).
In contrast, in dipole superfluids, the spatial derivative term is absent, resulting in a linear relationship between the external field and the induced field. Consequently, the external fields are amplified or suppressed by a factor of $(1-\eta)^{-1}$ in magnetic-dipole superfluids and
$ (1+\eta_p)^{-1}$ in electric-dipole superfluids,
where the opposite signs in front of $\eta$ and $\eta_p$
correspond to positive and negative feedback, respectively.

Finally, the origin of the opposite feedback effects in the magnetic and electric dipole systems is explained as follows.
Both systems share the parallel relationships described by
Eqs.~\eqref{equation:A_vs_B} and \eqref{equation:M_vs_j},
which lead to to the same anti-parallel orientations between the external field 
$\vb{E}_{\rm ext}\, / \, \vb{B}_{\rm ext}$ and the induced dipole density
$\vb{P}\, / \, \vb{M}$
\footnote{Note that the negative sign in $\vb{M}$ in Eq.~\eqref{equation:M_vs_j} cancels with the negative sign in $\vb{j}_\mathrm{p} \propto -\vb{A}_\mathrm{p}$}.
However, opposite signs come in the relations between the induced dipole density and the feedback fields:
$\vb{E}_{\rm int} = -\vb{P}/\varepsilon_0$ and
$\vb{B}_{\rm int} = + \mu_0\vb{M}$
[see, Secs.~\ref{sec:positive feedback} and \ref{sec:negative feedback}],
resulting in positive and negative feedback, respectively.

The domain formation with singular domain walls is a particularly remarkable feature of magnetic-dipole superfluids, with no counterparts in other physical systems.
In order to experimentally observe this,
we need a magnetic-dipole superfluid satisfying $\eta =g_{\rm AC}^2\Psi_\infty^2/(m^*\varepsilon_0) \gtrsim1$~[Eq.~\eqref{equation:eta}], which requires large spin-orbit coupling, high density of condensate, and small effective mass.
We provide a rough estimation of $\eta$ for a magnon BEC in Yttrium Iron Garnet (YIG), a representative example of currently available materials.
In magnetic insulators, $g_\mathrm{AC}^{}$ is typically $10^6$ times as large as in vacuum~\cite{Katsura2005,Liu2011}. The superfluid density can be assumed about $\Psi_\infty^2\sim 10^{21}~\mathrm{cm}^{-3}$~\cite{Demokritov2006}.
The effective mass is obtained by
$m^*=\hbar^2/(2JSa^2)$ by considering ferromagnetic Heisenberg model, where $J$ is the nearest-neighbor ferromagnetic exchange coupling, $S$ is the effective spin and $a$ is the lattice constant.
Using material parameters from experiments, we have $J\approx 1.37~\mathrm{K}$, $S\approx 14.3$, and $a\approx 12.4$ \AA ~\cite{Gilleo1958,Tupitsyn2008}.
Under these conditions, we have $\eta\sim0.36$.

\section{Conclusion}
\label{sec:conclusion}
We have theoretically studied magnetic and electric dipole superfluids subjected to a pseudo-magnetic field induced by a spatially modulated electromagnetic field. Using a Ginzburg-Landau phenomenological theory that incorporates the geometric AC/HMW phase, we obtained self-consistent solutions by fully accounting for the feedback effect of the electromagnetic field induced by the response current.
Under a pseudo-magnetic field, both electric and magnetic dipole superfluids spontaneously form vortex lattices but exhibit distinct field textures. When the electrostatic coupling parameter $\eta$ exceeds a critical value, magnetic dipole superfluids form a honeycomb lattice with a singular domain wall
with diverging pseudo-magnetic fields and polarization charges. This behavior contrasts starkly with conventional superconductors (monopole superfluids).
In electric dipole superfluids, the pseudo-magnetic field is concentrated only at vortex cores as in conventional superconductors, but it does not decay exponentially away from the vortices. Despite sharing parallel formulations, magnetic and electric dipole systems exhibit opposite features due to a sign difference in the feedback effect for the applied field.
These findings highlight novel field textures and self-organized structures unique to magnetic dipole superfluids, opening new directions in exploring superfluid systems with unconventional electromagnetic responses.

\section{Acknowledgement}
We acknowledge fruitful discussions with Yshai Avishai, Joji Nasu, Takeo Kato and Manato Fujimoto.
This work was supported in part by JSPS KAKENHI Grants No. JP20H01840, No. JP20K14415, No. JP21H05236, No. JP21H05232, No. 23KJ1518, No. 24K06921and by JST CREST Grant No. JPMJCR20T3, Japan.

\appendix
\section{Geometric phase for magnetic and electric dipoles}
In this section,  we provide parallel derivations of the
Aharonov-Casher (AC) phase~\cite{Aharanov1984} for magnetic dipoles,
and He-McKellar-Wilkens (HMW) phase~\cite{HeMcKellar1993,Wilkens1994} for electric dipoles.
\subsection{Aharonov-Casher phase}
\label{sec:app_AC}



We present a simplified derivation of the original Aharonov-Casher (AC) phase~\cite{Aharanov1984} in vacuum as a relativistic effect.
We consider a particle with a magnetic-dipole $\vb*{\mu} = \mu \, \vb{e}_z$, 
which is moving with velocity $\vb{v}$.
The magnetic-dipole is equivalent to a small loop electric current 
on the $xy$-plane.
When the loop current moves with velocity $\vb{v}$ relative to the rest frame, we (an observer in the rest frame) observe electric charge density $\rho=\vb{v}\vdot\vb{j}/c^2$ from Lorentz covariance.
Here $\vb{j}$ is the current density in the moving frame fixed to the loop current.
When a square loop current is moving along the $x$-axis, therefore, two sides along $x$ are positively and negatively charged as illustrated in Fig.~\ref{figure:AC_phase}(a),
resulting in an electric dipole 
\begin{equation}
\vb{p}=g_\mathrm{AC}^{}\vb{v}\times\vb{e}_\mathrm{z},
\label{equation:p_relativity}
\end{equation}
where $g_\mathrm{AC}^{} = \mu/c^2$.

If an external electric field $\vb{E}$ is present, the particle has an electrostatic potential energy $-\vb{p}\vdot\vb{E}$.
Thus, the Lagrangian for the particle is given by
\begin{equation}
    L=\frac{1}{2}m\vb{v}^2+\vb{p}\vdot\vb{E},
\end{equation}
where $m$ is a mass of the particle.
As the second term is transformed as
$    \vb{p}\vdot\vb{E}=g_\mathrm{AC}^{}(\vb{v}\times\vb{e}_\mathrm{z})\vdot\vb{E}
=-g_\mathrm{AC}^{}(\vb{E}\times\vb{e}_\mathrm{z})\vdot\vb{v} $,
the Lagrangian is written as
\begin{equation}
    L=\frac{1}{2}m\vb{v}^2-\vb{v} \vdot \vb{A}_\mathrm{m},
\label{eq:L_eff}
\end{equation}
where 
\begin{equation}
   \vb{A}_\mathrm{m}= g_\mathrm{AC}^{}\vb{E}\times\vb{e}_\mathrm{z}.
\end{equation}
Since Eq.~\eqref{eq:L_eff} is formally equivalent to the Lagrangian of an electron in a magnetic field,
$\vb{A}_\mathrm{m}$ plays the role of the effective vector potential for the magnetic-dipole.
Therefore, the moving magnetic-dipole acquires the AC phase [Eq.~\eqref{equation:AC phase}].
\subsection{He-McKellar-Wilkens phase}\label{sec:app_HMW}
\begin{figure}[t]
    \centering
    \includegraphics[width=1.\linewidth]{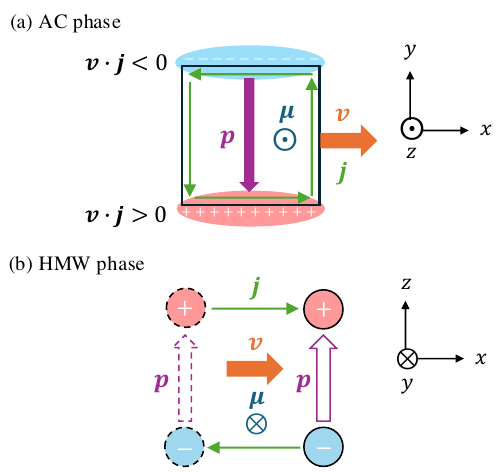}
    \caption{
    Schematic figure for (a) a moving magnetic-dipole (loop current) 
   with induced electric dipole, and (b) a moving electric-dipole with induced magnetic-dipole.
    }
    \label{figure:AC_phase}
\end{figure}

The He-McKellar-Wilkens (HMW) phase~\cite{HeMcKellar1993,Wilkens1994} can be derived in a similar manner to the AC phase.
We consider an electric-dipople constituting a pair of point charges $+q$ and $-q$ located at $z=+d/2$ and $z=-d/2$, where $p=qd$.
If the electric-dipole moves in the $x$-direction with velocity $\vb{v}$, the two point charges $\pm q$ generate counter-propagating electric currents along $x$, as illustrated in Fig.~\ref{figure:AC_phase}(b).
This gives rise to a magnetic moment in the $y$-direction as
\begin{equation}
    \vb*{\mu}=-g_\mathrm{HMW}^{}\vb{v}\times\vb{e}_z,
\end{equation}
where $g_\mathrm{HMW}^{}=p$.

If an external magnetic field $\vb{B}$ is present, the particle has an magnetostatic potential energy $-\vb*{\mu}\vdot\vb{B}$.
Thus, the Lagrangian for the particle is given by
\begin{equation}
    L=\frac{1}{2}m\vb{v}^2+\vb*{\mu}\vdot\vb{B},
\end{equation}
where $m$ is a mass of the particle.
As the second term is transformed as
$\vb*{\mu}\vdot\vb{B}=-g_\mathrm{HMW}^{}(\vb{v}\times\vb{e}_\mathrm{z})\vdot\vb{B}
=g_\mathrm{HMW}^{}(\vb{B}\times\vb{e}_\mathrm{z})\vdot\vb{v} $,
the Lagrangian is written as
\begin{equation}
    L=\frac{1}{2}m\vb{v}^2+\vb{v} \vdot \vb{A}_\mathrm{p},
\label{eq:L_eff_p}
\end{equation}
where 
\begin{equation}
   \vb{A}_\mathrm{p}= g_\mathrm{HMW}^{}\vb{B}\times\vb{e}_\mathrm{z}.
\end{equation}
Since Eq.~\eqref{eq:L_eff_p} is formally equivalent to the Lagrangian of an electron in a magnetic field,
$\vb{A}_\mathrm{p}$ plays the role of the effective vector potential for the electric-dipole.
Therefore, the moving electric-dipole acquires the HMW phase [Eq.~\eqref{equation:HMW phase}].

Altenatively, this HMW phase can directly be derived from the Aharonov-Bohm (AB) phase as follows.
When the dipole moves horizontally along the $x$-axis, from $x$ to $x + \Delta x$, the positive and negative charges acquire Aharonov-Bohm (AB) phases given by $\theta_\pm = (\pm q/\hbar) \int_{C_\pm} d\vb{r} \cdot \vb{A}(\vb{r})$, where $C_\pm$ are the paths followed by the positive and negative charges, respectively.
The total geometric phase for the dipole is given by $\theta_{\rm HMW} = \theta_+ + \theta_-$, and it becomes
 $(q/\hbar)[\int_{C_+} - \int_{C_-}] d\vb{r}\cdot \vb{A}(\vb{r}) = (p/\hbar)B_y(\vb{r})\Delta x$, where we used Stokes's theorem and $p=qd$.
The result is consistent with Eqs.~\eqref{equation:HMW phase} and \eqref{equation:Ap}.


\section{Derivation of total free energy for diople superfluids}\label{sec:toal_free_energy}

Here, we present the derivation of the Helmholtz free energy of magnetic dipole superfluids [\eqref{equation:free energy}], and the Gibbs free energy of electric dipole superfluids [Eq.~\eqref{equation:Gibbs}],
paralleling the argument for the conventional superconductor in Appendix \ref{sec:glsc}.


\subsection{Magnetic-dipole superfluid}
\label{sec:free_energy_magnetic_dipole}

We derive the expression of the Helmholtz free energy $F[\vb{D}(\vb{r}),T]$, Eq.~\eqref{equation:free energy}.
We consider a quasi-static, isothermal process where the external electric field,
or $\vb{D}(\vb{r})$, is slowly introduced to the magnetic-dipole superfluid with the temperature fixed.
Here, we consider a general situation where the external electric field $\vb{D}$ (and consequently $\vb{E}$ and $\vb{P}$) depends on position, to account for the system under the pseudo-magnetic field, which needs spatially dependent electric field [Eq.~\eqref{equation:Bm}].
We assume that the system is at the thermal equilibrium of given $\vb{D}(\vb{r})$ and $T$ at every moment of the process.
As in a general dielectric material, 
the change in the Helmholtz free energy from the initial state (i) with $\vb{D}=0$ to the finial state (f) with $\vb{D}=\vb{D}(\vb{r})$ is written as 
\begin{align}\label{equation:ede} 
    F[\vb{D}(\vb{r}),T]-F[0,T] 
    &=\int_{\rm{i}}^{\rm{f}}\int d{\vb{r}}\,\, \vb{E}(\vb{r})\cdot \delta \vb{D}(\vb{r})
    \nonumber\\
    & \hspace{-12mm}
    =\int_{\rm{i}}^{\rm{f}}\int d{\vb{r}}\, 
    \Biggl(
    \frac{\vb{D}\cdot \delta\vb{D}-\vb{P}\cdot \delta \vb{P}}{\varepsilon_0} - \vb{P}\cdot \delta\vb{E} 
     \Biggr)
     \nonumber\\
    & \hspace{-12mm}
    = \int d\vb{r}\,\,\frac{\vb{D}^2-\vb{P}^2}{2\varepsilon_0} -
    \int_{\rm{i}}^{\rm{f}}\int d{\vb{r}}\,\, \vb{P}\cdot \delta \vb{E} 
\end{align}
Here $\int_{\rm{i}}^{\rm{f}}\cdots \delta \vb{D}(\vb{r})$ represents the functional integral in $\vb{D}(\vb{r})$.

The last term in Eq.~\eqref{equation:ede} 
coincides with the change of GL energy functional $F_{\rm GL}[\Psi,\vb{A}_\mathrm{m}]$,
because the infinitesimal change in $F_{\rm GL}$ in the process is written as
\begin{equation}\label{equation:dhgl}
    \delta F_\mathrm{GL}[\Psi,\vb{A}_\mathrm{m}] = \int d\vb{r} \frac{\delta F_\mathrm{GL}}{\delta \vb{A}_\mathrm{m}}\cdot \delta \vb{A}_\mathrm{m}
    =-\int d\vb{r}~\vb{P}\cdot \delta\vb{E},
\end{equation}
where we used 
Eqs.~\eqref{equation:Am}, \eqref{equation:jm} and \eqref{equation:P}.
Here we note that the variation 
$\delta F_\mathrm{GL}$ should also include $\delta F_{\rm GL}/\delta \Psi$ and  $\delta F_{\rm GL}/\delta \Psi^*$, while
these terms just vanish because we assume that the GL equation Eq.~\eqref{equation:GL_equation} holds throughout the quasi-static process.
From Eq.~\eqref{equation:ede} and Eq.~\eqref{equation:dhgl}, 
we obtain Eq.~\eqref{equation:free energy}. 

We can readily check that the self-consistent equations are derived by the variation of the total Helmholtz free energy with $\vb{D}$ fixed.
Eq.~\eqref{equation:GL_equation} is given by $\delta F/\delta\Psi=0$, 
since $\delta F/\delta\Psi=\delta F_{\rm GL}/\delta\Psi$.
Eq.~\eqref{equation:atot} is obtained by $\delta F/\delta\vb{A}_\mathrm{m}=0$, noting
\begin{align}
    \fdv{F}{\vb{A}_\mathrm{m}}&=\frac{\delta F_\mathrm{GL}}{\delta \vb{A}_\mathrm{m}}+\fdv{\vb{A}_\mathrm{m}}\qty(\frac{\vb{D}^2-\vb{P}^2}{2\varepsilon_0})\notag\\
    &=\vb{j}_\mathrm{m}+\frac{\varepsilon_0}{g_\mathrm{AC}^{2}}\qty(\vb{A}_\mathrm{m}^\mathrm{ext}-\vb{A}_\mathrm{m})=0,
\end{align}
where we used relations $\vb{P}=\vb{D}-\varepsilon_0\vb{E}$ and 
$\vb{E} = g_\mathrm{AC}^{-1}(\vb{e}_z\times\vb{A}_\mathrm{m})+E_z\vb{e}_z$.



\subsection{Electric-dipole superfluid}
\label{sec:free_energy_electric_dipole}

We derive the expression of the Gibbs free energy $G[\vb{H},T]$, Eq.~\eqref{equation:Gibbs}. 
We consider a quasi-static, isothermal process where the external magnetic field $\vb{H}$ is slowly introduced to the electric-dipole superfluid with temperature fixed.
The change in the Gibbs free energy from the initial state (i) with $\vb{H}=0$ to the final state (f) with $\vb{H}$ is given by 
\begin{align}\label{equation:G_HMW_isothermal}
    &G[\vb{H}(\vb{r}),T]-G[0,T]
    =-\int_{\mathrm{i}}^{\mathrm{f}}\int d\vb{r}~\vb{B}(\vb{r})\cdot \delta \vb{H}(\vb{r})\nn\\ 
    &=\int_{\mathrm{i}}^{\mathrm{f}} \int d \vb{r} \left(\frac{\vb{B}\cdot\delta\vb{B}}{\mu_0} - \vb{H}\cdot \delta \vb{B}- \vb{B}\cdot \delta \vb{H}- \vb{M}\cdot \delta \vb{B}\right)\nn\\
    &=\int d \vb{r}~\qty(\frac{\vb{B}^2}{2\mu_0}-\vb{B}\cdot\vb{H}) - \int_{\mathrm{i}}^{\mathrm{f}}\int d\vb{r}~\vb{M}\cdot \delta \vb{B}
\end{align}
We can show that the last term in Eq.~\eqref{equation:G_HMW_isothermal} coincides with the change in $F_{\mathrm{GL}}$ by the following argument.
In a quasi-static process, the infinitesimal change in the GL free energy functional is given by
\begin{equation}\label{equation:dF_GL_HMW}
    \delta F_{\mathrm{GL}}[\Psi,\vb{A}_\mathrm{p}] = \int d\vb{r} \frac{\delta F_{\mathrm{GL}}}{\delta \vb{A}_\mathrm{p}}\cdot\delta \vb{A}_\mathrm{p} = -\int \vb{M}\cdot \delta \vb{B},
\end{equation}
where we used Eq.~\eqref{equation:jp},
and we noted that ${\delta F_{\mathrm{GL}}}/{\delta \Psi}={\delta F_{\mathrm{GL}}}/{\delta \Psi^\ast}=0$ throughout the process. Equations  \eqref{equation:G_HMW_isothermal} and \eqref{equation:dF_GL_HMW} lead to the Eq.~\eqref{equation:Gibbs}.

We check that self-consistent equations are derived by the variation of the total Gibbs free energy while fixing $\vb{H}$.
Eq.~\eqref{equation:gl_HMW} is straightforwardly derived from $\delta G/\delta\Psi=0$ since $\delta G/\delta\Psi=\delta F_\mathrm{GL}/\delta\Psi$. Eq.~\eqref{equation:Ap_tot} is obtained by $\delta G/\delta\vb{A}_\mathrm{p}=0$, noting
\begin{align}
    \fdv{G}{\vb{A}_\mathrm{p}}&=\fdv{F_\mathrm{GL}}{\vb{A}_\mathrm{p}}+\fdv{\vb{A}_\mathrm{p}}\qty(\frac{\vb{B}^2}{2\mu_0}-\vb{B}\cdot\vb{H})\notag\\
    &=-\vb{j}_\mathrm{p}+\frac{1}{\mu_0p^2}\qty(\vb{A}_\mathrm{p}-\vb{A}_\mathrm{p}^\mathrm{ext})=0,
\end{align}
where we used relations $\vb{B}=\mu_0(\vb{H}+\vb{M})$ and $\vb{B}=g_{\mathrm{HMW}}^{-1}(\vb{e}_z\times\vb{A}_\mathrm{p})+B_z\vb{e}_z$.

\section{Ginzburg-Landau theory for superconductors (electric-monopole superfluid)}\label{sec:glsc}

Here, we review the standard Ginzburg-Landau theory~\cite{tinkhambook} for superconductor,
to provide a basis for comparison with the corresponding theory for dipople superfluids.
The Ginzburg-Landau free energy functional for conventional superconductors is given by 
\begin{align}
    F_{\mathrm{GL}}[\Psi,\vb{A}] = \int d^3 r \Big[\frac{1}{4m_e}\big|(-i\hbar\vb{\grad}+2e \vb{A})\Psi(\vb{r})\big|^2 \nonumber\\ 
    - \alpha |\Psi(\vb{r})|^2 + \frac{\beta}{2} |\Psi(\vb{r})|^4 \Big]
    \label{equation:GP_Hamiltonian_superconductor}
\end{align}
where $m_e$ is the electron mass, and $e$ $(>\!\!0)$ is the elementary charge. 
$\Psi(\vb{r})$ represents the superconducting order parameter and $\vb{B}=\vb{\grad}\times\vb{A}$ represents the magnetic flux density, where $\vb{A}$ is the vector potential.
The Ginzburg-Landau equation for the order parameter is given by
\begin{equation}\label{equation:sc_gl}
    \frac{\delta F_{\mathrm{GL}}}{\delta\Psi^\ast}
    = \frac{1}{4m_e}(-i\hbar\grad+2e\vb{A})^2\Psi - \alpha\Psi+\beta|\Psi|^2\Psi=0.
\end{equation}
The superconducting current is described by the London equation, giving
\begin{equation}\label{equation:sc_j}
    \vb{j} = -\frac{e\hbar}{m_e}|\Psi|^2\qty(\grad\theta+\frac{2e}{\hbar}\vb{A})=-\frac{\delta F_{\mathrm{GL}}}{\delta \vb{A}},
\end{equation}
and this superconducting current produces the magnetization $\vb{M}$ by Maxwell equations, giving 
\begin{equation}
    \vb{j}=\grad\times \vb{M}.\label{equation:sc_M}
\end{equation}

The total Gibbs free energy including magnetic field contribution is given by
\begin{equation}
    G[\vb{H}(\vb{r}),T]=F_\mathrm{GL}+ \int d\vb{r}~ \qty(\frac{\vb{B}(\vb{r})^2}{2\mu_0}-\vb{B}(\vb{r})\cdot\vb{H}(\vb{r})),
\end{equation}
which can be derived by considerations similar to those in Appendix.~\ref{sec:toal_free_energy}.
We can readily check that self-consistent equations are derived by the variation of the total Gibbs free energy with $\vb{H}$ fixed:
Eq.~\eqref{equation:sc_gl} is obtained from $\delta G/\delta\Psi=0$ since $\delta G/\delta\Psi=\delta F_\mathrm{GL}/\delta\Psi$. Eq.~\eqref{equation:sc_M} is derived by $\delta G/\delta\vb{A}=0$, noting
\begin{align}
    \fdv{G}{\vb{A}}&=\fdv{F_\mathrm{GL}}{\vb{A}}+\fdv{\vb{A}}\qty(\frac{\vb{B}^2}{2\mu_0}-\vb{B}\cdot\vb{H})\notag\\
    &=-\vb{j}+\frac{1}{\mu_0}\curl\curl\qty(\vb{A}-\vb{A}^\mathrm{ext})=0,
\end{align}
where we used relations $\vb{B}=\mu_0(\vb{H}+\vb{M})$ and $\vb{B}=\curl\vb{A}$. 
Here, $\vb{A}^\mathrm{ext}$ is the vector potential of the external magnetic field defined by
$\mu_0\vb{H}=\curl\vb{A}^\mathrm{ext}$.

By combining Eq.~\eqref{equation:sc_j} with Eq.~\eqref{equation:sc_M}, we have
$\curl\curl\vb{A}=-\lambda^{-2}\vb{A}$, where $\lambda=\sqrt{m_e/(2\mu_0 e^2|\Psi|^2)}$ is the London's penetration depth.
Thus, we obtain $\grad^2\vb{B}=\lambda^{-2}\vb{B}$, which shows the exponential decay of the magnetic flux density
(Meissner effect).

\section{Technical details of self-consistent numerical calculations}
\label{sec:technical_detail}

We present the technical details of the numerical calculation. 
In the calculations in the main text, we numerically solve the GL equation [Eq.~\eqref{equation:GL_equation}]  
self-consistently with Eqs.~\eqref{equation:jm} and \eqref{equation:atot} under a given
external field, 
following the procedure outlined below.

First we set an initial configuration with $\Psi=\Psi^\mathrm{(0)}$ and $\vb{A}_\mathrm{m}=\vb{A}_\mathrm{m}^\mathrm{(0)}(\equiv\vb{A}_\mathrm{m}^\mathrm{ext})$.
Then, we calculate the supercurrent $\vb{j}_\mathrm{m}$ using Eq.~\eqref{equation:jm}.
From obtained $\vb{j}_\mathrm{m}$, we obtain the updated vector potential $\vb{A}_\mathrm{m}^\mathrm{(1)}$ using Eq.~\eqref{equation:atot}. We then obtain
the order parameter $\Psi^\mathrm{(1)}$ by solving GL equation [Eq.~\eqref{equation:GL_equation}] with $\vb{A}_\mathrm{m}=\vb{A}_\mathrm{m}^\mathrm{(1)}$. 
We go back to the first stage with $\Psi=\Psi^\mathrm{(1)}$ and $\vb{A}_\mathrm{m}=\vb{A}_\mathrm{m}^\mathrm{(1)}$, 
to obtain
$\Psi=\Psi^\mathrm{(2)}$ and $\vb{A}_\mathrm{m}^\mathrm{(2)}$.
We repeat the procedure until $|\Psi^{(n+1)}-\Psi^{(n)}|$ and $|\vb{A}^{(n+1)}_\mathrm{m}-\vb{A}^{(n)}_\mathrm{m}|$ become sufficiently small.

In the procedure to obtain $\Psi$ by the GL equation,
we minimize $F_\mathrm{GL}$ with respect to $\Psi$
instead of directly solving the GL equation, 
since the GL equation is equivalent to $\delta F_\mathrm{GL}/\delta \Psi^*=0$. 
Specifically, we optimize $\Psi$ by repeating steps of
\begin{equation}
\Psi \rightarrow \Psi-\fdv{F_\mathrm{GL}}{\Psi^*}\Delta t,
\end{equation}
where $\Delta t>0$ is a positive constant.
At every step, the $F_\mathrm{GL}$ consistently decreases
because the change of $F_\mathrm{GL}$ at the step is given by
\begin{equation}
    \delta F_\mathrm{GL} = \fdv{F_\mathrm{GL}}{\Psi}\delta\Psi=-\abs{\fdv{F_\mathrm{GL}}{\Psi^*}}^2\Delta t<0.
\end{equation}
We repeat this procedure until $|\delta F_\mathrm{GL}|\sim 0$, to obtain $\Psi$
satisfying the GL equation.



\bibliography{reference}

\begin{thebibliography}{86}%
\makeatletter
\providecommand \@ifxundefined [1]{%
 \@ifx{#1\undefined}
}%
\providecommand \@ifnum [1]{%
 \ifnum #1\expandafter \@firstoftwo
 \else \expandafter \@secondoftwo
 \fi
}%
\providecommand \@ifx [1]{%
 \ifx #1\expandafter \@firstoftwo
 \else \expandafter \@secondoftwo
 \fi
}%
\providecommand \natexlab [1]{#1}%
\providecommand \enquote  [1]{``#1''}%
\providecommand \bibnamefont  [1]{#1}%
\providecommand \bibfnamefont [1]{#1}%
\providecommand \citenamefont [1]{#1}%
\providecommand \href@noop [0]{\@secondoftwo}%
\providecommand \href [0]{\begingroup \@sanitize@url \@href}%
\providecommand \@href[1]{\@@startlink{#1}\@@href}%
\providecommand \@@href[1]{\endgroup#1\@@endlink}%
\providecommand \@sanitize@url [0]{\catcode `\\12\catcode `\$12\catcode `\&12\catcode `\#12\catcode `\^12\catcode `\_12\catcode `\%12\relax}%
\providecommand \@@startlink[1]{}%
\providecommand \@@endlink[0]{}%
\providecommand \url  [0]{\begingroup\@sanitize@url \@url }%
\providecommand \@url [1]{\endgroup\@href {#1}{\urlprefix }}%
\providecommand \urlprefix  [0]{URL }%
\providecommand \Eprint [0]{\href }%
\providecommand \doibase [0]{http://dx.doi.org/}%
\providecommand \selectlanguage [0]{\@gobble}%
\providecommand \bibinfo  [0]{\@secondoftwo}%
\providecommand \bibfield  [0]{\@secondoftwo}%
\providecommand \translation [1]{[#1]}%
\providecommand \BibitemOpen [0]{}%
\providecommand \bibitemStop [0]{}%
\providecommand \bibitemNoStop [0]{.\EOS\space}%
\providecommand \EOS [0]{\spacefactor3000\relax}%
\providecommand \BibitemShut  [1]{\csname bibitem#1\endcsname}%
\let\auto@bib@innerbib\@empty
\bibitem [{\citenamefont {Ikeda}\ \emph {et~al.}(2023)\citenamefont {Ikeda}, \citenamefont {Matsuda}, \citenamefont {Sato}, \citenamefont {Ishii}, \citenamefont {Sawabe}, \citenamefont {Nakamura}, \citenamefont {Takeyama},\ and\ \citenamefont {Nasu}}]{Ikeda2023}%
  \BibitemOpen
  \bibfield  {author} {\bibinfo {author} {\bibfnamefont {Akihiko}\ \bibnamefont {Ikeda}}, \bibinfo {author} {\bibfnamefont {Yasuhiro~H.}\ \bibnamefont {Matsuda}}, \bibinfo {author} {\bibfnamefont {Keisuke}\ \bibnamefont {Sato}}, \bibinfo {author} {\bibfnamefont {Yuto}\ \bibnamefont {Ishii}}, \bibinfo {author} {\bibfnamefont {Hironobu}\ \bibnamefont {Sawabe}}, \bibinfo {author} {\bibfnamefont {Daisuke}\ \bibnamefont {Nakamura}}, \bibinfo {author} {\bibfnamefont {Shojiro}\ \bibnamefont {Takeyama}}, \ and\ \bibinfo {author} {\bibfnamefont {Joji}\ \bibnamefont {Nasu}},\ }\bibfield  {title} {\enquote {\bibinfo {title} {Signature of spin-triplet exciton condensations in lacoo3 at ultrahigh magnetic fields up to 600 t},}\ }\href {\doibase 10.1038/s41467-023-37125-4} {\bibfield  {journal} {\bibinfo  {journal} {Nature Communications}\ }\textbf {\bibinfo {volume} {14}},\ \bibinfo {pages} {1744} (\bibinfo {year} {2023})}\BibitemShut {NoStop}%
\bibitem [{\citenamefont {Jiang}\ \emph {et~al.}(2020)\citenamefont {Jiang}, \citenamefont {Lou}, \citenamefont {Liu}, \citenamefont {Li}, \citenamefont {Song}, \citenamefont {Chang}, \citenamefont {Duan},\ and\ \citenamefont {Zhang}}]{Jiang2020}%
  \BibitemOpen
  \bibfield  {author} {\bibinfo {author} {\bibfnamefont {Zeyu}\ \bibnamefont {Jiang}}, \bibinfo {author} {\bibfnamefont {Wenkai}\ \bibnamefont {Lou}}, \bibinfo {author} {\bibfnamefont {Yu}~\bibnamefont {Liu}}, \bibinfo {author} {\bibfnamefont {Yuanchang}\ \bibnamefont {Li}}, \bibinfo {author} {\bibfnamefont {Haifeng}\ \bibnamefont {Song}}, \bibinfo {author} {\bibfnamefont {Kai}\ \bibnamefont {Chang}}, \bibinfo {author} {\bibfnamefont {Wenhui}\ \bibnamefont {Duan}}, \ and\ \bibinfo {author} {\bibfnamefont {Shengbai}\ \bibnamefont {Zhang}},\ }\bibfield  {title} {\enquote {\bibinfo {title} {Spin-triplet excitonic insulator: The case of semihydrogenated graphene},}\ }\href {\doibase 10.1103/PhysRevLett.124.166401} {\bibfield  {journal} {\bibinfo  {journal} {Phys. Rev. Lett.}\ }\textbf {\bibinfo {volume} {124}},\ \bibinfo {pages} {166401} (\bibinfo {year} {2020})}\BibitemShut {NoStop}%
\bibitem [{\citenamefont {Sun}\ and\ \citenamefont {Xie}(2013)}]{Sun2013}%
  \BibitemOpen
  \bibfield  {author} {\bibinfo {author} {\bibfnamefont {Qing-feng}\ \bibnamefont {Sun}}\ and\ \bibinfo {author} {\bibfnamefont {X.~C.}\ \bibnamefont {Xie}},\ }\bibfield  {title} {\enquote {\bibinfo {title} {Spin-polarized $\ensuremath{\nu}=0$ state of graphene: A spin superconductor},}\ }\href {\doibase 10.1103/PhysRevB.87.245427} {\bibfield  {journal} {\bibinfo  {journal} {Phys. Rev. B}\ }\textbf {\bibinfo {volume} {87}},\ \bibinfo {pages} {245427} (\bibinfo {year} {2013})}\BibitemShut {NoStop}%
\bibitem [{\citenamefont {Yang}\ \emph {et~al.}(2024)\citenamefont {Yang}, \citenamefont {Zeng}, \citenamefont {Shao}, \citenamefont {Xu}, \citenamefont {Dai},\ and\ \citenamefont {Li}}]{Yang2024}%
  \BibitemOpen
  \bibfield  {author} {\bibinfo {author} {\bibfnamefont {Huaiyuan}\ \bibnamefont {Yang}}, \bibinfo {author} {\bibfnamefont {Jiaxi}\ \bibnamefont {Zeng}}, \bibinfo {author} {\bibfnamefont {Yuelin}\ \bibnamefont {Shao}}, \bibinfo {author} {\bibfnamefont {Yuanfeng}\ \bibnamefont {Xu}}, \bibinfo {author} {\bibfnamefont {Xi}~\bibnamefont {Dai}}, \ and\ \bibinfo {author} {\bibfnamefont {Xin-Zheng}\ \bibnamefont {Li}},\ }\bibfield  {title} {\enquote {\bibinfo {title} {Spin-triplet topological excitonic insulators in two-dimensional materials},}\ }\href {\doibase 10.1103/PhysRevB.109.075167} {\bibfield  {journal} {\bibinfo  {journal} {Phys. Rev. B}\ }\textbf {\bibinfo {volume} {109}},\ \bibinfo {pages} {075167} (\bibinfo {year} {2024})}\BibitemShut {NoStop}%
\bibitem [{\citenamefont {Nishida}\ \emph {et~al.}(2019)\citenamefont {Nishida}, \citenamefont {Miyakoshi}, \citenamefont {Kaneko}, \citenamefont {Sugimoto},\ and\ \citenamefont {Ohta}}]{Nishida2019}%
  \BibitemOpen
  \bibfield  {author} {\bibinfo {author} {\bibfnamefont {Hisao}\ \bibnamefont {Nishida}}, \bibinfo {author} {\bibfnamefont {Shohei}\ \bibnamefont {Miyakoshi}}, \bibinfo {author} {\bibfnamefont {Tatsuya}\ \bibnamefont {Kaneko}}, \bibinfo {author} {\bibfnamefont {Koudai}\ \bibnamefont {Sugimoto}}, \ and\ \bibinfo {author} {\bibfnamefont {Yukinori}\ \bibnamefont {Ohta}},\ }\bibfield  {title} {\enquote {\bibinfo {title} {Spin texture and spin current in excitonic phases of the two-band hubbard model},}\ }\href {\doibase 10.1103/PhysRevB.99.035119} {\bibfield  {journal} {\bibinfo  {journal} {Phys. Rev. B}\ }\textbf {\bibinfo {volume} {99}},\ \bibinfo {pages} {035119} (\bibinfo {year} {2019})}\BibitemShut {NoStop}%
\bibitem [{\citenamefont {Wang}\ \emph {et~al.}(2019{\natexlab{a}})\citenamefont {Wang}, \citenamefont {Erten}, \citenamefont {Wang},\ and\ \citenamefont {Xing}}]{Wang2019}%
  \BibitemOpen
  \bibfield  {author} {\bibinfo {author} {\bibfnamefont {Rui}\ \bibnamefont {Wang}}, \bibinfo {author} {\bibfnamefont {Onur}\ \bibnamefont {Erten}}, \bibinfo {author} {\bibfnamefont {Baigeng}\ \bibnamefont {Wang}}, \ and\ \bibinfo {author} {\bibfnamefont {D.~Y.}\ \bibnamefont {Xing}},\ }\bibfield  {title} {\enquote {\bibinfo {title} {Prediction of a topological p + ip excitonic insulator with parity anomaly},}\ }\href {\doibase 10.1038/s41467-018-08203-9} {\bibfield  {journal} {\bibinfo  {journal} {Nature Communications}\ }\textbf {\bibinfo {volume} {10}},\ \bibinfo {pages} {210} (\bibinfo {year} {2019}{\natexlab{a}})}\BibitemShut {NoStop}%
\bibitem [{\citenamefont {Liu}\ \emph {et~al.}(2012)\citenamefont {Liu}, \citenamefont {Jiang}, \citenamefont {Xie},\ and\ \citenamefont {Sun}}]{Liu2012}%
  \BibitemOpen
  \bibfield  {author} {\bibinfo {author} {\bibfnamefont {Haiwen}\ \bibnamefont {Liu}}, \bibinfo {author} {\bibfnamefont {Hua}\ \bibnamefont {Jiang}}, \bibinfo {author} {\bibfnamefont {X.~C.}\ \bibnamefont {Xie}}, \ and\ \bibinfo {author} {\bibfnamefont {Qing-feng}\ \bibnamefont {Sun}},\ }\bibfield  {title} {\enquote {\bibinfo {title} {Spontaneous spin-triplet exciton condensation in abc-stacked trilayer graphene},}\ }\href {\doibase 10.1103/PhysRevB.86.085441} {\bibfield  {journal} {\bibinfo  {journal} {Phys. Rev. B}\ }\textbf {\bibinfo {volume} {86}},\ \bibinfo {pages} {085441} (\bibinfo {year} {2012})}\BibitemShut {NoStop}%
\bibitem [{\citenamefont {Sethi}\ \emph {et~al.}(2021)\citenamefont {Sethi}, \citenamefont {Zhou}, \citenamefont {Zhu}, \citenamefont {Yang},\ and\ \citenamefont {Liu}}]{Sethi2021}%
  \BibitemOpen
  \bibfield  {author} {\bibinfo {author} {\bibfnamefont {Gurjyot}\ \bibnamefont {Sethi}}, \bibinfo {author} {\bibfnamefont {Yinong}\ \bibnamefont {Zhou}}, \bibinfo {author} {\bibfnamefont {Linghan}\ \bibnamefont {Zhu}}, \bibinfo {author} {\bibfnamefont {Li}~\bibnamefont {Yang}}, \ and\ \bibinfo {author} {\bibfnamefont {Feng}\ \bibnamefont {Liu}},\ }\bibfield  {title} {\enquote {\bibinfo {title} {Flat-band-enabled triplet excitonic insulator in a diatomic kagome lattice},}\ }\href {\doibase 10.1103/PhysRevLett.126.196403} {\bibfield  {journal} {\bibinfo  {journal} {Phys. Rev. Lett.}\ }\textbf {\bibinfo {volume} {126}},\ \bibinfo {pages} {196403} (\bibinfo {year} {2021})}\BibitemShut {NoStop}%
\bibitem [{\citenamefont {Sethi}\ \emph {et~al.}(2023)\citenamefont {Sethi}, \citenamefont {Cuma},\ and\ \citenamefont {Liu}}]{Sethi2023}%
  \BibitemOpen
  \bibfield  {author} {\bibinfo {author} {\bibfnamefont {Gurjyot}\ \bibnamefont {Sethi}}, \bibinfo {author} {\bibfnamefont {Martin}\ \bibnamefont {Cuma}}, \ and\ \bibinfo {author} {\bibfnamefont {Feng}\ \bibnamefont {Liu}},\ }\bibfield  {title} {\enquote {\bibinfo {title} {Excitonic condensate in flat valence and conduction bands of opposite chirality},}\ }\href {\doibase 10.1103/PhysRevLett.130.186401} {\bibfield  {journal} {\bibinfo  {journal} {Phys. Rev. Lett.}\ }\textbf {\bibinfo {volume} {130}},\ \bibinfo {pages} {186401} (\bibinfo {year} {2023})}\BibitemShut {NoStop}%
\bibitem [{\citenamefont {Wei}\ \emph {et~al.}(2012)\citenamefont {Wei}, \citenamefont {Chao},\ and\ \citenamefont {Aji}}]{Wei2012}%
  \BibitemOpen
  \bibfield  {author} {\bibinfo {author} {\bibfnamefont {Huazhou}\ \bibnamefont {Wei}}, \bibinfo {author} {\bibfnamefont {Sung-Po}\ \bibnamefont {Chao}}, \ and\ \bibinfo {author} {\bibfnamefont {Vivek}\ \bibnamefont {Aji}},\ }\bibfield  {title} {\enquote {\bibinfo {title} {Excitonic phases from weyl semimetals},}\ }\href {\doibase 10.1103/PhysRevLett.109.196403} {\bibfield  {journal} {\bibinfo  {journal} {Phys. Rev. Lett.}\ }\textbf {\bibinfo {volume} {109}},\ \bibinfo {pages} {196403} (\bibinfo {year} {2012})}\BibitemShut {NoStop}%
\bibitem [{\citenamefont {Demokritov}\ \emph {et~al.}(2006)\citenamefont {Demokritov}, \citenamefont {Demidov}, \citenamefont {Dzyapko}, \citenamefont {Melkov}, \citenamefont {Serga}, \citenamefont {Hillebrands},\ and\ \citenamefont {Slavin}}]{Demokritov2006}%
  \BibitemOpen
  \bibfield  {author} {\bibinfo {author} {\bibfnamefont {S.~O.}\ \bibnamefont {Demokritov}}, \bibinfo {author} {\bibfnamefont {V.~E.}\ \bibnamefont {Demidov}}, \bibinfo {author} {\bibfnamefont {O.}~\bibnamefont {Dzyapko}}, \bibinfo {author} {\bibfnamefont {G.~A.}\ \bibnamefont {Melkov}}, \bibinfo {author} {\bibfnamefont {A.~A.}\ \bibnamefont {Serga}}, \bibinfo {author} {\bibfnamefont {B.}~\bibnamefont {Hillebrands}}, \ and\ \bibinfo {author} {\bibfnamefont {A.~N.}\ \bibnamefont {Slavin}},\ }\bibfield  {title} {\enquote {\bibinfo {title} {Bose--einstein condensation of quasi-equilibrium magnons at room temperature under pumping},}\ }\href {\doibase 10.1038/nature05117} {\bibfield  {journal} {\bibinfo  {journal} {Nature}\ }\textbf {\bibinfo {volume} {443}},\ \bibinfo {pages} {430--433} (\bibinfo {year} {2006})}\BibitemShut {NoStop}%
\bibitem [{\citenamefont {Bozhko}\ \emph {et~al.}(2016)\citenamefont {Bozhko}, \citenamefont {Serga}, \citenamefont {Clausen}, \citenamefont {Vasyuchka}, \citenamefont {Heussner}, \citenamefont {Melkov}, \citenamefont {Pomyalov}, \citenamefont {L'vov},\ and\ \citenamefont {Hillebrands}}]{Bozhko2016}%
  \BibitemOpen
  \bibfield  {author} {\bibinfo {author} {\bibfnamefont {Dmytro~A.}\ \bibnamefont {Bozhko}}, \bibinfo {author} {\bibfnamefont {Alexander~A.}\ \bibnamefont {Serga}}, \bibinfo {author} {\bibfnamefont {Peter}\ \bibnamefont {Clausen}}, \bibinfo {author} {\bibfnamefont {Vitaliy~I.}\ \bibnamefont {Vasyuchka}}, \bibinfo {author} {\bibfnamefont {Frank}\ \bibnamefont {Heussner}}, \bibinfo {author} {\bibfnamefont {Gennadii~A.}\ \bibnamefont {Melkov}}, \bibinfo {author} {\bibfnamefont {Anna}\ \bibnamefont {Pomyalov}}, \bibinfo {author} {\bibfnamefont {Victor~S.}\ \bibnamefont {L'vov}}, \ and\ \bibinfo {author} {\bibfnamefont {Burkard}\ \bibnamefont {Hillebrands}},\ }\bibfield  {title} {\enquote {\bibinfo {title} {Supercurrent in a room-temperature bose--einstein magnon condensate},}\ }\href {\doibase 10.1038/nphys3838} {\bibfield  {journal} {\bibinfo  {journal} {Nature Physics}\ }\textbf {\bibinfo {volume} {12}},\ \bibinfo {pages} {1057--1062} (\bibinfo {year} {2016})}\BibitemShut {NoStop}%
\bibitem [{\citenamefont {Bozhko}\ \emph {et~al.}(2019)\citenamefont {Bozhko}, \citenamefont {Kreil}, \citenamefont {Musiienko-Shmarova}, \citenamefont {Serga}, \citenamefont {Pomyalov}, \citenamefont {L'vov},\ and\ \citenamefont {Hillebrands}}]{Bozhko2019}%
  \BibitemOpen
  \bibfield  {author} {\bibinfo {author} {\bibfnamefont {Dmytro~A.}\ \bibnamefont {Bozhko}}, \bibinfo {author} {\bibfnamefont {Alexander J.~E.}\ \bibnamefont {Kreil}}, \bibinfo {author} {\bibfnamefont {Halyna~Yu.}\ \bibnamefont {Musiienko-Shmarova}}, \bibinfo {author} {\bibfnamefont {Alexander~A.}\ \bibnamefont {Serga}}, \bibinfo {author} {\bibfnamefont {Anna}\ \bibnamefont {Pomyalov}}, \bibinfo {author} {\bibfnamefont {Victor~S.}\ \bibnamefont {L'vov}}, \ and\ \bibinfo {author} {\bibfnamefont {Burkard}\ \bibnamefont {Hillebrands}},\ }\bibfield  {title} {\enquote {\bibinfo {title} {Bogoliubov waves and distant transport of magnon condensate at room temperature},}\ }\href {\doibase 10.1038/s41467-019-10118-y} {\bibfield  {journal} {\bibinfo  {journal} {Nature Communications}\ }\textbf {\bibinfo {volume} {10}},\ \bibinfo {pages} {2460} (\bibinfo {year} {2019})}\BibitemShut {NoStop}%
\bibitem [{\citenamefont {Olsson}\ \emph {et~al.}(2020)\citenamefont {Olsson}, \citenamefont {An}, \citenamefont {Fiete}, \citenamefont {Zhou}, \citenamefont {Shi},\ and\ \citenamefont {Li}}]{Olsson2020}%
  \BibitemOpen
  \bibfield  {author} {\bibinfo {author} {\bibfnamefont {Kevin~S.}\ \bibnamefont {Olsson}}, \bibinfo {author} {\bibfnamefont {Kyongmo}\ \bibnamefont {An}}, \bibinfo {author} {\bibfnamefont {Gregory~A.}\ \bibnamefont {Fiete}}, \bibinfo {author} {\bibfnamefont {Jianshi}\ \bibnamefont {Zhou}}, \bibinfo {author} {\bibfnamefont {Li}~\bibnamefont {Shi}}, \ and\ \bibinfo {author} {\bibfnamefont {Xiaoqin}\ \bibnamefont {Li}},\ }\bibfield  {title} {\enquote {\bibinfo {title} {Pure spin current and magnon chemical potential in a nonequilibrium magnetic insulator},}\ }\href {\doibase 10.1103/PhysRevX.10.021029} {\bibfield  {journal} {\bibinfo  {journal} {Phys. Rev. X}\ }\textbf {\bibinfo {volume} {10}},\ \bibinfo {pages} {021029} (\bibinfo {year} {2020})}\BibitemShut {NoStop}%
\bibitem [{\citenamefont {Divinskiy}\ \emph {et~al.}(2021)\citenamefont {Divinskiy}, \citenamefont {Merbouche}, \citenamefont {Demidov}, \citenamefont {Nikolaev}, \citenamefont {Soumah}, \citenamefont {Gou{\'e}r{\'e}}, \citenamefont {Lebrun}, \citenamefont {Cros}, \citenamefont {Youssef}, \citenamefont {Bortolotti}, \citenamefont {Anane},\ and\ \citenamefont {Demokritov}}]{Divinskiy2021}%
  \BibitemOpen
  \bibfield  {author} {\bibinfo {author} {\bibfnamefont {B.}~\bibnamefont {Divinskiy}}, \bibinfo {author} {\bibfnamefont {H.}~\bibnamefont {Merbouche}}, \bibinfo {author} {\bibfnamefont {V.~E.}\ \bibnamefont {Demidov}}, \bibinfo {author} {\bibfnamefont {K.~O.}\ \bibnamefont {Nikolaev}}, \bibinfo {author} {\bibfnamefont {L.}~\bibnamefont {Soumah}}, \bibinfo {author} {\bibfnamefont {D.}~\bibnamefont {Gou{\'e}r{\'e}}}, \bibinfo {author} {\bibfnamefont {R.}~\bibnamefont {Lebrun}}, \bibinfo {author} {\bibfnamefont {V.}~\bibnamefont {Cros}}, \bibinfo {author} {\bibfnamefont {Jamal~Ben}\ \bibnamefont {Youssef}}, \bibinfo {author} {\bibfnamefont {P.}~\bibnamefont {Bortolotti}}, \bibinfo {author} {\bibfnamefont {A.}~\bibnamefont {Anane}}, \ and\ \bibinfo {author} {\bibfnamefont {S.~O.}\ \bibnamefont {Demokritov}},\ }\bibfield  {title} {\enquote {\bibinfo {title} {Evidence for spin current driven bose-einstein condensation of magnons},}\ }\href {\doibase 10.1038/s41467-021-26790-y} {\bibfield  {journal} {\bibinfo
  {journal} {Nature Communications}\ }\textbf {\bibinfo {volume} {12}},\ \bibinfo {pages} {6541} (\bibinfo {year} {2021})}\BibitemShut {NoStop}%
\bibitem [{\citenamefont {Nikuni}\ \emph {et~al.}(2000)\citenamefont {Nikuni}, \citenamefont {Oshikawa}, \citenamefont {Oosawa},\ and\ \citenamefont {Tanaka}}]{Nikuni2000}%
  \BibitemOpen
  \bibfield  {author} {\bibinfo {author} {\bibfnamefont {T.}~\bibnamefont {Nikuni}}, \bibinfo {author} {\bibfnamefont {M.}~\bibnamefont {Oshikawa}}, \bibinfo {author} {\bibfnamefont {A.}~\bibnamefont {Oosawa}}, \ and\ \bibinfo {author} {\bibfnamefont {H.}~\bibnamefont {Tanaka}},\ }\bibfield  {title} {\enquote {\bibinfo {title} {Bose-einstein condensation of dilute magnons in ${\mathrm{tlcucl}}_{3}$},}\ }\href {\doibase 10.1103/PhysRevLett.84.5868} {\bibfield  {journal} {\bibinfo  {journal} {Phys. Rev. Lett.}\ }\textbf {\bibinfo {volume} {84}},\ \bibinfo {pages} {5868--5871} (\bibinfo {year} {2000})}\BibitemShut {NoStop}%
\bibitem [{\citenamefont {R{\"u}egg}\ \emph {et~al.}(2003)\citenamefont {R{\"u}egg}, \citenamefont {Cavadini}, \citenamefont {Furrer}, \citenamefont {G{\"u}del}, \citenamefont {Kr{\"a}mer}, \citenamefont {Mutka}, \citenamefont {Wildes}, \citenamefont {Habicht},\ and\ \citenamefont {Vorderwisch}}]{Ruegg2003}%
  \BibitemOpen
  \bibfield  {author} {\bibinfo {author} {\bibfnamefont {Ch.}\ \bibnamefont {R{\"u}egg}}, \bibinfo {author} {\bibfnamefont {N.}~\bibnamefont {Cavadini}}, \bibinfo {author} {\bibfnamefont {A.}~\bibnamefont {Furrer}}, \bibinfo {author} {\bibfnamefont {H.-U.}\ \bibnamefont {G{\"u}del}}, \bibinfo {author} {\bibfnamefont {K.}~\bibnamefont {Kr{\"a}mer}}, \bibinfo {author} {\bibfnamefont {H.}~\bibnamefont {Mutka}}, \bibinfo {author} {\bibfnamefont {A.}~\bibnamefont {Wildes}}, \bibinfo {author} {\bibfnamefont {K.}~\bibnamefont {Habicht}}, \ and\ \bibinfo {author} {\bibfnamefont {P.}~\bibnamefont {Vorderwisch}},\ }\bibfield  {title} {\enquote {\bibinfo {title} {Bose--einstein condensation of the triplet states in the magnetic insulator tlcucl3},}\ }\href {\doibase 10.1038/nature01617} {\bibfield  {journal} {\bibinfo  {journal} {Nature}\ }\textbf {\bibinfo {volume} {423}},\ \bibinfo {pages} {62--65} (\bibinfo {year} {2003})}\BibitemShut {NoStop}%
\bibitem [{\citenamefont {Giamarchi}\ \emph {et~al.}(2008)\citenamefont {Giamarchi}, \citenamefont {R{\"u}egg},\ and\ \citenamefont {Tchernyshyov}}]{Giamarchi2008}%
  \BibitemOpen
  \bibfield  {author} {\bibinfo {author} {\bibfnamefont {Thierry}\ \bibnamefont {Giamarchi}}, \bibinfo {author} {\bibfnamefont {Christian}\ \bibnamefont {R{\"u}egg}}, \ and\ \bibinfo {author} {\bibfnamefont {Oleg}\ \bibnamefont {Tchernyshyov}},\ }\bibfield  {title} {\enquote {\bibinfo {title} {Bose--einstein condensation in magnetic insulators},}\ }\href {\doibase 10.1038/nphys893} {\bibfield  {journal} {\bibinfo  {journal} {Nature Physics}\ }\textbf {\bibinfo {volume} {4}},\ \bibinfo {pages} {198--204} (\bibinfo {year} {2008})}\BibitemShut {NoStop}%
\bibitem [{\citenamefont {Aczel}\ \emph {et~al.}(2009)\citenamefont {Aczel}, \citenamefont {Kohama}, \citenamefont {Marcenat}, \citenamefont {Weickert}, \citenamefont {Jaime}, \citenamefont {Ayala-Valenzuela}, \citenamefont {McDonald}, \citenamefont {Selesnic}, \citenamefont {Dabkowska},\ and\ \citenamefont {Luke}}]{Aczel2009}%
  \BibitemOpen
  \bibfield  {author} {\bibinfo {author} {\bibfnamefont {A.~A.}\ \bibnamefont {Aczel}}, \bibinfo {author} {\bibfnamefont {Y.}~\bibnamefont {Kohama}}, \bibinfo {author} {\bibfnamefont {C.}~\bibnamefont {Marcenat}}, \bibinfo {author} {\bibfnamefont {F.}~\bibnamefont {Weickert}}, \bibinfo {author} {\bibfnamefont {M.}~\bibnamefont {Jaime}}, \bibinfo {author} {\bibfnamefont {O.~E.}\ \bibnamefont {Ayala-Valenzuela}}, \bibinfo {author} {\bibfnamefont {R.~D.}\ \bibnamefont {McDonald}}, \bibinfo {author} {\bibfnamefont {S.~D.}\ \bibnamefont {Selesnic}}, \bibinfo {author} {\bibfnamefont {H.~A.}\ \bibnamefont {Dabkowska}}, \ and\ \bibinfo {author} {\bibfnamefont {G.~M.}\ \bibnamefont {Luke}},\ }\bibfield  {title} {\enquote {\bibinfo {title} {Field-induced bose-einstein condensation of triplons up to 8 k in ${\mathrm{sr}}_{3}{\mathrm{cr}}_{2}{\mathbf{o}}_{8}$},}\ }\href {\doibase 10.1103/PhysRevLett.103.207203} {\bibfield  {journal} {\bibinfo  {journal} {Phys. Rev. Lett.}\ }\textbf {\bibinfo {volume} {103}},\ \bibinfo
  {pages} {207203} (\bibinfo {year} {2009})}\BibitemShut {NoStop}%
\bibitem [{\citenamefont {Sonin}(2010)}]{Sonin2010}%
  \BibitemOpen
  \bibfield  {author} {\bibinfo {author} {\bibfnamefont {E.B.}\ \bibnamefont {Sonin}},\ }\bibfield  {title} {\enquote {\bibinfo {title} {Spin currents and spin superfluidity},}\ }\href {\doibase 10.1080/00018731003739943} {\bibfield  {journal} {\bibinfo  {journal} {Advances in Physics}\ }\textbf {\bibinfo {volume} {59}},\ \bibinfo {pages} {181--255} (\bibinfo {year} {2010})},\ \Eprint {http://arxiv.org/abs/https://doi.org/10.1080/00018731003739943} {https://doi.org/10.1080/00018731003739943} \BibitemShut {NoStop}%
\bibitem [{\citenamefont {Zapf}\ \emph {et~al.}(2014)\citenamefont {Zapf}, \citenamefont {Jaime},\ and\ \citenamefont {Batista}}]{Zapf2014}%
  \BibitemOpen
  \bibfield  {author} {\bibinfo {author} {\bibfnamefont {Vivien}\ \bibnamefont {Zapf}}, \bibinfo {author} {\bibfnamefont {Marcelo}\ \bibnamefont {Jaime}}, \ and\ \bibinfo {author} {\bibfnamefont {C.~D.}\ \bibnamefont {Batista}},\ }\bibfield  {title} {\enquote {\bibinfo {title} {Bose-einstein condensation in quantum magnets},}\ }\href {\doibase 10.1103/RevModPhys.86.563} {\bibfield  {journal} {\bibinfo  {journal} {Rev. Mod. Phys.}\ }\textbf {\bibinfo {volume} {86}},\ \bibinfo {pages} {563--614} (\bibinfo {year} {2014})}\BibitemShut {NoStop}%
\bibitem [{\citenamefont {Yuan}\ \emph {et~al.}(2018)\citenamefont {Yuan}, \citenamefont {Zhu}, \citenamefont {Su}, \citenamefont {Yao}, \citenamefont {Xing}, \citenamefont {Chen}, \citenamefont {Ma}, \citenamefont {Lin}, \citenamefont {Shi}, \citenamefont {Shindou}, \citenamefont {Xie},\ and\ \citenamefont {Han}}]{Yuan2018}%
  \BibitemOpen
  \bibfield  {author} {\bibinfo {author} {\bibfnamefont {Wei}\ \bibnamefont {Yuan}}, \bibinfo {author} {\bibfnamefont {Qiong}\ \bibnamefont {Zhu}}, \bibinfo {author} {\bibfnamefont {Tang}\ \bibnamefont {Su}}, \bibinfo {author} {\bibfnamefont {Yunyan}\ \bibnamefont {Yao}}, \bibinfo {author} {\bibfnamefont {Wenyu}\ \bibnamefont {Xing}}, \bibinfo {author} {\bibfnamefont {Yangyang}\ \bibnamefont {Chen}}, \bibinfo {author} {\bibfnamefont {Yang}\ \bibnamefont {Ma}}, \bibinfo {author} {\bibfnamefont {Xi}~\bibnamefont {Lin}}, \bibinfo {author} {\bibfnamefont {Jing}\ \bibnamefont {Shi}}, \bibinfo {author} {\bibfnamefont {Ryuichi}\ \bibnamefont {Shindou}}, \bibinfo {author} {\bibfnamefont {X.~C.}\ \bibnamefont {Xie}}, \ and\ \bibinfo {author} {\bibfnamefont {Wei}\ \bibnamefont {Han}},\ }\bibfield  {title} {\enquote {\bibinfo {title} {Experimental signatures of spin superfluid ground state in canted antiferromagnet cr<sub>2</sub>o<sub>3</sub> via nonlocal spin transport},}\ }\href {\doibase 10.1126/sciadv.aat1098}
  {\bibfield  {journal} {\bibinfo  {journal} {Science Advances}\ }\textbf {\bibinfo {volume} {4}},\ \bibinfo {pages} {eaat1098} (\bibinfo {year} {2018})},\ \Eprint {http://arxiv.org/abs/https://www.science.org/doi/pdf/10.1126/sciadv.aat1098} {https://www.science.org/doi/pdf/10.1126/sciadv.aat1098} \BibitemShut {NoStop}%
\bibitem [{\citenamefont {Esaki}\ \emph {et~al.}(2024)\citenamefont {Esaki}, \citenamefont {Akagi},\ and\ \citenamefont {Katsura}}]{Esaki2024}%
  \BibitemOpen
  \bibfield  {author} {\bibinfo {author} {\bibfnamefont {Nanse}\ \bibnamefont {Esaki}}, \bibinfo {author} {\bibfnamefont {Yutaka}\ \bibnamefont {Akagi}}, \ and\ \bibinfo {author} {\bibfnamefont {Hosho}\ \bibnamefont {Katsura}},\ }\bibfield  {title} {\enquote {\bibinfo {title} {Electric field induced thermal hall effect of triplons in the quantum dimer magnets $\mathit{X}{\mathrm{cucl}}_{3}$ ($\mathit{X}=\mathrm{Tl},\mathrm{K}$)},}\ }\href {\doibase 10.1103/PhysRevResearch.6.L032050} {\bibfield  {journal} {\bibinfo  {journal} {Phys. Rev. Res.}\ }\textbf {\bibinfo {volume} {6}},\ \bibinfo {pages} {L032050} (\bibinfo {year} {2024})}\BibitemShut {NoStop}%
\bibitem [{\citenamefont {Kimura}\ \emph {et~al.}(2016)\citenamefont {Kimura}, \citenamefont {Kakihata}, \citenamefont {Sawada}, \citenamefont {Watanabe}, \citenamefont {Matsumoto}, \citenamefont {Hagiwara},\ and\ \citenamefont {Tanaka}}]{Kimura2016}%
  \BibitemOpen
  \bibfield  {author} {\bibinfo {author} {\bibfnamefont {S.}~\bibnamefont {Kimura}}, \bibinfo {author} {\bibfnamefont {K.}~\bibnamefont {Kakihata}}, \bibinfo {author} {\bibfnamefont {Y.}~\bibnamefont {Sawada}}, \bibinfo {author} {\bibfnamefont {K.}~\bibnamefont {Watanabe}}, \bibinfo {author} {\bibfnamefont {M.}~\bibnamefont {Matsumoto}}, \bibinfo {author} {\bibfnamefont {M.}~\bibnamefont {Hagiwara}}, \ and\ \bibinfo {author} {\bibfnamefont {H.}~\bibnamefont {Tanaka}},\ }\bibfield  {title} {\enquote {\bibinfo {title} {Ferroelectricity by bose--einstein condensation in a quantum magnet},}\ }\href {\doibase 10.1038/ncomms12822} {\bibfield  {journal} {\bibinfo  {journal} {Nature Communications}\ }\textbf {\bibinfo {volume} {7}},\ \bibinfo {pages} {12822} (\bibinfo {year} {2016})}\BibitemShut {NoStop}%
\bibitem [{\citenamefont {Kimura}\ \emph {et~al.}(2017)\citenamefont {Kimura}, \citenamefont {Kakihata}, \citenamefont {Sawada}, \citenamefont {Watanabe}, \citenamefont {Matsumoto}, \citenamefont {Hagiwara},\ and\ \citenamefont {Tanaka}}]{Kimura2017}%
  \BibitemOpen
  \bibfield  {author} {\bibinfo {author} {\bibfnamefont {Shojiro}\ \bibnamefont {Kimura}}, \bibinfo {author} {\bibfnamefont {Kento}\ \bibnamefont {Kakihata}}, \bibinfo {author} {\bibfnamefont {Yuya}\ \bibnamefont {Sawada}}, \bibinfo {author} {\bibfnamefont {Kazuo}\ \bibnamefont {Watanabe}}, \bibinfo {author} {\bibfnamefont {Masashige}\ \bibnamefont {Matsumoto}}, \bibinfo {author} {\bibfnamefont {Masayuki}\ \bibnamefont {Hagiwara}}, \ and\ \bibinfo {author} {\bibfnamefont {Hidekazu}\ \bibnamefont {Tanaka}},\ }\bibfield  {title} {\enquote {\bibinfo {title} {Magnetoelectric effect in the quantum spin gap system ${\mathrm{tlcucl}}_{3}$},}\ }\href {\doibase 10.1103/PhysRevB.95.184420} {\bibfield  {journal} {\bibinfo  {journal} {Phys. Rev. B}\ }\textbf {\bibinfo {volume} {95}},\ \bibinfo {pages} {184420} (\bibinfo {year} {2017})}\BibitemShut {NoStop}%
\bibitem [{\citenamefont {Kimura}\ \emph {et~al.}(2020)\citenamefont {Kimura}, \citenamefont {Matsumoto},\ and\ \citenamefont {Tanaka}}]{Kimura2020}%
  \BibitemOpen
  \bibfield  {author} {\bibinfo {author} {\bibfnamefont {Shojiro}\ \bibnamefont {Kimura}}, \bibinfo {author} {\bibfnamefont {Masashige}\ \bibnamefont {Matsumoto}}, \ and\ \bibinfo {author} {\bibfnamefont {Hidekazu}\ \bibnamefont {Tanaka}},\ }\bibfield  {title} {\enquote {\bibinfo {title} {Electrical switching of the nonreciprocal directional microwave response in a triplon bose-einstein condensate},}\ }\href {\doibase 10.1103/PhysRevLett.124.217401} {\bibfield  {journal} {\bibinfo  {journal} {Phys. Rev. Lett.}\ }\textbf {\bibinfo {volume} {124}},\ \bibinfo {pages} {217401} (\bibinfo {year} {2020})}\BibitemShut {NoStop}%
\bibitem [{\citenamefont {Sakurai}\ \emph {et~al.}(2020)\citenamefont {Sakurai}, \citenamefont {Kimura}, \citenamefont {Awaji}, \citenamefont {Matsumoto},\ and\ \citenamefont {Tanaka}}]{Sakurai2020}%
  \BibitemOpen
  \bibfield  {author} {\bibinfo {author} {\bibfnamefont {Kyosuke}\ \bibnamefont {Sakurai}}, \bibinfo {author} {\bibfnamefont {Shojiro}\ \bibnamefont {Kimura}}, \bibinfo {author} {\bibfnamefont {Staoshi}\ \bibnamefont {Awaji}}, \bibinfo {author} {\bibfnamefont {Masashige}\ \bibnamefont {Matsumoto}}, \ and\ \bibinfo {author} {\bibfnamefont {Hidekazu}\ \bibnamefont {Tanaka}},\ }\bibfield  {title} {\enquote {\bibinfo {title} {Spin-driven ferroelectricity in the quantum magnet ${\mathrm{tlcucl}}_{3}$ under high pressure},}\ }\href {\doibase 10.1103/PhysRevB.102.064104} {\bibfield  {journal} {\bibinfo  {journal} {Phys. Rev. B}\ }\textbf {\bibinfo {volume} {102}},\ \bibinfo {pages} {064104} (\bibinfo {year} {2020})}\BibitemShut {NoStop}%
\bibitem [{\citenamefont {Aharonov}\ and\ \citenamefont {Casher}(1984)}]{Aharanov1984}%
  \BibitemOpen
  \bibfield  {author} {\bibinfo {author} {\bibfnamefont {Y.}~\bibnamefont {Aharonov}}\ and\ \bibinfo {author} {\bibfnamefont {A.}~\bibnamefont {Casher}},\ }\bibfield  {title} {\enquote {\bibinfo {title} {Topological quantum effects for neutral particles},}\ }\href {\doibase 10.1103/PhysRevLett.53.319} {\bibfield  {journal} {\bibinfo  {journal} {Phys. Rev. Lett.}\ }\textbf {\bibinfo {volume} {53}},\ \bibinfo {pages} {319--321} (\bibinfo {year} {1984})}\BibitemShut {NoStop}%
\bibitem [{\citenamefont {Meier}\ and\ \citenamefont {Loss}(2003)}]{Meier2003}%
  \BibitemOpen
  \bibfield  {author} {\bibinfo {author} {\bibfnamefont {Florian}\ \bibnamefont {Meier}}\ and\ \bibinfo {author} {\bibfnamefont {Daniel}\ \bibnamefont {Loss}},\ }\bibfield  {title} {\enquote {\bibinfo {title} {Magnetization transport and quantized spin conductance},}\ }\href {\doibase 10.1103/PhysRevLett.90.167204} {\bibfield  {journal} {\bibinfo  {journal} {Phys. Rev. Lett.}\ }\textbf {\bibinfo {volume} {90}},\ \bibinfo {pages} {167204} (\bibinfo {year} {2003})}\BibitemShut {NoStop}%
\bibitem [{\citenamefont {Nakata}\ \emph {et~al.}(2017{\natexlab{a}})\citenamefont {Nakata}, \citenamefont {Kim}, \citenamefont {Klinovaja},\ and\ \citenamefont {Loss}}]{Nakata2017}%
  \BibitemOpen
  \bibfield  {author} {\bibinfo {author} {\bibfnamefont {Kouki}\ \bibnamefont {Nakata}}, \bibinfo {author} {\bibfnamefont {Se~Kwon}\ \bibnamefont {Kim}}, \bibinfo {author} {\bibfnamefont {Jelena}\ \bibnamefont {Klinovaja}}, \ and\ \bibinfo {author} {\bibfnamefont {Daniel}\ \bibnamefont {Loss}},\ }\bibfield  {title} {\enquote {\bibinfo {title} {Magnonic topological insulators in antiferromagnets},}\ }\href {\doibase 10.1103/PhysRevB.96.224414} {\bibfield  {journal} {\bibinfo  {journal} {Phys. Rev. B}\ }\textbf {\bibinfo {volume} {96}},\ \bibinfo {pages} {224414} (\bibinfo {year} {2017}{\natexlab{a}})}\BibitemShut {NoStop}%
\bibitem [{\citenamefont {Nakata}\ \emph {et~al.}(2017{\natexlab{b}})\citenamefont {Nakata}, \citenamefont {Klinovaja},\ and\ \citenamefont {Loss}}]{Nakata2017_Landu_level}%
  \BibitemOpen
  \bibfield  {author} {\bibinfo {author} {\bibfnamefont {Kouki}\ \bibnamefont {Nakata}}, \bibinfo {author} {\bibfnamefont {Jelena}\ \bibnamefont {Klinovaja}}, \ and\ \bibinfo {author} {\bibfnamefont {Daniel}\ \bibnamefont {Loss}},\ }\bibfield  {title} {\enquote {\bibinfo {title} {Magnonic quantum hall effect and wiedemann-franz law},}\ }\href {\doibase 10.1103/PhysRevB.95.125429} {\bibfield  {journal} {\bibinfo  {journal} {Phys. Rev. B}\ }\textbf {\bibinfo {volume} {95}},\ \bibinfo {pages} {125429} (\bibinfo {year} {2017}{\natexlab{b}})}\BibitemShut {NoStop}%
\bibitem [{\citenamefont {Wang}\ \emph {et~al.}(2024)\citenamefont {Wang}, \citenamefont {Zhu},\ and\ \citenamefont {Su}}]{Wang2024}%
  \BibitemOpen
  \bibfield  {author} {\bibinfo {author} {\bibfnamefont {YuanDong}\ \bibnamefont {Wang}}, \bibinfo {author} {\bibfnamefont {Zhen-Gang}\ \bibnamefont {Zhu}}, \ and\ \bibinfo {author} {\bibfnamefont {Gang}\ \bibnamefont {Su}},\ }\bibfield  {title} {\enquote {\bibinfo {title} {Magnon spin photogalvanic effect induced by aharonov-casher phase},}\ }\href {\doibase 10.1103/PhysRevB.110.054434} {\bibfield  {journal} {\bibinfo  {journal} {Phys. Rev. B}\ }\textbf {\bibinfo {volume} {110}},\ \bibinfo {pages} {054434} (\bibinfo {year} {2024})}\BibitemShut {NoStop}%
\bibitem [{\citenamefont {Su}\ and\ \citenamefont {Wang}(2017)}]{SuWang2017}%
  \BibitemOpen
  \bibfield  {author} {\bibinfo {author} {\bibfnamefont {Ying}\ \bibnamefont {Su}}\ and\ \bibinfo {author} {\bibfnamefont {X.~R.}\ \bibnamefont {Wang}},\ }\bibfield  {title} {\enquote {\bibinfo {title} {Chiral anomaly of weyl magnons in stacked honeycomb ferromagnets},}\ }\href {\doibase 10.1103/PhysRevB.96.104437} {\bibfield  {journal} {\bibinfo  {journal} {Phys. Rev. B}\ }\textbf {\bibinfo {volume} {96}},\ \bibinfo {pages} {104437} (\bibinfo {year} {2017})}\BibitemShut {NoStop}%
\bibitem [{\citenamefont {Wagh}\ and\ \citenamefont {Rakhecha}(1996)}]{Wagh1996}%
  \BibitemOpen
  \bibfield  {author} {\bibinfo {author} {\bibfnamefont {A.G.}\ \bibnamefont {Wagh}}\ and\ \bibinfo {author} {\bibfnamefont {V.C.}\ \bibnamefont {Rakhecha}},\ }\bibfield  {title} {\enquote {\bibinfo {title} {Quantum physics with neutrons},}\ }\href {\doibase https://doi.org/10.1016/0146-6410(96)00056-7} {\bibfield  {journal} {\bibinfo  {journal} {Progress in Particle and Nuclear Physics}\ }\textbf {\bibinfo {volume} {37}},\ \bibinfo {pages} {485--563} (\bibinfo {year} {1996})}\BibitemShut {NoStop}%
\bibitem [{\citenamefont {Sangster}\ \emph {et~al.}(1995)\citenamefont {Sangster}, \citenamefont {Hinds}, \citenamefont {Barnett}, \citenamefont {Riis},\ and\ \citenamefont {Sinclair}}]{Sangster1995}%
  \BibitemOpen
  \bibfield  {author} {\bibinfo {author} {\bibfnamefont {Karin}\ \bibnamefont {Sangster}}, \bibinfo {author} {\bibfnamefont {E.~A.}\ \bibnamefont {Hinds}}, \bibinfo {author} {\bibfnamefont {Stephen~M.}\ \bibnamefont {Barnett}}, \bibinfo {author} {\bibfnamefont {Erling}\ \bibnamefont {Riis}}, \ and\ \bibinfo {author} {\bibfnamefont {A.~G.}\ \bibnamefont {Sinclair}},\ }\bibfield  {title} {\enquote {\bibinfo {title} {Aharonov-casher phase in an atomic system},}\ }\href {\doibase 10.1103/PhysRevA.51.1776} {\bibfield  {journal} {\bibinfo  {journal} {Phys. Rev. A}\ }\textbf {\bibinfo {volume} {51}},\ \bibinfo {pages} {1776--1786} (\bibinfo {year} {1995})}\BibitemShut {NoStop}%
\bibitem [{\citenamefont {Avishai}\ \emph {et~al.}(2019)\citenamefont {Avishai}, \citenamefont {Totsuka},\ and\ \citenamefont {Nagaosa}}]{Avishai2019}%
  \BibitemOpen
  \bibfield  {author} {\bibinfo {author} {\bibfnamefont {Yshai}\ \bibnamefont {Avishai}}, \bibinfo {author} {\bibfnamefont {Keisuke}\ \bibnamefont {Totsuka}}, \ and\ \bibinfo {author} {\bibfnamefont {Naoto}\ \bibnamefont {Nagaosa}},\ }\bibfield  {title} {\enquote {\bibinfo {title} {Non-abelian aharonov–casher phase factor in mesoscopic systems},}\ }\href {\doibase 10.7566/JPSJ.88.084705} {\bibfield  {journal} {\bibinfo  {journal} {Journal of the Physical Society of Japan}\ }\textbf {\bibinfo {volume} {88}},\ \bibinfo {pages} {084705} (\bibinfo {year} {2019})},\ \Eprint {http://arxiv.org/abs/https://doi.org/10.7566/JPSJ.88.084705} {https://doi.org/10.7566/JPSJ.88.084705} \BibitemShut {NoStop}%
\bibitem [{\citenamefont {Avishai}\ and\ \citenamefont {Band}(2023)}]{Avishai2023_review}%
  \BibitemOpen
  \bibfield  {author} {\bibinfo {author} {\bibfnamefont {Y.}~\bibnamefont {Avishai}}\ and\ \bibinfo {author} {\bibfnamefont {Y.~B.}\ \bibnamefont {Band}},\ }\href@noop {} {\enquote {\bibinfo {title} {Aharonov--bohm and aharonov--casher effects in meso-scopic physics: A brief review},}\ } (\bibinfo {year} {2023}),\ \Eprint {http://arxiv.org/abs/2302.06300} {arXiv:2302.06300 [cond-mat.mes-hall]} \BibitemShut {NoStop}%
\bibitem [{\citenamefont {K\"onig}\ \emph {et~al.}(2006)\citenamefont {K\"onig}, \citenamefont {Tschetschetkin}, \citenamefont {Hankiewicz}, \citenamefont {Sinova}, \citenamefont {Hock}, \citenamefont {Daumer}, \citenamefont {Sch\"afer}, \citenamefont {Becker}, \citenamefont {Buhmann},\ and\ \citenamefont {Molenkamp}}]{Konig2006}%
  \BibitemOpen
  \bibfield  {author} {\bibinfo {author} {\bibfnamefont {M.}~\bibnamefont {K\"onig}}, \bibinfo {author} {\bibfnamefont {A.}~\bibnamefont {Tschetschetkin}}, \bibinfo {author} {\bibfnamefont {E.~M.}\ \bibnamefont {Hankiewicz}}, \bibinfo {author} {\bibfnamefont {Jairo}\ \bibnamefont {Sinova}}, \bibinfo {author} {\bibfnamefont {V.}~\bibnamefont {Hock}}, \bibinfo {author} {\bibfnamefont {V.}~\bibnamefont {Daumer}}, \bibinfo {author} {\bibfnamefont {M.}~\bibnamefont {Sch\"afer}}, \bibinfo {author} {\bibfnamefont {C.~R.}\ \bibnamefont {Becker}}, \bibinfo {author} {\bibfnamefont {H.}~\bibnamefont {Buhmann}}, \ and\ \bibinfo {author} {\bibfnamefont {L.~W.}\ \bibnamefont {Molenkamp}},\ }\bibfield  {title} {\enquote {\bibinfo {title} {Direct observation of the aharonov-casher phase},}\ }\href {\doibase 10.1103/PhysRevLett.96.076804} {\bibfield  {journal} {\bibinfo  {journal} {Phys. Rev. Lett.}\ }\textbf {\bibinfo {volume} {96}},\ \bibinfo {pages} {076804} (\bibinfo {year} {2006})}\BibitemShut {NoStop}%
\bibitem [{\citenamefont {Shekhter}\ \emph {et~al.}(2022)\citenamefont {Shekhter}, \citenamefont {Entin-Wohlman}, \citenamefont {Jonson},\ and\ \citenamefont {Aharony}}]{Shekhter2022}%
  \BibitemOpen
  \bibfield  {author} {\bibinfo {author} {\bibfnamefont {R.~I.}\ \bibnamefont {Shekhter}}, \bibinfo {author} {\bibfnamefont {O.}~\bibnamefont {Entin-Wohlman}}, \bibinfo {author} {\bibfnamefont {M.}~\bibnamefont {Jonson}}, \ and\ \bibinfo {author} {\bibfnamefont {A.}~\bibnamefont {Aharony}},\ }\bibfield  {title} {\enquote {\bibinfo {title} {Magnetoconductance anisotropies and aharonov-casher phases},}\ }\href {\doibase 10.1103/PhysRevLett.129.037704} {\bibfield  {journal} {\bibinfo  {journal} {Phys. Rev. Lett.}\ }\textbf {\bibinfo {volume} {129}},\ \bibinfo {pages} {037704} (\bibinfo {year} {2022})}\BibitemShut {NoStop}%
\bibitem [{\citenamefont {Yamane}\ \emph {et~al.}(2022)\citenamefont {Yamane}, \citenamefont {Fukami},\ and\ \citenamefont {Ieda}}]{Yamane2022}%
  \BibitemOpen
  \bibfield  {author} {\bibinfo {author} {\bibfnamefont {Yuta}\ \bibnamefont {Yamane}}, \bibinfo {author} {\bibfnamefont {Shunsuke}\ \bibnamefont {Fukami}}, \ and\ \bibinfo {author} {\bibfnamefont {Jun'ichi}\ \bibnamefont {Ieda}},\ }\bibfield  {title} {\enquote {\bibinfo {title} {Theory of emergent inductance with spin-orbit coupling effects},}\ }\href {\doibase 10.1103/PhysRevLett.128.147201} {\bibfield  {journal} {\bibinfo  {journal} {Phys. Rev. Lett.}\ }\textbf {\bibinfo {volume} {128}},\ \bibinfo {pages} {147201} (\bibinfo {year} {2022})}\BibitemShut {NoStop}%
\bibitem [{\citenamefont {Balatsky}\ and\ \citenamefont {Altshuler}(1993)}]{Balatsky1993}%
  \BibitemOpen
  \bibfield  {author} {\bibinfo {author} {\bibfnamefont {A.~V.}\ \bibnamefont {Balatsky}}\ and\ \bibinfo {author} {\bibfnamefont {B.~L.}\ \bibnamefont {Altshuler}},\ }\bibfield  {title} {\enquote {\bibinfo {title} {Persistent spin and mass currents and aharonov-casher effect},}\ }\href {\doibase 10.1103/PhysRevLett.70.1678} {\bibfield  {journal} {\bibinfo  {journal} {Phys. Rev. Lett.}\ }\textbf {\bibinfo {volume} {70}},\ \bibinfo {pages} {1678--1681} (\bibinfo {year} {1993})}\BibitemShut {NoStop}%
\bibitem [{\citenamefont {Grosfeld}\ and\ \citenamefont {Stern}(2011)}]{Grosfeld2011}%
  \BibitemOpen
  \bibfield  {author} {\bibinfo {author} {\bibfnamefont {Eytan}\ \bibnamefont {Grosfeld}}\ and\ \bibinfo {author} {\bibfnamefont {Ady}\ \bibnamefont {Stern}},\ }\bibfield  {title} {\enquote {\bibinfo {title} {Observing majorana bound states of josephson vortices in topological superconductors},}\ }\href {\doibase 10.1073/pnas.1101469108} {\bibfield  {journal} {\bibinfo  {journal} {Proceedings of the National Academy of Sciences}\ }\textbf {\bibinfo {volume} {108}},\ \bibinfo {pages} {11810--11814} (\bibinfo {year} {2011})},\ \Eprint {http://arxiv.org/abs/https://www.pnas.org/doi/pdf/10.1073/pnas.1101469108} {https://www.pnas.org/doi/pdf/10.1073/pnas.1101469108} \BibitemShut {NoStop}%
\bibitem [{\citenamefont {Elion}\ \emph {et~al.}(1993)\citenamefont {Elion}, \citenamefont {Wachters}, \citenamefont {Sohn},\ and\ \citenamefont {Mooij}}]{Elion1993}%
  \BibitemOpen
  \bibfield  {author} {\bibinfo {author} {\bibfnamefont {W.~J.}\ \bibnamefont {Elion}}, \bibinfo {author} {\bibfnamefont {J.~J.}\ \bibnamefont {Wachters}}, \bibinfo {author} {\bibfnamefont {L.~L.}\ \bibnamefont {Sohn}}, \ and\ \bibinfo {author} {\bibfnamefont {J.~E.}\ \bibnamefont {Mooij}},\ }\bibfield  {title} {\enquote {\bibinfo {title} {Observation of the aharonov-casher effect for vortices in josephson-junction arrays},}\ }\href {\doibase 10.1103/PhysRevLett.71.2311} {\bibfield  {journal} {\bibinfo  {journal} {Phys. Rev. Lett.}\ }\textbf {\bibinfo {volume} {71}},\ \bibinfo {pages} {2311--2314} (\bibinfo {year} {1993})}\BibitemShut {NoStop}%
\bibitem [{\citenamefont {Seidov}\ and\ \citenamefont {Fistul}(2021)}]{Seidov2021}%
  \BibitemOpen
  \bibfield  {author} {\bibinfo {author} {\bibfnamefont {S.~S.}\ \bibnamefont {Seidov}}\ and\ \bibinfo {author} {\bibfnamefont {M.~V.}\ \bibnamefont {Fistul}},\ }\bibfield  {title} {\enquote {\bibinfo {title} {Quantum dynamics of a single fluxon in josephson-junction parallel arrays with large kinetic inductances},}\ }\href {\doibase 10.1103/PhysRevA.103.062410} {\bibfield  {journal} {\bibinfo  {journal} {Phys. Rev. A}\ }\textbf {\bibinfo {volume} {103}},\ \bibinfo {pages} {062410} (\bibinfo {year} {2021})}\BibitemShut {NoStop}%
\bibitem [{\citenamefont {Zhang}\ \emph {et~al.}(2014)\citenamefont {Zhang}, \citenamefont {Liu}, \citenamefont {Flatt\'e},\ and\ \citenamefont {Tang}}]{Zhang2014}%
  \BibitemOpen
  \bibfield  {author} {\bibinfo {author} {\bibfnamefont {Xufeng}\ \bibnamefont {Zhang}}, \bibinfo {author} {\bibfnamefont {Tianyu}\ \bibnamefont {Liu}}, \bibinfo {author} {\bibfnamefont {Michael~E.}\ \bibnamefont {Flatt\'e}}, \ and\ \bibinfo {author} {\bibfnamefont {Hong~X.}\ \bibnamefont {Tang}},\ }\bibfield  {title} {\enquote {\bibinfo {title} {Electric-field coupling to spin waves in a centrosymmetric ferrite},}\ }\href {\doibase 10.1103/PhysRevLett.113.037202} {\bibfield  {journal} {\bibinfo  {journal} {Phys. Rev. Lett.}\ }\textbf {\bibinfo {volume} {113}},\ \bibinfo {pages} {037202} (\bibinfo {year} {2014})}\BibitemShut {NoStop}%
\bibitem [{\citenamefont {Serha}\ \emph {et~al.}(2023)\citenamefont {Serha}, \citenamefont {Vasyuchka}, \citenamefont {Serga},\ and\ \citenamefont {Hillebrands}}]{Serha2023}%
  \BibitemOpen
  \bibfield  {author} {\bibinfo {author} {\bibfnamefont {Rostyslav~O.}\ \bibnamefont {Serha}}, \bibinfo {author} {\bibfnamefont {Vitaliy~I.}\ \bibnamefont {Vasyuchka}}, \bibinfo {author} {\bibfnamefont {Alexander~A.}\ \bibnamefont {Serga}}, \ and\ \bibinfo {author} {\bibfnamefont {Burkard}\ \bibnamefont {Hillebrands}},\ }\bibfield  {title} {\enquote {\bibinfo {title} {Towards an experimental proof of the magnonic aharonov-casher effect},}\ }\href {\doibase 10.1103/PhysRevB.108.L220404} {\bibfield  {journal} {\bibinfo  {journal} {Phys. Rev. B}\ }\textbf {\bibinfo {volume} {108}},\ \bibinfo {pages} {L220404} (\bibinfo {year} {2023})}\BibitemShut {NoStop}%
\bibitem [{\citenamefont {Katsura}\ \emph {et~al.}(2005)\citenamefont {Katsura}, \citenamefont {Nagaosa},\ and\ \citenamefont {Balatsky}}]{Katsura2005}%
  \BibitemOpen
  \bibfield  {author} {\bibinfo {author} {\bibfnamefont {Hosho}\ \bibnamefont {Katsura}}, \bibinfo {author} {\bibfnamefont {Naoto}\ \bibnamefont {Nagaosa}}, \ and\ \bibinfo {author} {\bibfnamefont {Alexander~V.}\ \bibnamefont {Balatsky}},\ }\bibfield  {title} {\enquote {\bibinfo {title} {Spin current and magnetoelectric effect in noncollinear magnets},}\ }\href {\doibase 10.1103/PhysRevLett.95.057205} {\bibfield  {journal} {\bibinfo  {journal} {Phys. Rev. Lett.}\ }\textbf {\bibinfo {volume} {95}},\ \bibinfo {pages} {057205} (\bibinfo {year} {2005})}\BibitemShut {NoStop}%
\bibitem [{\citenamefont {Liu}\ and\ \citenamefont {Vignale}(2011)}]{Liu2011}%
  \BibitemOpen
  \bibfield  {author} {\bibinfo {author} {\bibfnamefont {Tianyu}\ \bibnamefont {Liu}}\ and\ \bibinfo {author} {\bibfnamefont {G.}~\bibnamefont {Vignale}},\ }\bibfield  {title} {\enquote {\bibinfo {title} {Electric control of spin currents and spin-wave logic},}\ }\href {\doibase 10.1103/PhysRevLett.106.247203} {\bibfield  {journal} {\bibinfo  {journal} {Phys. Rev. Lett.}\ }\textbf {\bibinfo {volume} {106}},\ \bibinfo {pages} {247203} (\bibinfo {year} {2011})}\BibitemShut {NoStop}%
\bibitem [{\citenamefont {Go}\ \emph {et~al.}(2024)\citenamefont {Go}, \citenamefont {An}, \citenamefont {Lee},\ and\ \citenamefont {Kim}}]{Go2024}%
  \BibitemOpen
  \bibfield  {author} {\bibinfo {author} {\bibfnamefont {Gyungchoon}\ \bibnamefont {Go}}, \bibinfo {author} {\bibfnamefont {Daehyeon}\ \bibnamefont {An}}, \bibinfo {author} {\bibfnamefont {Hyun-Woo}\ \bibnamefont {Lee}}, \ and\ \bibinfo {author} {\bibfnamefont {Se~Kwon}\ \bibnamefont {Kim}},\ }\bibfield  {title} {\enquote {\bibinfo {title} {Magnon orbital nernst effect in honeycomb antiferromagnets without spin--orbit coupling},}\ }\href {\doibase 10.1021/acs.nanolett.4c00430} {\bibfield  {journal} {\bibinfo  {journal} {Nano Letters}\ }\textbf {\bibinfo {volume} {24}},\ \bibinfo {pages} {5968--5974} (\bibinfo {year} {2024})}\BibitemShut {NoStop}%
\bibitem [{\citenamefont {Hirsch}(1999)}]{Hirsch1999}%
  \BibitemOpen
  \bibfield  {author} {\bibinfo {author} {\bibfnamefont {J.~E.}\ \bibnamefont {Hirsch}},\ }\bibfield  {title} {\enquote {\bibinfo {title} {Overlooked contribution to the hall effect in ferromagnetic metals},}\ }\href {\doibase 10.1103/PhysRevB.60.14787} {\bibfield  {journal} {\bibinfo  {journal} {Phys. Rev. B}\ }\textbf {\bibinfo {volume} {60}},\ \bibinfo {pages} {14787--14792} (\bibinfo {year} {1999})}\BibitemShut {NoStop}%
\bibitem [{\citenamefont {Sun}\ \emph {et~al.}(2004)\citenamefont {Sun}, \citenamefont {Guo},\ and\ \citenamefont {Wang}}]{Sun2004}%
  \BibitemOpen
  \bibfield  {author} {\bibinfo {author} {\bibfnamefont {Qing-feng}\ \bibnamefont {Sun}}, \bibinfo {author} {\bibfnamefont {Hong}\ \bibnamefont {Guo}}, \ and\ \bibinfo {author} {\bibfnamefont {Jian}\ \bibnamefont {Wang}},\ }\bibfield  {title} {\enquote {\bibinfo {title} {Spin-current-induced electric field},}\ }\href {\doibase 10.1103/PhysRevB.69.054409} {\bibfield  {journal} {\bibinfo  {journal} {Phys. Rev. B}\ }\textbf {\bibinfo {volume} {69}},\ \bibinfo {pages} {054409} (\bibinfo {year} {2004})}\BibitemShut {NoStop}%
\bibitem [{\citenamefont {Tokura}\ \emph {et~al.}(2014)\citenamefont {Tokura}, \citenamefont {Seki},\ and\ \citenamefont {Nagaosa}}]{Tokura2014}%
  \BibitemOpen
  \bibfield  {author} {\bibinfo {author} {\bibfnamefont {Yoshinori}\ \bibnamefont {Tokura}}, \bibinfo {author} {\bibfnamefont {Shinichiro}\ \bibnamefont {Seki}}, \ and\ \bibinfo {author} {\bibfnamefont {Naoto}\ \bibnamefont {Nagaosa}},\ }\bibfield  {title} {\enquote {\bibinfo {title} {Multiferroics of spin origin},}\ }\href {\doibase 10.1088/0034-4885/77/7/076501} {\bibfield  {journal} {\bibinfo  {journal} {Reports on Progress in Physics}\ }\textbf {\bibinfo {volume} {77}},\ \bibinfo {pages} {076501} (\bibinfo {year} {2014})}\BibitemShut {NoStop}%
\bibitem [{\citenamefont {Solovyev}\ \emph {et~al.}(2021)\citenamefont {Solovyev}, \citenamefont {Ono},\ and\ \citenamefont {Nikolaev}}]{Solovyev2021}%
  \BibitemOpen
  \bibfield  {author} {\bibinfo {author} {\bibfnamefont {Igor}\ \bibnamefont {Solovyev}}, \bibinfo {author} {\bibfnamefont {Ryota}\ \bibnamefont {Ono}}, \ and\ \bibinfo {author} {\bibfnamefont {Sergey}\ \bibnamefont {Nikolaev}},\ }\bibfield  {title} {\enquote {\bibinfo {title} {Magnetically induced polarization in centrosymmetric bonds},}\ }\href {\doibase 10.1103/PhysRevLett.127.187601} {\bibfield  {journal} {\bibinfo  {journal} {Phys. Rev. Lett.}\ }\textbf {\bibinfo {volume} {127}},\ \bibinfo {pages} {187601} (\bibinfo {year} {2021})}\BibitemShut {NoStop}%
\bibitem [{\citenamefont {Nakata}\ \emph {et~al.}(2014)\citenamefont {Nakata}, \citenamefont {van Hoogdalem}, \citenamefont {Simon},\ and\ \citenamefont {Loss}}]{Nakata2014}%
  \BibitemOpen
  \bibfield  {author} {\bibinfo {author} {\bibfnamefont {Kouki}\ \bibnamefont {Nakata}}, \bibinfo {author} {\bibfnamefont {Kevin~A.}\ \bibnamefont {van Hoogdalem}}, \bibinfo {author} {\bibfnamefont {Pascal}\ \bibnamefont {Simon}}, \ and\ \bibinfo {author} {\bibfnamefont {Daniel}\ \bibnamefont {Loss}},\ }\bibfield  {title} {\enquote {\bibinfo {title} {Josephson and persistent spin currents in bose-einstein condensates of magnons},}\ }\href {\doibase 10.1103/PhysRevB.90.144419} {\bibfield  {journal} {\bibinfo  {journal} {Phys. Rev. B}\ }\textbf {\bibinfo {volume} {90}},\ \bibinfo {pages} {144419} (\bibinfo {year} {2014})}\BibitemShut {NoStop}%
\bibitem [{\citenamefont {Nakata}(2021)}]{Nakata2021}%
  \BibitemOpen
  \bibfield  {author} {\bibinfo {author} {\bibfnamefont {Kouki}\ \bibnamefont {Nakata}},\ }\bibfield  {title} {\enquote {\bibinfo {title} {Optomagnonic josephson effect in antiferromagnets},}\ }\href {\doibase 10.1103/PhysRevB.104.104402} {\bibfield  {journal} {\bibinfo  {journal} {Phys. Rev. B}\ }\textbf {\bibinfo {volume} {104}},\ \bibinfo {pages} {104402} (\bibinfo {year} {2021})}\BibitemShut {NoStop}%
\bibitem [{\citenamefont {Wu}\ \emph {et~al.}(2022)\citenamefont {Wu}, \citenamefont {Jiang}, \citenamefont {Chen}, \citenamefont {Liu}, \citenamefont {Liu},\ and\ \citenamefont {Xie}}]{Wu2022}%
  \BibitemOpen
  \bibfield  {author} {\bibinfo {author} {\bibfnamefont {Yijia}\ \bibnamefont {Wu}}, \bibinfo {author} {\bibfnamefont {Hua}\ \bibnamefont {Jiang}}, \bibinfo {author} {\bibfnamefont {Hua}\ \bibnamefont {Chen}}, \bibinfo {author} {\bibfnamefont {Haiwen}\ \bibnamefont {Liu}}, \bibinfo {author} {\bibfnamefont {Jie}\ \bibnamefont {Liu}}, \ and\ \bibinfo {author} {\bibfnamefont {X.~C.}\ \bibnamefont {Xie}},\ }\bibfield  {title} {\enquote {\bibinfo {title} {Non-abelian braiding in spin superconductors utilizing the aharonov-casher effect},}\ }\href {\doibase 10.1103/PhysRevLett.128.106804} {\bibfield  {journal} {\bibinfo  {journal} {Phys. Rev. Lett.}\ }\textbf {\bibinfo {volume} {128}},\ \bibinfo {pages} {106804} (\bibinfo {year} {2022})}\BibitemShut {NoStop}%
\bibitem [{\citenamefont {Wang}\ \emph {et~al.}(2013)\citenamefont {Wang}, \citenamefont {Sun},\ and\ \citenamefont {Xie}}]{Wang2013}%
  \BibitemOpen
  \bibfield  {author} {\bibinfo {author} {\bibfnamefont {Zi-bo}\ \bibnamefont {Wang}}, \bibinfo {author} {\bibfnamefont {Qing-feng}\ \bibnamefont {Sun}}, \ and\ \bibinfo {author} {\bibfnamefont {X.~C.}\ \bibnamefont {Xie}},\ }\bibfield  {title} {\enquote {\bibinfo {title} {The electric ``meissner effect'' in spin superconductor},}\ }\href {\doibase 10.1140/epjb/e2013-40798-2} {\bibfield  {journal} {\bibinfo  {journal} {The European Physical Journal B}\ }\textbf {\bibinfo {volume} {86}},\ \bibinfo {pages} {496} (\bibinfo {year} {2013})}\BibitemShut {NoStop}%
\bibitem [{\citenamefont {Bao}\ \emph {et~al.}(2013)\citenamefont {Bao}, \citenamefont {Xie},\ and\ \citenamefont {Sun}}]{Bao2013}%
  \BibitemOpen
  \bibfield  {author} {\bibinfo {author} {\bibfnamefont {Zhi-qiang}\ \bibnamefont {Bao}}, \bibinfo {author} {\bibfnamefont {X.~C.}\ \bibnamefont {Xie}}, \ and\ \bibinfo {author} {\bibfnamefont {Qing-feng}\ \bibnamefont {Sun}},\ }\bibfield  {title} {\enquote {\bibinfo {title} {Ginzburg--landau-type theory of spin superconductivity},}\ }\href {\doibase 10.1038/ncomms3951} {\bibfield  {journal} {\bibinfo  {journal} {Nature Communications}\ }\textbf {\bibinfo {volume} {4}},\ \bibinfo {pages} {2951} (\bibinfo {year} {2013})}\BibitemShut {NoStop}%
\bibitem [{\citenamefont {Sun}\ \emph {et~al.}(2011)\citenamefont {Sun}, \citenamefont {Jiang}, \citenamefont {Yu},\ and\ \citenamefont {Xie}}]{Sun2011}%
  \BibitemOpen
  \bibfield  {author} {\bibinfo {author} {\bibfnamefont {Qing-feng}\ \bibnamefont {Sun}}, \bibinfo {author} {\bibfnamefont {Zhao-tan}\ \bibnamefont {Jiang}}, \bibinfo {author} {\bibfnamefont {Yue}\ \bibnamefont {Yu}}, \ and\ \bibinfo {author} {\bibfnamefont {X.~C.}\ \bibnamefont {Xie}},\ }\bibfield  {title} {\enquote {\bibinfo {title} {Spin superconductor in ferromagnetic graphene},}\ }\href {\doibase 10.1103/PhysRevB.84.214501} {\bibfield  {journal} {\bibinfo  {journal} {Phys. Rev. B}\ }\textbf {\bibinfo {volume} {84}},\ \bibinfo {pages} {214501} (\bibinfo {year} {2011})}\BibitemShut {NoStop}%
\bibitem [{\citenamefont {J\'erome}\ \emph {et~al.}(1967)\citenamefont {J\'erome}, \citenamefont {Rice},\ and\ \citenamefont {Kohn}}]{Jerome1967}%
  \BibitemOpen
  \bibfield  {author} {\bibinfo {author} {\bibfnamefont {D.}~\bibnamefont {J\'erome}}, \bibinfo {author} {\bibfnamefont {T.~M.}\ \bibnamefont {Rice}}, \ and\ \bibinfo {author} {\bibfnamefont {W.}~\bibnamefont {Kohn}},\ }\bibfield  {title} {\enquote {\bibinfo {title} {Excitonic insulator},}\ }\href {\doibase 10.1103/PhysRev.158.462} {\bibfield  {journal} {\bibinfo  {journal} {Phys. Rev.}\ }\textbf {\bibinfo {volume} {158}},\ \bibinfo {pages} {462--475} (\bibinfo {year} {1967})}\BibitemShut {NoStop}%
\bibitem [{\citenamefont {Kohn}(1967)}]{Kohn1967}%
  \BibitemOpen
  \bibfield  {author} {\bibinfo {author} {\bibfnamefont {W.}~\bibnamefont {Kohn}},\ }\bibfield  {title} {\enquote {\bibinfo {title} {Excitonic phases},}\ }\href {\doibase 10.1103/PhysRevLett.19.439} {\bibfield  {journal} {\bibinfo  {journal} {Phys. Rev. Lett.}\ }\textbf {\bibinfo {volume} {19}},\ \bibinfo {pages} {439--442} (\bibinfo {year} {1967})}\BibitemShut {NoStop}%
\bibitem [{\citenamefont {Butov}\ \emph {et~al.}(1994)\citenamefont {Butov}, \citenamefont {Zrenner}, \citenamefont {Abstreiter}, \citenamefont {B\"ohm},\ and\ \citenamefont {Weimann}}]{Butov1994}%
  \BibitemOpen
  \bibfield  {author} {\bibinfo {author} {\bibfnamefont {L.~V.}\ \bibnamefont {Butov}}, \bibinfo {author} {\bibfnamefont {A.}~\bibnamefont {Zrenner}}, \bibinfo {author} {\bibfnamefont {G.}~\bibnamefont {Abstreiter}}, \bibinfo {author} {\bibfnamefont {G.}~\bibnamefont {B\"ohm}}, \ and\ \bibinfo {author} {\bibfnamefont {G.}~\bibnamefont {Weimann}},\ }\bibfield  {title} {\enquote {\bibinfo {title} {Condensation of indirect excitons in coupled alas/gaas quantum wells},}\ }\href {\doibase 10.1103/PhysRevLett.73.304} {\bibfield  {journal} {\bibinfo  {journal} {Phys. Rev. Lett.}\ }\textbf {\bibinfo {volume} {73}},\ \bibinfo {pages} {304--307} (\bibinfo {year} {1994})}\BibitemShut {NoStop}%
\bibitem [{\citenamefont {Khveshchenko}(2001)}]{Khveshchenko2001}%
  \BibitemOpen
  \bibfield  {author} {\bibinfo {author} {\bibfnamefont {D.~V.}\ \bibnamefont {Khveshchenko}},\ }\bibfield  {title} {\enquote {\bibinfo {title} {Ghost excitonic insulator transition in layered graphite},}\ }\href {\doibase 10.1103/PhysRevLett.87.246802} {\bibfield  {journal} {\bibinfo  {journal} {Phys. Rev. Lett.}\ }\textbf {\bibinfo {volume} {87}},\ \bibinfo {pages} {246802} (\bibinfo {year} {2001})}\BibitemShut {NoStop}%
\bibitem [{\citenamefont {Eisenstein}\ and\ \citenamefont {MacDonald}(2004)}]{Eisenstein2004}%
  \BibitemOpen
  \bibfield  {author} {\bibinfo {author} {\bibfnamefont {J.~P.}\ \bibnamefont {Eisenstein}}\ and\ \bibinfo {author} {\bibfnamefont {A.~H.}\ \bibnamefont {MacDonald}},\ }\bibfield  {title} {\enquote {\bibinfo {title} {Bose--einstein condensation of excitons in bilayer electron systems},}\ }\href {\doibase 10.1038/nature03081} {\bibfield  {journal} {\bibinfo  {journal} {Nature}\ }\textbf {\bibinfo {volume} {432}},\ \bibinfo {pages} {691--694} (\bibinfo {year} {2004})}\BibitemShut {NoStop}%
\bibitem [{\citenamefont {Cercellier}\ \emph {et~al.}(2007)\citenamefont {Cercellier}, \citenamefont {Monney}, \citenamefont {Clerc}, \citenamefont {Battaglia}, \citenamefont {Despont}, \citenamefont {Garnier}, \citenamefont {Beck}, \citenamefont {Aebi}, \citenamefont {Patthey}, \citenamefont {Berger},\ and\ \citenamefont {Forr\'o}}]{Cercellier2007}%
  \BibitemOpen
  \bibfield  {author} {\bibinfo {author} {\bibfnamefont {H.}~\bibnamefont {Cercellier}}, \bibinfo {author} {\bibfnamefont {C.}~\bibnamefont {Monney}}, \bibinfo {author} {\bibfnamefont {F.}~\bibnamefont {Clerc}}, \bibinfo {author} {\bibfnamefont {C.}~\bibnamefont {Battaglia}}, \bibinfo {author} {\bibfnamefont {L.}~\bibnamefont {Despont}}, \bibinfo {author} {\bibfnamefont {M.~G.}\ \bibnamefont {Garnier}}, \bibinfo {author} {\bibfnamefont {H.}~\bibnamefont {Beck}}, \bibinfo {author} {\bibfnamefont {P.}~\bibnamefont {Aebi}}, \bibinfo {author} {\bibfnamefont {L.}~\bibnamefont {Patthey}}, \bibinfo {author} {\bibfnamefont {H.}~\bibnamefont {Berger}}, \ and\ \bibinfo {author} {\bibfnamefont {L.}~\bibnamefont {Forr\'o}},\ }\bibfield  {title} {\enquote {\bibinfo {title} {Evidence for an excitonic insulator phase in $1t\mathrm{\text{\ensuremath{-}}}{\mathrm{tise}}_{2}$},}\ }\href {\doibase 10.1103/PhysRevLett.99.146403} {\bibfield  {journal} {\bibinfo  {journal} {Phys. Rev. Lett.}\ }\textbf {\bibinfo {volume} {99}},\
  \bibinfo {pages} {146403} (\bibinfo {year} {2007})}\BibitemShut {NoStop}%
\bibitem [{\citenamefont {Min}\ \emph {et~al.}(2008)\citenamefont {Min}, \citenamefont {Bistritzer}, \citenamefont {Su},\ and\ \citenamefont {MacDonald}}]{Min2008}%
  \BibitemOpen
  \bibfield  {author} {\bibinfo {author} {\bibfnamefont {Hongki}\ \bibnamefont {Min}}, \bibinfo {author} {\bibfnamefont {Rafi}\ \bibnamefont {Bistritzer}}, \bibinfo {author} {\bibfnamefont {Jung-Jung}\ \bibnamefont {Su}}, \ and\ \bibinfo {author} {\bibfnamefont {A.~H.}\ \bibnamefont {MacDonald}},\ }\bibfield  {title} {\enquote {\bibinfo {title} {Room-temperature superfluidity in graphene bilayers},}\ }\href {\doibase 10.1103/PhysRevB.78.121401} {\bibfield  {journal} {\bibinfo  {journal} {Phys. Rev. B}\ }\textbf {\bibinfo {volume} {78}},\ \bibinfo {pages} {121401} (\bibinfo {year} {2008})}\BibitemShut {NoStop}%
\bibitem [{\citenamefont {Kogar}\ \emph {et~al.}(2017)\citenamefont {Kogar}, \citenamefont {Rak}, \citenamefont {Vig}, \citenamefont {Husain}, \citenamefont {Flicker}, \citenamefont {Joe}, \citenamefont {Venema}, \citenamefont {MacDougall}, \citenamefont {Chiang}, \citenamefont {Fradkin}, \citenamefont {van Wezel},\ and\ \citenamefont {Abbamonte}}]{Anshul2017}%
  \BibitemOpen
  \bibfield  {author} {\bibinfo {author} {\bibfnamefont {Anshul}\ \bibnamefont {Kogar}}, \bibinfo {author} {\bibfnamefont {Melinda~S.}\ \bibnamefont {Rak}}, \bibinfo {author} {\bibfnamefont {Sean}\ \bibnamefont {Vig}}, \bibinfo {author} {\bibfnamefont {Ali~A.}\ \bibnamefont {Husain}}, \bibinfo {author} {\bibfnamefont {Felix}\ \bibnamefont {Flicker}}, \bibinfo {author} {\bibfnamefont {Young~Il}\ \bibnamefont {Joe}}, \bibinfo {author} {\bibfnamefont {Luc}\ \bibnamefont {Venema}}, \bibinfo {author} {\bibfnamefont {Greg~J.}\ \bibnamefont {MacDougall}}, \bibinfo {author} {\bibfnamefont {Tai~C.}\ \bibnamefont {Chiang}}, \bibinfo {author} {\bibfnamefont {Eduardo}\ \bibnamefont {Fradkin}}, \bibinfo {author} {\bibfnamefont {Jasper}\ \bibnamefont {van Wezel}}, \ and\ \bibinfo {author} {\bibfnamefont {Peter}\ \bibnamefont {Abbamonte}},\ }\bibfield  {title} {\enquote {\bibinfo {title} {Signatures of exciton condensation in a transition metal dichalcogenide},}\ }\href {\doibase 10.1126/science.aam6432} {\bibfield
  {journal} {\bibinfo  {journal} {Science}\ }\textbf {\bibinfo {volume} {358}},\ \bibinfo {pages} {1314--1317} (\bibinfo {year} {2017})},\ \Eprint {http://arxiv.org/abs/https://www.science.org/doi/pdf/10.1126/science.aam6432} {https://www.science.org/doi/pdf/10.1126/science.aam6432} \BibitemShut {NoStop}%
\bibitem [{\citenamefont {Li}\ \emph {et~al.}(2017)\citenamefont {Li}, \citenamefont {Taniguchi}, \citenamefont {Watanabe}, \citenamefont {Hone},\ and\ \citenamefont {Dean}}]{Li2017}%
  \BibitemOpen
  \bibfield  {author} {\bibinfo {author} {\bibfnamefont {J.~I.~A.}\ \bibnamefont {Li}}, \bibinfo {author} {\bibfnamefont {T.}~\bibnamefont {Taniguchi}}, \bibinfo {author} {\bibfnamefont {K.}~\bibnamefont {Watanabe}}, \bibinfo {author} {\bibfnamefont {J.}~\bibnamefont {Hone}}, \ and\ \bibinfo {author} {\bibfnamefont {C.~R.}\ \bibnamefont {Dean}},\ }\bibfield  {title} {\enquote {\bibinfo {title} {Excitonic superfluid phase in double bilayer graphene},}\ }\href {\doibase 10.1038/nphys4140} {\bibfield  {journal} {\bibinfo  {journal} {Nature Physics}\ }\textbf {\bibinfo {volume} {13}},\ \bibinfo {pages} {751--755} (\bibinfo {year} {2017})}\BibitemShut {NoStop}%
\bibitem [{\citenamefont {Wang}\ \emph {et~al.}(2019{\natexlab{b}})\citenamefont {Wang}, \citenamefont {Rhodes}, \citenamefont {Watanabe}, \citenamefont {Taniguchi}, \citenamefont {Hone}, \citenamefont {Shan},\ and\ \citenamefont {Mak}}]{WangZefang2019}%
  \BibitemOpen
  \bibfield  {author} {\bibinfo {author} {\bibfnamefont {Zefang}\ \bibnamefont {Wang}}, \bibinfo {author} {\bibfnamefont {Daniel~A.}\ \bibnamefont {Rhodes}}, \bibinfo {author} {\bibfnamefont {Kenji}\ \bibnamefont {Watanabe}}, \bibinfo {author} {\bibfnamefont {Takashi}\ \bibnamefont {Taniguchi}}, \bibinfo {author} {\bibfnamefont {James~C.}\ \bibnamefont {Hone}}, \bibinfo {author} {\bibfnamefont {Jie}\ \bibnamefont {Shan}}, \ and\ \bibinfo {author} {\bibfnamefont {Kin~Fai}\ \bibnamefont {Mak}},\ }\bibfield  {title} {\enquote {\bibinfo {title} {Evidence of high-temperature exciton condensation in two-dimensional atomic double layers},}\ }\href {\doibase 10.1038/s41586-019-1591-7} {\bibfield  {journal} {\bibinfo  {journal} {Nature}\ }\textbf {\bibinfo {volume} {574}},\ \bibinfo {pages} {76--80} (\bibinfo {year} {2019}{\natexlab{b}})}\BibitemShut {NoStop}%
\bibitem [{\citenamefont {Jauregui}\ \emph {et~al.}(2019)\citenamefont {Jauregui}, \citenamefont {Joe}, \citenamefont {Pistunova}, \citenamefont {Wild}, \citenamefont {High}, \citenamefont {Zhou}, \citenamefont {Scuri}, \citenamefont {Greve}, \citenamefont {Sushko}, \citenamefont {Yu}, \citenamefont {Taniguchi}, \citenamefont {Watanabe}, \citenamefont {Needleman}, \citenamefont {Lukin}, \citenamefont {Park},\ and\ \citenamefont {Kim}}]{Luis2019}%
  \BibitemOpen
  \bibfield  {author} {\bibinfo {author} {\bibfnamefont {Luis~A.}\ \bibnamefont {Jauregui}}, \bibinfo {author} {\bibfnamefont {Andrew~Y.}\ \bibnamefont {Joe}}, \bibinfo {author} {\bibfnamefont {Kateryna}\ \bibnamefont {Pistunova}}, \bibinfo {author} {\bibfnamefont {Dominik~S.}\ \bibnamefont {Wild}}, \bibinfo {author} {\bibfnamefont {Alexander~A.}\ \bibnamefont {High}}, \bibinfo {author} {\bibfnamefont {You}\ \bibnamefont {Zhou}}, \bibinfo {author} {\bibfnamefont {Giovanni}\ \bibnamefont {Scuri}}, \bibinfo {author} {\bibfnamefont {Kristiaan~De}\ \bibnamefont {Greve}}, \bibinfo {author} {\bibfnamefont {Andrey}\ \bibnamefont {Sushko}}, \bibinfo {author} {\bibfnamefont {Che-Hang}\ \bibnamefont {Yu}}, \bibinfo {author} {\bibfnamefont {Takashi}\ \bibnamefont {Taniguchi}}, \bibinfo {author} {\bibfnamefont {Kenji}\ \bibnamefont {Watanabe}}, \bibinfo {author} {\bibfnamefont {Daniel~J.}\ \bibnamefont {Needleman}}, \bibinfo {author} {\bibfnamefont {Mikhail~D.}\ \bibnamefont {Lukin}}, \bibinfo {author} {\bibfnamefont
  {Hongkun}\ \bibnamefont {Park}}, \ and\ \bibinfo {author} {\bibfnamefont {Philip}\ \bibnamefont {Kim}},\ }\bibfield  {title} {\enquote {\bibinfo {title} {Electrical control of interlayer exciton dynamics in atomically thin heterostructures},}\ }\href {\doibase 10.1126/science.aaw4194} {\bibfield  {journal} {\bibinfo  {journal} {Science}\ }\textbf {\bibinfo {volume} {366}},\ \bibinfo {pages} {870--875} (\bibinfo {year} {2019})},\ \Eprint {http://arxiv.org/abs/https://www.science.org/doi/pdf/10.1126/science.aaw4194} {https://www.science.org/doi/pdf/10.1126/science.aaw4194} \BibitemShut {NoStop}%
\bibitem [{\citenamefont {Ma}\ \emph {et~al.}(2021)\citenamefont {Ma}, \citenamefont {Nguyen}, \citenamefont {Wang}, \citenamefont {Zeng}, \citenamefont {Watanabe}, \citenamefont {Taniguchi}, \citenamefont {MacDonald}, \citenamefont {Mak},\ and\ \citenamefont {Shan}}]{Ma2021}%
  \BibitemOpen
  \bibfield  {author} {\bibinfo {author} {\bibfnamefont {Liguo}\ \bibnamefont {Ma}}, \bibinfo {author} {\bibfnamefont {Phuong~X.}\ \bibnamefont {Nguyen}}, \bibinfo {author} {\bibfnamefont {Zefang}\ \bibnamefont {Wang}}, \bibinfo {author} {\bibfnamefont {Yongxin}\ \bibnamefont {Zeng}}, \bibinfo {author} {\bibfnamefont {Kenji}\ \bibnamefont {Watanabe}}, \bibinfo {author} {\bibfnamefont {Takashi}\ \bibnamefont {Taniguchi}}, \bibinfo {author} {\bibfnamefont {Allan~H.}\ \bibnamefont {MacDonald}}, \bibinfo {author} {\bibfnamefont {Kin~Fai}\ \bibnamefont {Mak}}, \ and\ \bibinfo {author} {\bibfnamefont {Jie}\ \bibnamefont {Shan}},\ }\bibfield  {title} {\enquote {\bibinfo {title} {Strongly correlated excitonic insulator in atomic double layers},}\ }\href {\doibase 10.1038/s41586-021-03947-9} {\bibfield  {journal} {\bibinfo  {journal} {Nature}\ }\textbf {\bibinfo {volume} {598}},\ \bibinfo {pages} {585--589} (\bibinfo {year} {2021})}\BibitemShut {NoStop}%
\bibitem [{\citenamefont {Huang}\ \emph {et~al.}(2024)\citenamefont {Huang}, \citenamefont {Jiang}, \citenamefont {Yao}, \citenamefont {Yan}, \citenamefont {Lei}, \citenamefont {Gao}, \citenamefont {Guo}, \citenamefont {Jin}, \citenamefont {Li}, \citenamefont {Yuan}, \citenamefont {Chai}, \citenamefont {Sheng}, \citenamefont {Pan}, \citenamefont {Chen}, \citenamefont {Liu}, \citenamefont {Gao}, \citenamefont {Qu}, \citenamefont {Liu}, \citenamefont {Jiang}, \citenamefont {Liu}, \citenamefont {Ma}, \citenamefont {Zhou}, \citenamefont {Huang}, \citenamefont {Yun}, \citenamefont {Zhang}, \citenamefont {Li}, \citenamefont {Jin}, \citenamefont {Ding}, \citenamefont {Shen}, \citenamefont {Su}, \citenamefont {Shi}, \citenamefont {Wang},\ and\ \citenamefont {Qian}}]{Huang2024}%
  \BibitemOpen
  \bibfield  {author} {\bibinfo {author} {\bibfnamefont {Jierui}\ \bibnamefont {Huang}}, \bibinfo {author} {\bibfnamefont {Bei}\ \bibnamefont {Jiang}}, \bibinfo {author} {\bibfnamefont {Jingyu}\ \bibnamefont {Yao}}, \bibinfo {author} {\bibfnamefont {Dayu}\ \bibnamefont {Yan}}, \bibinfo {author} {\bibfnamefont {Xincheng}\ \bibnamefont {Lei}}, \bibinfo {author} {\bibfnamefont {Jiacheng}\ \bibnamefont {Gao}}, \bibinfo {author} {\bibfnamefont {Zhaopeng}\ \bibnamefont {Guo}}, \bibinfo {author} {\bibfnamefont {Feng}\ \bibnamefont {Jin}}, \bibinfo {author} {\bibfnamefont {Yupeng}\ \bibnamefont {Li}}, \bibinfo {author} {\bibfnamefont {Zhenyu}\ \bibnamefont {Yuan}}, \bibinfo {author} {\bibfnamefont {Congcong}\ \bibnamefont {Chai}}, \bibinfo {author} {\bibfnamefont {Haohao}\ \bibnamefont {Sheng}}, \bibinfo {author} {\bibfnamefont {Mojun}\ \bibnamefont {Pan}}, \bibinfo {author} {\bibfnamefont {Famin}\ \bibnamefont {Chen}}, \bibinfo {author} {\bibfnamefont {Junde}\ \bibnamefont {Liu}}, \bibinfo {author} {\bibfnamefont
  {Shunye}\ \bibnamefont {Gao}}, \bibinfo {author} {\bibfnamefont {Gexing}\ \bibnamefont {Qu}}, \bibinfo {author} {\bibfnamefont {Bo}~\bibnamefont {Liu}}, \bibinfo {author} {\bibfnamefont {Zhicheng}\ \bibnamefont {Jiang}}, \bibinfo {author} {\bibfnamefont {Zhengtai}\ \bibnamefont {Liu}}, \bibinfo {author} {\bibfnamefont {Xiaoyan}\ \bibnamefont {Ma}}, \bibinfo {author} {\bibfnamefont {Shiming}\ \bibnamefont {Zhou}}, \bibinfo {author} {\bibfnamefont {Yaobo}\ \bibnamefont {Huang}}, \bibinfo {author} {\bibfnamefont {Chenxia}\ \bibnamefont {Yun}}, \bibinfo {author} {\bibfnamefont {Qingming}\ \bibnamefont {Zhang}}, \bibinfo {author} {\bibfnamefont {Shiliang}\ \bibnamefont {Li}}, \bibinfo {author} {\bibfnamefont {Shifeng}\ \bibnamefont {Jin}}, \bibinfo {author} {\bibfnamefont {Hong}\ \bibnamefont {Ding}}, \bibinfo {author} {\bibfnamefont {Jie}\ \bibnamefont {Shen}}, \bibinfo {author} {\bibfnamefont {Dong}\ \bibnamefont {Su}}, \bibinfo {author} {\bibfnamefont {Youguo}\ \bibnamefont {Shi}}, \bibinfo {author}
  {\bibfnamefont {Zhijun}\ \bibnamefont {Wang}}, \ and\ \bibinfo {author} {\bibfnamefont {Tian}\ \bibnamefont {Qian}},\ }\bibfield  {title} {\enquote {\bibinfo {title} {Evidence for an excitonic insulator state in ${\mathrm{ta}}_{2}{\mathrm{pd}}_{3}{\mathrm{te}}_{5}$},}\ }\href {\doibase 10.1103/PhysRevX.14.011046} {\bibfield  {journal} {\bibinfo  {journal} {Phys. Rev. X}\ }\textbf {\bibinfo {volume} {14}},\ \bibinfo {pages} {011046} (\bibinfo {year} {2024})}\BibitemShut {NoStop}%
\bibitem [{\citenamefont {Balatsky}\ \emph {et~al.}(2004)\citenamefont {Balatsky}, \citenamefont {Joglekar},\ and\ \citenamefont {Littlewood}}]{Balatsky2004}%
  \BibitemOpen
  \bibfield  {author} {\bibinfo {author} {\bibfnamefont {Alexander~V.}\ \bibnamefont {Balatsky}}, \bibinfo {author} {\bibfnamefont {Yogesh~N.}\ \bibnamefont {Joglekar}}, \ and\ \bibinfo {author} {\bibfnamefont {Peter~B.}\ \bibnamefont {Littlewood}},\ }\bibfield  {title} {\enquote {\bibinfo {title} {Dipolar superfluidity in electron-hole bilayer systems},}\ }\href {\doibase 10.1103/PhysRevLett.93.266801} {\bibfield  {journal} {\bibinfo  {journal} {Phys. Rev. Lett.}\ }\textbf {\bibinfo {volume} {93}},\ \bibinfo {pages} {266801} (\bibinfo {year} {2004})}\BibitemShut {NoStop}%
\bibitem [{\citenamefont {Dubinkin}\ \emph {et~al.}(2021)\citenamefont {Dubinkin}, \citenamefont {May-Mann},\ and\ \citenamefont {Hughes}}]{Dubinkin2021}%
  \BibitemOpen
  \bibfield  {author} {\bibinfo {author} {\bibfnamefont {Oleg}\ \bibnamefont {Dubinkin}}, \bibinfo {author} {\bibfnamefont {Julian}\ \bibnamefont {May-Mann}}, \ and\ \bibinfo {author} {\bibfnamefont {Taylor~L.}\ \bibnamefont {Hughes}},\ }\bibfield  {title} {\enquote {\bibinfo {title} {Theory of dipole insulators},}\ }\href {\doibase 10.1103/PhysRevB.103.125129} {\bibfield  {journal} {\bibinfo  {journal} {Phys. Rev. B}\ }\textbf {\bibinfo {volume} {103}},\ \bibinfo {pages} {125129} (\bibinfo {year} {2021})}\BibitemShut {NoStop}%
\bibitem [{\citenamefont {Jiang}\ \emph {et~al.}(2015)\citenamefont {Jiang}, \citenamefont {Bao}, \citenamefont {Sun},\ and\ \citenamefont {Xie}}]{Jiang2015}%
  \BibitemOpen
  \bibfield  {author} {\bibinfo {author} {\bibfnamefont {Qing-Dong}\ \bibnamefont {Jiang}}, \bibinfo {author} {\bibfnamefont {Zhi-qiang}\ \bibnamefont {Bao}}, \bibinfo {author} {\bibfnamefont {Qing-Feng}\ \bibnamefont {Sun}}, \ and\ \bibinfo {author} {\bibfnamefont {X.~C.}\ \bibnamefont {Xie}},\ }\bibfield  {title} {\enquote {\bibinfo {title} {Theory for electric dipole superconductivity with an application for bilayer excitons},}\ }\href {\doibase 10.1038/srep11925} {\bibfield  {journal} {\bibinfo  {journal} {Scientific Reports}\ }\textbf {\bibinfo {volume} {5}},\ \bibinfo {pages} {11925} (\bibinfo {year} {2015})}\BibitemShut {NoStop}%
\bibitem [{\citenamefont {He}\ and\ \citenamefont {McKellar}(1993)}]{HeMcKellar1993}%
  \BibitemOpen
  \bibfield  {author} {\bibinfo {author} {\bibfnamefont {Xiao-Gang}\ \bibnamefont {He}}\ and\ \bibinfo {author} {\bibfnamefont {Bruce H.~J.}\ \bibnamefont {McKellar}},\ }\bibfield  {title} {\enquote {\bibinfo {title} {Topological phase due to electric dipole moment and magnetic monopole interaction},}\ }\href {\doibase 10.1103/PhysRevA.47.3424} {\bibfield  {journal} {\bibinfo  {journal} {Phys. Rev. A}\ }\textbf {\bibinfo {volume} {47}},\ \bibinfo {pages} {3424--3425} (\bibinfo {year} {1993})}\BibitemShut {NoStop}%
\bibitem [{\citenamefont {Wilkens}(1994)}]{Wilkens1994}%
  \BibitemOpen
  \bibfield  {author} {\bibinfo {author} {\bibfnamefont {Martin}\ \bibnamefont {Wilkens}},\ }\bibfield  {title} {\enquote {\bibinfo {title} {Quantum phase of a moving dipole},}\ }\href {\doibase 10.1103/PhysRevLett.72.5} {\bibfield  {journal} {\bibinfo  {journal} {Phys. Rev. Lett.}\ }\textbf {\bibinfo {volume} {72}},\ \bibinfo {pages} {5--8} (\bibinfo {year} {1994})}\BibitemShut {NoStop}%
\bibitem [{\citenamefont {Tinkham}(1996)}]{tinkhambook}%
  \BibitemOpen
  \bibfield  {author} {\bibinfo {author} {\bibfnamefont {M.}~\bibnamefont {Tinkham}},\ }\href@noop {} {\emph {\bibinfo {title} {Introduction to Superconductivity}}}\ (\bibinfo  {publisher} {Dover},\ \bibinfo {year} {1996})\BibitemShut {NoStop}%
\bibitem [{\citenamefont {Wei}\ \emph {et~al.}(1995)\citenamefont {Wei}, \citenamefont {Han},\ and\ \citenamefont {Wei}}]{Wei1995}%
  \BibitemOpen
  \bibfield  {author} {\bibinfo {author} {\bibfnamefont {Haiqing}\ \bibnamefont {Wei}}, \bibinfo {author} {\bibfnamefont {Rushan}\ \bibnamefont {Han}}, \ and\ \bibinfo {author} {\bibfnamefont {Xiuqing}\ \bibnamefont {Wei}},\ }\bibfield  {title} {\enquote {\bibinfo {title} {Quantum phase of induced dipoles moving in a magnetic field},}\ }\href {\doibase 10.1103/PhysRevLett.75.2071} {\bibfield  {journal} {\bibinfo  {journal} {Phys. Rev. Lett.}\ }\textbf {\bibinfo {volume} {75}},\ \bibinfo {pages} {2071--2073} (\bibinfo {year} {1995})}\BibitemShut {NoStop}%
\bibitem [{\citenamefont {Dowling}\ \emph {et~al.}(1999)\citenamefont {Dowling}, \citenamefont {Williams},\ and\ \citenamefont {Franson}}]{Dowling1999}%
  \BibitemOpen
  \bibfield  {author} {\bibinfo {author} {\bibfnamefont {Jonathan~P.}\ \bibnamefont {Dowling}}, \bibinfo {author} {\bibfnamefont {Colin~P.}\ \bibnamefont {Williams}}, \ and\ \bibinfo {author} {\bibfnamefont {J.~D.}\ \bibnamefont {Franson}},\ }\bibfield  {title} {\enquote {\bibinfo {title} {Maxwell duality, lorentz invariance, and topological phase},}\ }\href {\doibase 10.1103/PhysRevLett.83.2486} {\bibfield  {journal} {\bibinfo  {journal} {Phys. Rev. Lett.}\ }\textbf {\bibinfo {volume} {83}},\ \bibinfo {pages} {2486--2489} (\bibinfo {year} {1999})}\BibitemShut {NoStop}%
\bibitem [{\citenamefont {Lepoutre}\ \emph {et~al.}(2012)\citenamefont {Lepoutre}, \citenamefont {Gauguet}, \citenamefont {Tr\'enec}, \citenamefont {B\"uchner},\ and\ \citenamefont {Vigu\'e}}]{Lepoutre2012}%
  \BibitemOpen
  \bibfield  {author} {\bibinfo {author} {\bibfnamefont {S.}~\bibnamefont {Lepoutre}}, \bibinfo {author} {\bibfnamefont {A.}~\bibnamefont {Gauguet}}, \bibinfo {author} {\bibfnamefont {G.}~\bibnamefont {Tr\'enec}}, \bibinfo {author} {\bibfnamefont {M.}~\bibnamefont {B\"uchner}}, \ and\ \bibinfo {author} {\bibfnamefont {J.}~\bibnamefont {Vigu\'e}},\ }\bibfield  {title} {\enquote {\bibinfo {title} {He-mckellar-wilkens topological phase in atom interferometry},}\ }\href {\doibase 10.1103/PhysRevLett.109.120404} {\bibfield  {journal} {\bibinfo  {journal} {Phys. Rev. Lett.}\ }\textbf {\bibinfo {volume} {109}},\ \bibinfo {pages} {120404} (\bibinfo {year} {2012})}\BibitemShut {NoStop}%
\bibitem [{\citenamefont {Chen}\ \emph {et~al.}(2013)\citenamefont {Chen}, \citenamefont {Horsch},\ and\ \citenamefont {Manske}}]{Chen2013}%
  \BibitemOpen
  \bibfield  {author} {\bibinfo {author} {\bibfnamefont {Wei}\ \bibnamefont {Chen}}, \bibinfo {author} {\bibfnamefont {Peter}\ \bibnamefont {Horsch}}, \ and\ \bibinfo {author} {\bibfnamefont {Dirk}\ \bibnamefont {Manske}},\ }\bibfield  {title} {\enquote {\bibinfo {title} {Flux quantization due to monopole and dipole currents},}\ }\href {\doibase 10.1103/PhysRevB.87.214502} {\bibfield  {journal} {\bibinfo  {journal} {Phys. Rev. B}\ }\textbf {\bibinfo {volume} {87}},\ \bibinfo {pages} {214502} (\bibinfo {year} {2013})}\BibitemShut {NoStop}%
\bibitem [{\citenamefont {Wood}\ \emph {et~al.}(2016)\citenamefont {Wood}, \citenamefont {McKellar},\ and\ \citenamefont {Martin}}]{Wood2016}%
  \BibitemOpen
  \bibfield  {author} {\bibinfo {author} {\bibfnamefont {A.~A.}\ \bibnamefont {Wood}}, \bibinfo {author} {\bibfnamefont {B.~H.~J.}\ \bibnamefont {McKellar}}, \ and\ \bibinfo {author} {\bibfnamefont {A.~M.}\ \bibnamefont {Martin}},\ }\bibfield  {title} {\enquote {\bibinfo {title} {Persistent superfluid flow arising from the he-mckellar-wilkens effect in molecular dipolar condensates},}\ }\href {\doibase 10.1103/PhysRevLett.116.250403} {\bibfield  {journal} {\bibinfo  {journal} {Phys. Rev. Lett.}\ }\textbf {\bibinfo {volume} {116}},\ \bibinfo {pages} {250403} (\bibinfo {year} {2016})}\BibitemShut {NoStop}%
\bibitem [{Note1()}]{Note1}%
  \BibitemOpen
  \bibinfo {note} {Note that the negative sign in $\vb {M}$ in Eq.~\protect \eqref {equation:M_vs_j} cancels with the negative sign in $\vb {j}_\protect \mathrm {p} \propto -\vb {A}_\protect \mathrm {p}$}\BibitemShut {NoStop}%
\bibitem [{\citenamefont {Gilleo}\ and\ \citenamefont {Geller}(1958)}]{Gilleo1958}%
  \BibitemOpen
  \bibfield  {author} {\bibinfo {author} {\bibfnamefont {M.~A.}\ \bibnamefont {Gilleo}}\ and\ \bibinfo {author} {\bibfnamefont {S.}~\bibnamefont {Geller}},\ }\bibfield  {title} {\enquote {\bibinfo {title} {Magnetic and crystallographic properties of substituted yttrium-iron garnet, $3{\mathrm{y}}_{2}{\mathrm{o}}_{3}\ifmmode\cdot\else\textperiodcentered\fi{}x{\mathrm{m}}_{2}{\mathrm{o}}_{3}\ifmmode\cdot\else\textperiodcentered\fi{}(5\ensuremath{-}x){\mathrm{fe}}_{2}{\mathrm{o}}_{3}$},}\ }\href {\doibase 10.1103/PhysRev.110.73} {\bibfield  {journal} {\bibinfo  {journal} {Phys. Rev.}\ }\textbf {\bibinfo {volume} {110}},\ \bibinfo {pages} {73--78} (\bibinfo {year} {1958})}\BibitemShut {NoStop}%
\bibitem [{\citenamefont {Tupitsyn}\ \emph {et~al.}(2008)\citenamefont {Tupitsyn}, \citenamefont {Stamp},\ and\ \citenamefont {Burin}}]{Tupitsyn2008}%
  \BibitemOpen
  \bibfield  {author} {\bibinfo {author} {\bibfnamefont {I.~S.}\ \bibnamefont {Tupitsyn}}, \bibinfo {author} {\bibfnamefont {P.~C.~E.}\ \bibnamefont {Stamp}}, \ and\ \bibinfo {author} {\bibfnamefont {A.~L.}\ \bibnamefont {Burin}},\ }\bibfield  {title} {\enquote {\bibinfo {title} {Stability of bose-einstein condensates of hot magnons in yttrium iron garnet films},}\ }\href {\doibase 10.1103/PhysRevLett.100.257202} {\bibfield  {journal} {\bibinfo  {journal} {Phys. Rev. Lett.}\ }\textbf {\bibinfo {volume} {100}},\ \bibinfo {pages} {257202} (\bibinfo {year} {2008})}\BibitemShut {NoStop}%
\end{thebibliography}%


\end{document}